\documentclass{jfm}

\usepackage{graphicx}
\usepackage{newtxtext}
\usepackage{newtxmath}
\usepackage{natbib}
\usepackage{hyperref}
\usepackage{xcolor}
\usepackage[normalem]{ulem}

\hypersetup{
    colorlinks = true,
    urlcolor   = blue,
    citecolor  = black,
}

\newcommand{\RomanNumeralCaps}[1]
\linenumbers

\title{Energy transfer and budget analysis for transient process with phase-averaged reduced-order model}

\author{Yuto Nakamura
  \corresp{\email{yuto.nakamura.t4@dc.tohoku.ac.jp}},
  Yuma Kuroda,
  Shintaro Sato,
 \and Naofumi Ohnishi}

\affiliation{Department of Aerospace Engineering, Tohoku University,
Sendai, 980-8579, Japan}

\begin{document}
\maketitle

\begin{abstract} 
We derive a phase-averaged representation of transient flows based on the eigenmodes of a data-driven linear operator that approximates the Navier–Stokes dynamics.  In performing phase averaging, it is assumed that, at each instant during the transient evolution, the eigenmode amplitude remains invariant, while only the complex phase angle differs among distinct realizations of the transient process. From this modal-phase perspective, the linear operator is defined as the best-fit operator that represents phase-different transient evolutions. By introducing a time-varying dynamic mode decomposition with a phase-control strategy formulated from this modal-phase perspective, time-varying eigenmodes are extracted from numerical simulations. In this formulation, the transient process is decomposed into time-varying eigenmodes, phase-shift angles, and amplitude coefficients. Furthermore, by averaging the Navier–Stokes equations over the phase-shift angle, a frequency-domain form of the equations can be derived at any given instant, assuming that the phase-shift angle is time-independent. This frequency-domain representation reveals the instantaneous energy budget and the presence of energy transfer through triadic interactions. The proposed analysis is demonstrated using a canonical example of two-dimensional flow around a circular cylinder transitioning from a steady to an unsteady state. The time-varying dynamic mode decomposition with phase control is shown to capture the transient evolution of the frequency components accurately. In addition, the temporal evolution of the energy budget and transfer distribution reveals that transient growth processes exhibit different time-dependent characteristics of energy transfer, even in cylinder flows at Reynolds numbers that eventually lead to a periodic state.
\end{abstract}



\section{Introduction}
Energy transfer due to the nonlinearity and viscous diffusion effects plays a fundamental role in fluid dynamics, governing phenomena such as vortex dynamics \citep{Biswas_2024, freeman_2023, mypaper4}, bifurcation instability \citep{shift, Deng_2019, Mittal_2009}, transient development \citep{Zhong_2025, ballouz2024transient}, and turbulence \citep{yamada1991identification, Nekkanti_2025, yeung_2025}. 
Consequently, the budget of energy diffusion and transfer across different scales has been investigated through various approaches. 
Recent advances in data-driven science \citep{POD1,POD15,schmidt_2020,SPOD,DMD,fukami2024data} have enabled the extraction of coherent structures embedded in complex fluid flows and their decomposition into different scales. 
Furthermore, these advancements have clarified the energy transfer mechanisms between different scales. 
However, for transient flows undergoing nonlinear development, even decomposing flow structures into distinct scales remains a challenge. This paper presents the decomposition of transient flow into components of different scales, characterized by their frequencies, and analyzes the energy budget, considering both viscous diffusion and energy transfer. 

Several modal decomposition methods exist for extracting coherent structures that reveal energy transfer and the energy budget in fluid flow. 
One approach is to decompose the flow field into orthonormal modes. 
By projecting the governing equations onto a low-dimensional subspace spanned by these orthonormal modes, the viscous diffusion of the modes and their interactions can be studied. 
This approach is referred to as the Galerkin projection approach or Galerkin model \citep{Noack_1994, POD_Galerkin}, and the low-dimensional model is referred to as a reduced-order model (ROM). 
A well-known method for finding an orthonormal basis is proper orthogonal decomposition (POD) \citep{POD2,Holmes,lumley1967structure}. 
POD extracts the most energetic modes from a flow dataset. \citet{shift} demonstrated that the energy budget of each mode can be computed in a projection-based ROM using POD modes for the periodic flow around a two-dimensional cylinder. 
However, POD assumes a simple representation in terms of orthonormal bases, which means that a single mode may contain multiple frequency components. 
This limitation makes POD unsuitable for analyzing energy transfer between different frequencies.

Because of its ability to decompose into frequency-wise modes, decomposition in the spectral domain using wavelet analysis \citep{morlet1982wave,goupillaud1984cycle,Rinoshika_2020,ballouz2024transient} and Fourier decomposition \citep{freeman_2024,SPOD} are the most fundamental and widely used methods for energy transfer and budget analysis. Wavelet analysis has been used to analyze the energy budget and transfer related to high-frequency spectra in turbulent flows because of its ability to achieve high temporal resolution in the frequency domain \citep{yamada1991identification}. In contrast, the Fourier decomposition is effective in resolving the low-frequency region. Since the low-frequency components characterize the large-scale structure of the flow fields, Fourier decomposition is used to study the energy transfer related to the primary structure.

To extract large-scale coherent structures in turbulence, spectral proper orthogonal decomposition (SPOD) \citep{SPOD,SPOD2} has been developed to identify statistically significant modes at each frequency from Fourier modes. SPOD addresses the limitation that POD modes may mix structures of different frequencies by performing POD on a frequency-by-frequency basis, thus ensuring that each mode corresponds to a single frequency component. The introduction of SPOD has significantly advanced energy transfer analysis by enabling the efficient extraction of dominant frequency components, even in turbulent flows.

\citet{schmidt_2020} quantified the interaction between different frequencies, which drives energy transfer, using the bispectrum and proposed bispectral mode decomposition (BMD), which extracts the mode that maximizes the bispectrum and identifies the spatial regions where the interaction is strong. BMD has been applied to numerous flow situations, such as the wake flow of a turbine blade \citep{kinjangi2023characterization}, cylinder flow \citep{mypaper4}, and jet flow \citep{Nekkanti_2025,yeung_2025}. The relationship among the spatial distribution of interactions, the bispectrum, and energy transfer has been discussed.

Although the bispectrum characterizes the strength of nonlinear interaction, it does not quantify the amount of energy transfer. To address this limitation, \citet{freeman_2024} uses frequency domain Navier–Stokes equations  with Fourier modes, providing a way to physically interpret the bispectrum maximized by BMD. Based on the frequency domain equation, the kinetic energy budget for the frequency component $f_k (k = 1, 2, \cdots)$ is given by
\begin{eqnarray}
\text{Real}\left(-\sum_{l=-\infty}^\infty {\int{\boldsymbol{\hat{u}}}_{f_{k}}^H(\boldsymbol{\hat{u}}_{f_k-f_l}\cdot\nabla)\boldsymbol{\hat{u}}_{f_l}d\boldsymbol{x}}+{{\frac{1}{\Rey}}\int{\boldsymbol{\hat{u}}}_{f_{k}}^H{\nabla^2}\boldsymbol{\hat{u}}_{f_k}d\boldsymbol{x}}\right)=0,
\label{ballance1}
\end{eqnarray}
where $\boldsymbol{\hat{u}}_{f_k}$ is the spectrum of the $f_k$-frequency component, and $\text{Real}$ denotes the real part of a complex value.  
The first term represents the triadic energy transfer related to $f_k$, $f_l$, and $f_k - f_l$, while the second term represents the viscous diffusion effect. Through the first term, the frequency $f_l$ component transfers energy to the $f_k$ component. Computing the values of these terms from the Fourier modes, i.e., the spectrum, reveals the energy transfer relationship between the $f_k$- and $f_l$-components, as well as the diffusion effect on the $f_k$-component.

\citet{yeung_2024} focused on the triadic energy transfer term obtained from the frequency domain Navier--Stokes equation. They proposed triadic orthogonal decomposition (TOD) to extract the spatial structure that maximizes this triadic energy transfer term. By employing TOD, the frequency components that maximize their transfer relations and the local spatial structures in which energy transfer occurs are revealed for frequencies with arbitrary triadic relations.  
However, none of BMD, frequency domain Navier--Stokes equation using Fourier modes, or TOD is directly applicable to transiently developing flows, as they all rely on decomposition into Fourier modes, which assumes a statistically stationary flow without long-term growing or decaying components. Applying mode decomposition optimized for weakly stationary flows to transient flows fails to capture the short-term behavior of the transient process \citep{Schmid_2007}. Because many natural and engineering flows exhibit transient development, extending energy transfer analysis to transient flows is crucial for understanding fluid dynamics.

To handle the triadic energy transfer between different frequencies in transient processes and viscous diffusion, a modal decomposition technique is required for the transient process. 
To account for transient flows, \citet{Amiri_Margavi_2024} proposed the use of optimally time-dependent (OTD) modes, computed at each time step, which explicitly incorporate the time dependence of the modal structures. By enabling the time variation of the spatial distribution during the transient process, the transiently developing flow field can be optimally decomposed. 
The OTD formulation is similar to extracting POD modes for a specific time in a transient process rather than decomposing it in the frequency domain. 
\citet{VMD} extended the variational mode decomposition, which can extract signals with variation in a specific frequency bandwidth, to flow fields and succeeded in decomposing transient flows into time-dependent modes with a specific center frequency.

Another approach for mode extraction is to consider the operators derived from the Navier–Stokes equations. 
Linear stability analysis (LSA) \citep{LSA1, variousshapes1, Ohmichi_D, Ranjan_2020} extracts global modes that can grow linearly from a base flow. 
To examine the eigenvalues of operators linearized around the base flow, the growth and decay of the corresponding eigenmodes, along with their frequencies, are determined. Dynamic mode decomposition (DMD) \citep{DMD, DMD3, Tu_2013} is a method that can extract eigenmodes of a discrete dataset based on a linear operator that best approximates the temporal evolution of the dataset. 
Strictly speaking, DMD is not a method to extract eigenmodes from an operator, but DMD modes are comparable to eigenmodes of the operator \citep{DMD2}. 
However, these methods are specialized for dealing with linear growth or decay from base flow. 
They are not suitable for direct application to transient processes because linear operators around a particular base flow cannot describe the nonlinear time evolution associated with variations in the base flow. 

In a broader context, several extensions of operator-based frameworks for nonlinear dynamical systems have been proposed, such as the Koopman mode decomposition \citep{mezic2005spectral, Bagheri_2013} and the Mori–Zwanzig formalism \citep{Mori_1965, Zwanzig_1973}. The Koopman mode decomposition defines the linear evolution of observables in a nonlinear system and characterizes the dynamics through spectral analysis of the associated Koopman operator \citep{koopman1931hamiltonian, koopman1932dynamical}. DMD can be regarded as a numerical approximation to the Koopman decomposition, and the two coincide when the snapshots approximate Koopman modes. The Mori–Zwanzig–based modal decomposition extends this framework by introducing a memory kernel into the linear-operator formulation of DMD, thereby explicitly accounting for the influence of unresolved variables. Similarly, higher-order DMD \citep{le2017higher} and Hankel DMD \citep{brunton2017chaos} incorporate temporal nonlocality (memory effects) by utilizing past information \citep{Woodward_2023}. Residual DMD \citep{colbrook2024rigorous,colbrook2023residual} extends the DMD framework by quantifying the residual between DMD modes and the spectrum of the infinite-dimensional Koopman operator. Recursive DMD \citep{noack2016recursive}, on the other hand, improves robustness to nonlinear processes by combining the reconstruction capability of POD with the frequency-wise decomposition capability of DMD. In principle, these approaches can be extended to the decomposition of transient flows; however, with only a finite number of snapshots, nonlinear and memory effects cannot be fully captured, and further refinement is required for accurate representation.

To decompose a transient process into eigenmodes of an operator, one approach is to consider the time variation of the base flow.  
Using a time-varying base flow, the operator can be linearized around the base flow at each time. Thus, in this case, the linear operator and the base flow are time-varying. 
For instance, time-varying DMD \citep{Zhang_2019} addresses the time variation of the linear operator by applying weights to the dataset, considering only certain times or using only data from specific time ranges. \citet{stankiewicz2017need} proposed a method to continuously interpolate the DMD modes of early development, comparable to eigenmodes from LSA, with those of fully developed periodic flows. The change in base flow during the transient process is accomplished by correcting the mean field of stable periodic flow with the base flow at initial development \citep{shift}, which is the steady-state flow satisfying the time-independent Navier–Stokes equation \citep{Fornberg_1980}.  Note that rather than providing a model that aligns completely with the transient development of the actual flow, this is an approximation through interpolation. These methods can potentially describe the time variation of linear operators and their eigenmodes in transient processes. However, these methods are not extended to the energy transfer and budget analysis between different frequencies and viscous diffusion.

Resolvent analysis \citep{trefethen1993hydrodynamic} or input–output analysis \citep{JOVANOVI__2005} are also a method for dealing with the development caused by energy amplification in transient flows. Considering the situation where the flow linearly develops from a specific base flow, resolvent analysis determines the optimal forcing that will maximize the energy amplification of the developed perturbation over a specific time span. An appropriate choice of time span and base flow can provide insights into energy transfer during the transient process. However, in practice, properly extracting the time-dependent variation in energy transfer during the growing process is difficult because the time-varying base flow and time span cannot be determined during the transient process.

Mode extraction based on time-varying linear operators provides a natural foundation for energy transfer analysis, as the decomposition is inherently frequency-based. Since DMD is formulated on frequency, it differs from POD in that a single mode cannot include multiple frequency components within a single mode. As a result, it is more suitable for analyzing energy transfer and budget across different scales. In this study, we focus on energy transfer and budget analysis in transient flow processes using a ROM constructed from the eigenmodes of time-varying linear operators. While the direction of this analysis is similar to the energy budget approach of \citet{shift}, which uses a POD-based ROM, our emphasis is on analyzing actual transient flow fields and developing an operator-based framework.
To achieve this, we adopt a projection-based model that incorporates the eigenmodes of the time-varying  linear operator. This formulation enables a more accurate assessment of energy interactions among dynamically evolving modes during the transient process. 

A canonical example of modeling a transient process using a projection-based model is the two-dimensional cylinder wake, in which a steady flow transitions to an unsteady regime and eventually saturates into a periodic flow \citep{Fornberg_1980, shift, barkley2006linear, GIANNETTI_2007}.
This transition, known as a Hopf bifurcation \citep{Hopf}, can be characterized by the dynamics near the steady flow (a fixed point) and a single dominant oscillator in its vicinity.
As the amplitude of the oscillator grows from the steady flow, nonlinear effects induce the appearance of higher harmonics, leading to saturation into a periodic flow.
After saturation, the oscillation center corresponds to the mean flow of the post-transient periodic state, in which the recirculation region behind the cylinder becomes smaller than that of the steady flow.
\citet{shift} demonstrated that, by constructing a ROM incorporating the steady flow mode, the dominant oscillator, and the mean flow, the saturated periodic state can be successfully reproduced.
However, several studies \citep{Sengupta_2010, Manti_Lugo_2015} have reported that the saturation process obtained from numerical simulations of the Navier–Stokes equations cannot be fully represented by the canonical Hopf bifurcation model, known as the Landau equation.
For instance, \citet{Sengupta_2010} showed that applying POD to the nonlinearly growing stage reveals abnormal oscillation modes that deviate from the standard Hopf bifurcation model.
Despite extensive efforts to model this process, there have been few examples that accurately capture the saturation dynamics as a simple superposition of multiple oscillators with different frequencies.

This paper explores the possibility of energy transfer analysis using linear operator eigenmodes with the projection-based model and analyzes the energy transfer of transient flows. As an example of a transient process, we consider the two-dimensional flow around a circular cylinder, which evolves from a steady flow to an unsteady regime and eventually saturates into a periodic flow. The manuscript is organized as follows. Section \ref{sec2} describes the preparation strategy for the dataset used in mode extraction. Additionally, a novel method is introduced for efficiently extracting modes from time-varying linear operators in transient flows. In Section \ref{sec4}, the modal extraction method for transient processes introduced in Section \ref{sec2} is applied to the flow past a circular cylinder. Section \ref{sec3} derives a phase-averaged ROM based on the eigenmodes of the time-varying linear operators and formulates energy transfer equations for analyzing the energy transfer and budget of transient flows. In Section \ref{sec5}, the phase-averaged ROM derived in Section \ref{sec3} is applied to analyze the energy budget in both linear and nonlinear transient processes of the flow past a circular cylinder.
Section \ref{sec6} summarizes the main findings of this study.

\section{Data preparation}\label{sec2}

\subsection{Numerical method for solving governing equation}
The flow around a circular cylinder was obtained from a numerical simulation of the incompressible Navier--Stokes equations. The governing equations are presented below.
\begin{equation}
\nabla\cdot\boldsymbol{u}=0,
   \label{eqcont}	
\end{equation}
\begin{equation}
\frac{\partial{\boldsymbol{u}}}{\partial{t}}=-(\boldsymbol{u}\cdot\nabla)\boldsymbol{u}-\frac{1}{\rho}\nabla{p}+\frac{1}{\Rey}{\nabla^2}\boldsymbol{u},
   \label{eqnavi}
\end{equation}
where $\boldsymbol{u}$ represents the velocity vector (bold symbols represent vectors), ${p}$ is the pressure, and $\rho$ is the fluid density. The $Re$ is the Reynolds number defined as
\begin{equation}
Re  \stackrel{\mathrm{def}}{=}  \frac{U_{\infty}D}{\nu},
   \label{eqRe}
\end{equation}
where $U_{\infty}$ denotes the free-stream velocity, $\nu$ is the kinematic viscosity, and $D$ is the cylinder diameter. 

The governing equations are solved by the fractional step method proposed by \citet{RKincomp}. The time step size is determined based on our previous validation \citep{12ICCFD}. 
The second-order central difference \citep{poisson} and the QUICK method \citep{QUICK} were used for evaluating the spatial differences. The details of these numerical procedures, including boundary conditions, are described in \citet{mypaper2}. 

This study employs two computational grids with different domains, as depicted in figure \ref{fig_grid}.  The grid convergence is provided in Appendix \ref{apena}.
 To ensure the numerical results remained unaffected by boundary effects, the far-field boundaries were extended to $100D$ for the regular grid and $200D$ for the long grid \citep{Kumar_2006}. In the wall-normal direction, the number of grid points was $240$ for the regular grid, and $360$ for the long grid. In the wall-parallel direction, the number of grid points was $590$  for the regular grid, and $730$ for the long grid. The height of the first layer next to the cylinder was set at $1.0 \times 10^{-3}D$ based on the DNS of \citet{cylinder4}.

\begin{figure}
	\centering\includegraphics[width=12cm,keepaspectratio]{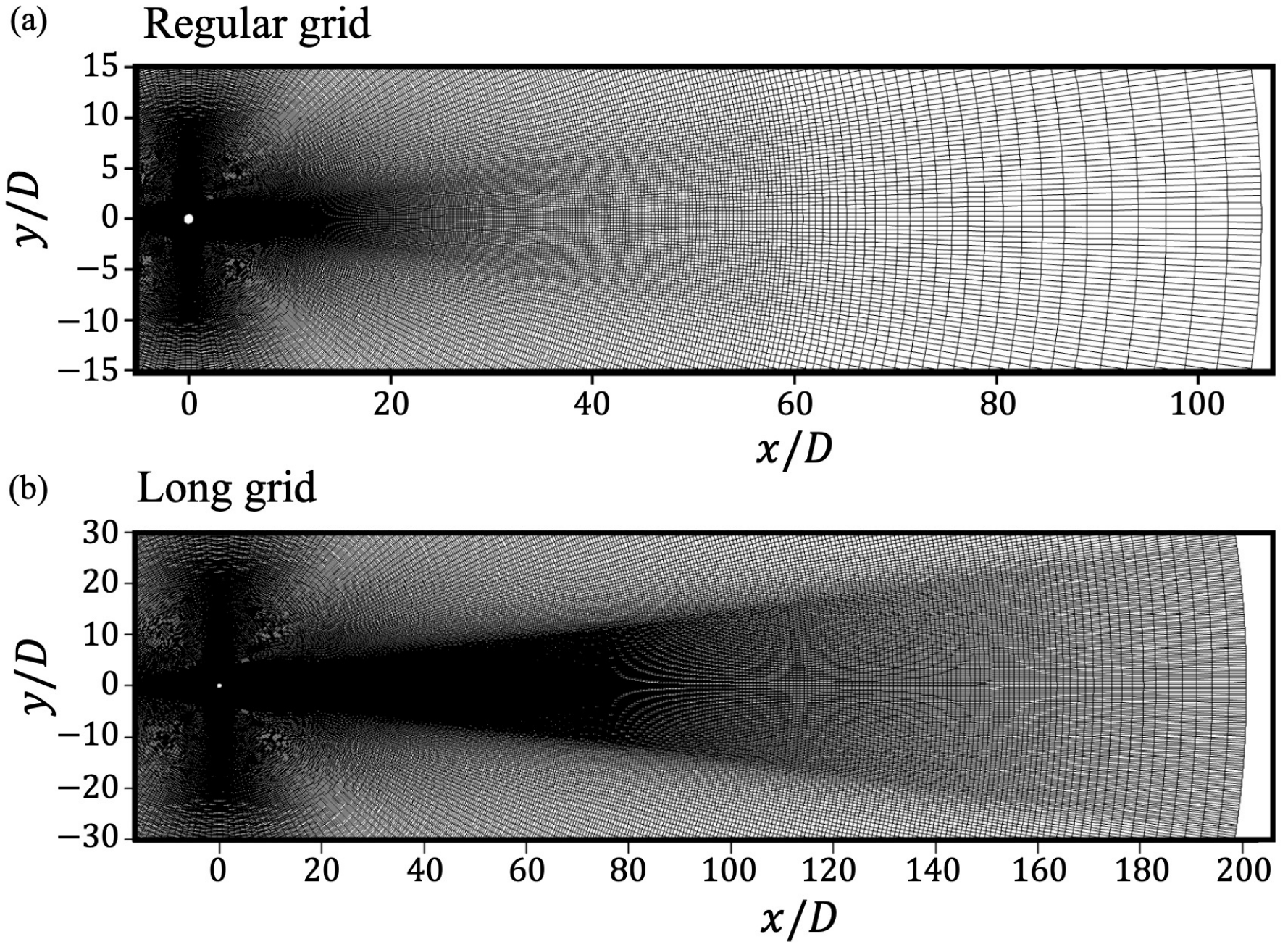}
  \captionsetup{justification=raggedright,singlelinecheck=false}
  \caption{Computational grids around a circular cylinder.}
 \label{fig_grid}
\end{figure}

\subsection{General LSA formulation}\label{sec11}
We consider the general governing equation
\begin{equation}
\frac{\partial \boldsymbol{u}}{\partial t}= \boldsymbol{\mathcal{F}}(\boldsymbol{u}),
\label{general_eq_new}
\end{equation}
where $\boldsymbol{\mathcal{F}}(\boldsymbol{u})$ denotes a generalized operator. Calligraphic letters are used to represent operators in continuous-time dynamical systems.
In general, $\boldsymbol{u}$ represents the state vector; however, in the present study, the velocity vector is employed as the state variable.

LSA considers the time evolution of a perturbation $\boldsymbol{u}'$ about a given base flow $\boldsymbol{u}_b$, governed by the linearized operator as
\begin{equation}
\frac{\partial \boldsymbol{u}'}{\partial t}=\left. \frac{\partial \boldsymbol{\mathcal{F}}}{\partial \boldsymbol{u}}\right|_{\boldsymbol{u}=\boldsymbol{u}_{b}}\boldsymbol{u}',
\label{perturb_general_new}
\end{equation}
where the prime denotes a perturbation and the subscript $b$ refers to the base flow.

The base flow $\boldsymbol{u}_b$ is not uniquely defined. In the present study, we adopt the phase-averaged flow field (introduced later) as a representative flow without instantaneous frequency components. In the linear growth regime considered in LSA, the phase-averaged flow corresponds to a fixed point of the Navier–Stokes equations, i.e., a steady flow. For clarity, we denote the phase-averaged flow as $\boldsymbol{u}_t$ and the steady flow as $\boldsymbol{u}_s$, while $\boldsymbol{u}_b$ is used to indicate a general base flow.

In discrete form, LSA examines the time evolution of a perturbation $\boldsymbol{u}'$, governed by the continuous-time linear operator $\mathcal{A}'$ derived from the Navier–Stokes equations linearized around a base flow:
\begin{equation}
\frac{\partial \boldsymbol{u}'}{\partial t} = \mathcal{A}'\boldsymbol{u}'.
\label{evoLSA_new}
\end{equation}
Here, the prime indicates quantities associated with the linearized Navier–Stokes equations.

The temporal behavior of the perturbation is determined by the eigenvalue problem associated with the operator $\mathcal{A}'$:
\begin{equation}
(\sigma + 2\pi i f)\boldsymbol{\varphi} = \mathcal{A}'\boldsymbol{\varphi},
\label{LSAeigenvalueproblem_new}
\end{equation}
where $i  \stackrel{\mathrm{def}}{=}  \sqrt{-1}$, $\boldsymbol{\varphi}$ denotes the eigenmode, and $\sigma + 2\pi i f$ represents the corresponding eigenvalue. The real part $\sigma$ indicates the growth rate, while $f$ represents the oscillation frequency. Any perturbation can therefore be expressed as a superposition of such eigenmodes, each evolving according to its own growth rate and frequency.
To obtain these eigenmodes, we employ the time-stepping LSA method, which numerically approximates the linear operator $\mathcal{A}'$, that is, the discrete-time operator $A'$, through direct time integration. The details of this procedure are provided in Appendix \ref{apenb}.

\subsection{DMD for mode extraction}
DMD was proposed by \citet{DMD} to extract coherent structures from time series data of flow fields. This study computes eigenmodes from matrices $X$ and $Y$ based on an exact DMD algorithm \citep{Tu_2013}. 
We refer to mode extraction using the exact DMD algorithm simply as DMD. Here, $X$ and $Y$ are matrices whose columns contain flow snapshots. Each column of $Y$ corresponds to the flow field obtained at a time interval $\Delta T$ after the snapshot represented by the corresponding column of $X$ presented below
\begin{eqnarray}
X&=&\left[\boldsymbol{u}(\boldsymbol{x}, t_1\,\,\,\,\,\,\,\,\,\,\,\,\,\,),\,\boldsymbol{u}(\boldsymbol{x}, t_2\,\,\,\,\,\,\,\,\,\,\,\,\,\,),\,\cdots,\boldsymbol{u}(\boldsymbol{x}, t_M\,\,\,\,\,\,\,\,\,\,\,\,\,)\right] \in \mathbb{R}^{N \times M}.\\
Y&=&\left[\boldsymbol{u}(\boldsymbol{x}, t_1+\Delta T),\,\boldsymbol{u}(\boldsymbol{x}, t_2+\Delta T),\,\cdots,\boldsymbol{u}(\boldsymbol{x}, t_M+\Delta T)\right] \in \mathbb{R}^{N \times M},
   \label{matrixX_general}
\end{eqnarray}
where $\boldsymbol{u}(\boldsymbol{x},t)$ denotes the flow snapshot at time $t$,  
$N$ is the number of state variables multiplied by the number of spatial degrees of freedom, $M$ is the number of snapshots,
and $\Delta T$ is the time interval between $X$ and $Y$.

In exact DMD, the linear operator is defined as the solution to a minimization problem of the form
\begin{equation}
A = \underset{A \in \mathbb{R}^{N \times N}}{\text{argmin}} \, \| Y - AX \|_F,
\label{Aargmin}
\end{equation}
where $\| \cdot \|_F$ denotes the Frobenius norm. 
The Frobenius norm is based on the inner product in the $N$-dimensional vector space, defined as
\begin{equation}
\left\langle \boldsymbol{u},\, \boldsymbol{v} \right\rangle \stackrel{\mathrm{def}}{=} \int_{\Omega} \boldsymbol{v}^H(\boldsymbol{x}) \boldsymbol{u}(\boldsymbol{x}) \, d\boldsymbol{x},
\label{eq:inner}
\end{equation}
where $\boldsymbol{u}(\boldsymbol{x}), \boldsymbol{v}(\boldsymbol{x}) \in \mathbb{C}^{N}$, $\langle \cdot, \cdot \rangle$ represents the inner product, and $\Omega$ denotes the spatial domain. The superscript $H$ indicates the Hermitian transpose.  For two-dimensional cases, this inner product reduces to
\begin{equation}
\boldsymbol{v}^H \boldsymbol{u} \stackrel{\mathrm{def}}{=} (\boldsymbol{v}^*)_x (\boldsymbol{u})_x + (\boldsymbol{v}^*)_y (\boldsymbol{u})_y ,
\label{eq:inner}
\end{equation}
where $*$ represents the complex conjugate, and $(\cdot)_x$ and $(\cdot)_y$ denote the $x$- and $y$-components of the vector variable, respectively. 
 It should be noted that DMD estimates a discrete-time linear operator $A$ rather than a continuous-time one $\mathcal{A}$.
Unlike LSA, this operator is not necessarily obtained by linearizing the governing equations around a base flow, but rather represents the optimal linear operator that describes the temporal evolution of the dataset. 
However, as shown in Appendix \ref{apenb}, in the time-stepping method, the application of DMD to datasets obtained from numerical simulations of the governing equations linearized around the base flow yields eigenvalues and eigenvectors consistent with those obtained from the LSA formulation.

By solving the minimization problem, the matrix $A$ is computed from $X$ and $Y$ as 
 \begin{equation}
{A}=Y{X}^{\dagger},
  \label{Aapproximation}
\end{equation}
where the superscript $\dagger$ denotes the Moore–Penrose pseudoinverse. In the numerical processing, $X$ is decomposed by singular value decomposition (SVD) $X=USV^T$, where $U \in \mathbb{R}^{N \times M}$ and $V  \in \mathbb{R}^{M \times M}$ are the left and right singular vectors, respectively, and $S  \in \mathbb{R}^{M \times M}$ is the diagonal matrix with non-negative diagonal elements (the singular values of $X$). By truncating to the leading $r$ singular values, $X$ can be approximated as
 \begin{equation}
X \approx U_rS_rV_r^{T},
  \label{SVD}
\end{equation}
where subscript $r$ denotes rank $r$ truncation of each matrix, $U_r  \in \mathbb{R}^{N \times r}$, $S_r  \in \mathbb{R}^{r \times r}$, and $V_r  \in \mathbb{R}^{M \times r}$. The matrix $A$ is approximated by a low-rank matrix
 \begin{equation}
A_r = U_r^{T}YV_rS_r^{-1},
  \label{AapproximationA}
\end{equation}
where $A_r = U^{T}_rAU_r \in \mathbb{R}^{r \times r}$. 

The eigenvalue of the matrix $A_r$ represents the temporal evolution of the corresponding eigenmodes. The growth rate and the frequency are computed from
\begin{equation}
\sigma_k = \frac{\text{Real}\{\log({\lambda}_k)\}}{\Delta T},
  \label{DMDgrowth}
\end{equation}
\begin{equation}
f_k = \frac{\text{Imag}\{\log({\lambda}_k)\}}{2\pi \Delta T},
  \label{DMDfreq}
\end{equation}
where $\text{Real}(\cdot)$ and  $\text{Imag}(\cdot)$ represent the real and imaginary parts of the complex values, respectively, and $\lambda_k$ is the eigenvalue of $A_r$ (Ritz value) corresponding to the $k$th eigenmode. The function $\log(\cdot)$ denotes the complex logarithm, with the argument (phase angle) defined in the $-\pi$ to $\pi$. 
 In this paper, we denote the eigenmode for frequency $f_k$ by $\boldsymbol{\varphi}_{f_k}  \in \mathbb{C}^{N}$. 
In the exact DMD, the eigenmodes $\boldsymbol{\varphi}_{f_k}$ is computed from the eigenmode $\boldsymbol{\varphi}^{\text{low}}_{f_k}  \in \mathbb{C}^{r}$ of $A_r$ as follows:
 \begin{equation}
\boldsymbol{\varphi}_{f_k}=YV_rS_r^{-1}\boldsymbol{\varphi}^{\text{low}}_{f_k}.
  \label{AapproximationA}
\end{equation}
In this paper, $f_k$ is indexed from $k=1$ in decreasing order of $|f_k|$, and negative frequency is denoted by $f_{-k}$. Here, $| \cdot |$ denotes the absolute value of a real or complex value. In general, the magnitude of eigenmodes associated with a linear operator is not prescribed. In the present study, however, they are normalized to unity:
\begin{equation}
\| \boldsymbol{\varphi}_{f_k} \| = 1,
\label{eq:inner}
\end{equation}
where $\| \cdot \|$ denotes the vector norm induced by the inner product $\langle \cdot,\cdot \rangle$, defined as
\begin{equation}
\| \boldsymbol{u} \| \stackrel{\mathrm{def}}{=} \sqrt{ \left\langle \boldsymbol{u},\, \boldsymbol{u} \right\rangle }
\;=\; \left( \int_{\Omega} \boldsymbol{u}^H(\boldsymbol{x}) \cdot \boldsymbol{u}(\boldsymbol{x}) \, d\boldsymbol{x} \right)^{1/2}.
\label{eq:inner}
\end{equation}

\subsection{Mode extraction method for transient process using phase control}

\subsubsection{Introduction of time-varying operator}
To capture the temporal evolution of eigenmodes during their nonlinear growth, we extend the linear-operator-based modal extraction framework to time-dependent systems. 
We consider the instantaneous dynamics at $t=\tau$ in the general governing equation (\ref{general_eq_new}) as
\begin{equation}
\left.\frac{\partial \boldsymbol{u}}{\partial t}\right|_{t=\tau} = \boldsymbol{\mathcal{F}}\left\{\boldsymbol{u}(\boldsymbol{x}, t=\tau)\right\}.
\end{equation}
In this expression, we approximate the instantaneous dynamics using a linear operator. As an example of the linear operator governing the instantaneous dynamics of the system, \citet{Sapsis_2009} introduced a linear operator  for an orthogonal set of basis functions (referred to as OTD modes), based on the minimization problem \citep{babaee2016minimization}
\begin{equation}
\mathcal{A}_{\text{OTD}}(\tau) = \underset{\mathcal{A}_{\text{OTD}}}{\text{argmin}} \, \left\| \left.\frac{dU_{\text{OTD}}}{dt}\right|_{t=\tau} - \mathcal{A}_{\text{OTD}}U_{\text{OTD}} \right\|_F,
\label{AOTD}
\end{equation}
where $U_{\text{OTD}} \in \mathbb{R}^{N \times r}$ represents sets of orthonomal basis, $\mathcal{A}_{\text{OTD}}$ is linear operator for  $U_{\text{OTD}}$

Although the previous OTD framework focuses on the time evolution of an orthonormal basis, the present framework directly estimates the temporal evolution of the solution using time-dependent linear operators. In other words, we apply the DMD-based minimization problem to the instantaneous flow field, defined as
\begin{equation}
A(t=\tau) = \underset{A(\tau)}{\text{argmin}} \, \left\| Y(\tau) - A(\tau)X(\tau) \right\|_F,
\label{Aargmin_tDMDpc}
\end{equation}
where the linear operator $A(t)$ and the flow datasets $X(t)$ and $Y(t)$ are time dependent.
To avoid ambiguity, the explicit time dependence $(t)$ is retained throughout the paper whenever a time-dependent linear operator is considered.

\subsubsection{Dataset preparation for time-varying linear operator}
In the context of DMD, constructing a linear operator that best fits the dynamics at a specific instant requires a dataset restricted to that particular dynamical state. However, during a transient process, the flow field continuously evolves in time, making it difficult to prepare a dataset that is strictly limited to a specific dynamical regime.
In contrast, within the framework of resolvent analysis, it has been shown that the original input–output formulation, which was designed for neutrally stable input and output modes, can also be extended to cases involving growing modes by applying a time-window filter \citep{Rolandi_2024, jovanovic2004modeling}:
\begin{equation}
W_t \boldsymbol{u} = e^{-s \Delta T} \boldsymbol{u},
\label{window_filter}
\end{equation}
where $s$ is a parameter introduced to suppress flow growth within finite-time-scale dynamics. If the parameter is set such that $s > \sigma$, the growth of eigenmodes is bounded over time. 

The core of the time-window filter is to confine the analysis to the instantaneous dynamics by applying a filter that suppresses the growth of modal amplitudes. From a fluid-dynamical perspective, the amplitude resulting from modal growth determines the extent of interactions between frequency components that characterize the dynamics. Therefore, appropriately suppressing the modal amplitude through the time-window filter indicates that the same dynamical behavior can be maintained even as the flow evolves in time.
Therefore, by applying this time-window filter to the dataset, one can effectively isolate a subset of data that represents the instantaneous dynamics, even when the flow contains growing structures.

Using this filter, the dataset at $t=\tau$ is expressed as
\begin{eqnarray}
X(\tau)
&=&
\left[
\boldsymbol{u}(\tau),\,
W_t \boldsymbol{u}(\tau+\Delta T),\,
\cdots,\,
(W_t)^{M-1} \boldsymbol{u}(\tau+(M-1)\Delta T)
\right],
\label{matrixX_filter}
\end{eqnarray}
where, for $t > \tau $, a time-window filter of duration $\Delta T$ is applied to suppress the growth of the dynamics beyond the time $t=\tau$. The flow field at $t=\tau$ is then represented in terms of time-varying eigenmodes as
\begin{equation}
\boldsymbol{u}(\boldsymbol{x}, \tau)
=
\sum_{l=-\infty}^{\infty}
a_{f_l}(\tau)\,
\boldsymbol{\varphi}_{f_l}(\boldsymbol{x}, \tau),
\end{equation}
where $a_{f_l}(\tau)$ denotes the amplitude coefficient of the $f_l$-frequency eigenmode $\boldsymbol{\varphi}_{f_l}(\boldsymbol{x}, \tau)$. 
In other words, the negative index denotes the negative frequency and satisfies
\begin{eqnarray}
f_{-l}& \stackrel{\mathrm{def}}{=} & -f_l,\\
a_{f_{-l}} & \stackrel{\mathrm{def}}{=} & a^*_{f_l}.
   \label{negativerepresent_new}
\end{eqnarray}
The negative frequency is required to represent the real-valued velocity field by the complex-valued eigenmode. The $f_0$-frequency mode denotes the zero-frequency mode. 
In practice, only a finite number of eigenmodes can be obtained numerically; however, the formulation is expressed as a summation extending formally to infinity.  
For cases represented by a finite set of eigenmodes, the corresponding frequencies may be regarded as nonexistent, with their amplitude coefficients set to zero.

Time-evolving this representation using the instantaneous best-fit linear operator yields
\begin{eqnarray}
W_t \boldsymbol{u}(\boldsymbol{x}, \tau+\Delta T) 
&\approx& W_t A(\tau) \boldsymbol{u}(\boldsymbol{x}, \tau) \nonumber \\ 
&=& \sum_{l=-\infty}^{\infty} W_t A(\tau) a_{f_l}(\tau) \boldsymbol{\varphi}_{f_l}(\boldsymbol{x}, \tau) \nonumber \\ 
&=& \sum_{l=-\infty}^{\infty} W_t e^{\sigma_{f_l} \Delta T + 2 \pi f_l \Delta T i} a_{f_l}(\tau) \boldsymbol{\varphi}_{f_l}(\boldsymbol{x}, \tau)\nonumber \\ 
&=& \sum_{l=-\infty}^{\infty} e^{(\sigma_{f_l} - s)\Delta T + 2 \pi f_l \Delta T i} a_{f_l}(\tau) \boldsymbol{\varphi}_{f_l}(\boldsymbol{x}, \tau),
\end{eqnarray}
where the first line represents an approximation of the time evolution at $t=\tau$ using the best-fit linear operator.  
When we assume that the subsequent columns of the dataset are constrained by the dynamics at $t=\tau$ through the time-window filter, the dataset can be expressed as
\begin{eqnarray}
X(\tau)
&=&
\Bigg[
\sum_{l=-\infty}^{\infty} a_{f_l} \boldsymbol{\varphi}_{f_l},\,
\sum_{l=-\infty}^{\infty} a_{f_l} e^{(\sigma_{f_l} - s)\Delta T + 2 \pi f_l \Delta T i} \boldsymbol{\varphi}_{f_l},\, \nonumber \\
&& \,\,\,\,\,\,\,\,\,\,\,\,\,\,\,\,\,\,\,\,\,\,\,\,\,\,\,\,\,\,\,\,\,\,\,\,\,\,\,\,\,\,\,\,\,
\cdots,\,
\sum_{l=-\infty}^{\infty} a_{f_l} e^{(\sigma_{f_l} - s)(M-1)\Delta T + 2 \pi f_l (M-1)\Delta T i} \boldsymbol{\varphi}_{f_l}
\Bigg].
\label{matrixX_filter}
\end{eqnarray}
The difference between adjacent snapshots in the dataset is characterized by the factor $e^{(\sigma_{f_l} - s)\Delta T + 2 \pi f_l \Delta T i}$.  
Returning to the role of the time-window filter, since its purpose is to confine the dynamics within a finite time span by suppressing growth, the ideal condition satisfies $\sigma_{f_l} - s \leq 0$.  
Consequently, the variation among adjacent snapshots should originate solely from the phase rotation term $e^{2 \pi f_l \Delta T i}$. 
From this modal-phase perspective, the dataset ideal for computing $A(t)$ is a sequence of uniformly spaced snapshots in which the time-window filter neutralizes growth, such that
\begin{eqnarray}
X(\tau)
&=&
\Bigg[
\sum_{l=-\infty}^{\infty} a_{f_l}(\tau) \boldsymbol{\varphi}_{f_l},\,
\sum_{l=-\infty}^{\infty} a_{f_l}(\tau) \mathcal{R}(2 \pi f_l \Delta T) \boldsymbol{\varphi}_{f_l},\, \nonumber \\
&& \,\,\,\,\,\,\,\,\,\,\,\,\,\,\,\,\,\,\,\,\,\,\,\,\,\,\,\,\,\,\,\,\,\,\,\,\,\,\,\,\,\,\,\,\,
\cdots,\,
\sum_{l=-\infty}^{\infty} a_{f_l}(\tau) \mathcal{R}(2 \pi f_l (M-1)\Delta T) \boldsymbol{\varphi}_{f_l}
\Bigg],
\label{matrixX_filter_rotation}
\end{eqnarray}
where $\mathcal{R}(\theta) = e^{\theta i}$ is a $2\pi$-periodic function.  
Focusing on the rotation induced by $\mathcal{R}(\theta)$, the rotation angle increases linearly with frequency $f_l$; thus, for a fixed time increment $\Delta T$, higher frequencies correspond to larger phase changes.

\subsubsection{Utilizing time-stepping approach}
To obtain the dataset $X$ from a modal-phase perspective at arbitrary times during the transient process, we introduce a phase-controlled data acquisition strategy using CFD.  
In the first step, the initial flow fields are defined as
\begin{equation}
\boldsymbol{u}(\boldsymbol{x},0, \alpha)
= \boldsymbol{u}_{b}(\boldsymbol{x})
+ \sum_{l=-\infty}^{\infty}
\epsilon_{f_l}
\mathcal{R}\!\left(\frac{f_l}{f_c}\alpha\right)
\boldsymbol{\varphi}_{f_l}(\boldsymbol{x},0),
\label{tDMDpc_inital_new}
\end{equation}
where $0 \leq \alpha < 2\pi$ denotes the phase-shift angle from the basic process $\boldsymbol{u}(\boldsymbol{x}, t, 0)$, $\epsilon_{f_l}$ is the complex initial amplitude of the eigenmode with frequency $f_l$, and $f_c$ is the target frequency.  
In this formulation, the phase-shift angle is rotated such that the eigenmode at frequency $f_c$ completes exactly one full cycle.  
As a result, the dataset is guaranteed to include more than one full cycle of the flow field for any frequency higher than $f_c$.  
Therefore, it is preferable to select frequencies equal to or lower than the frequency of interest. 

The base flow and the eigenmodes (or spectrum) should be chosen such that the initial field satisfies the Navier--Stokes equations.  
One approach is to adopt a fixed point as the base flow and use the eigenmodes computed from the LSA around this fixed point.  
In that case, the amplitude $\epsilon_{f_l}$ of each $f_l$-frequency eigenmode must be specified individually.  
When two or more frequency eigenmodes are present, the relationships among their amplitudes are not straightforward and depend on the specific problem.  
Alternatively, the initial field can be constructed using the time-averaged flow of a statistically steady state without external disturbances as the base flow, and the Fourier spectrum of that state as the set of eigenmodes.

To prepare the dataset with different phases, the parameter $\alpha$ is discretized into $j_{\text{max}}$ points defined as
\begin{equation}
\alpha_j = 2\pi \frac{j-1}{j_{\text{max}}}\,\,\,(j = 1, 2, \cdots, ).
\label{alpha_DMDpc}
\end{equation}
After preparing the initial flow fields, each initial condition is advanced independently in time using CFD.  
As a result, $j_{\text{max}}$ sets of time-series data are obtained.  
The eigenmodes of the best-fit linear operator at time $t$ are then determined by applying DMD to the following matrices:
\begin{eqnarray}
X(t)
&=&
[\boldsymbol{u}(\boldsymbol{x}, t, \alpha_1),\, \boldsymbol{u}(\boldsymbol{x}, t, \alpha_2),\, \cdots,\, \boldsymbol{u}(\boldsymbol{x}, t, \alpha_{j_{\text{max}}})]
\in \mathbb{R}^{N \times j_{\text{max}}}, \\
Y(t)
&=&
[\boldsymbol{u}(\boldsymbol{x}, t+\Delta T, \alpha_1),\, \boldsymbol{u}(\boldsymbol{x}, t+\Delta T, \alpha_2),\, \cdots,\, \boldsymbol{u}(\boldsymbol{x}, t+\Delta T, \alpha_{j_{\text{max}}})]
\in \mathbb{R}^{N \times j_{\text{max}}}.
\label{matrixX}
\end{eqnarray}
 In this approach, the CFD simulations do not neglect the quadratic nonlinearity of perturbations, as is done in LSA, but instead solve the full nonlinear equations that include both the perturbation and the base flow.  
Consequently, the matrices $X(t)$ and $Y(t)$ are not strictly related through a linear operator, and the objective is to approximate their temporal evolution using a best-fit linear operator.

In this framework, DMD is performed independently at each time step, and thus the continuity of eigenmodes between consecutive time-varying linear operators is not necessarily guaranteed. In the present case, although the eigenmodes are normalized to have unit magnitude, their complex phase angles can take arbitrary values. Consequently, discontinuities in the complex phase may appear as discontinuities in the eigenmodes.
For flow reconstruction, these phase discontinuities are compensated by corresponding adjustments in the amplitude coefficients and therefore do not affect the reconstructed velocity field. Nevertheless, suppressing such discontinuities at the algorithmic level is essential for improving the interpretability of the time-varying eigenmodes. To this end, the eigenmodes are corrected by solving the following constrained minimization problem:
\begin{equation} 
	z_0=1,\\ z_n = \underset{z_n \in \mathbb{C}}{\text{argmin}} \, \left\| e^{z_{n-1}}\boldsymbol{\varphi}_{f_k}(\boldsymbol{x},(n-1)\Delta T) - e^{z_{n}}\boldsymbol{\varphi}_{f_k}(\boldsymbol{x},n\Delta T) \right\| \,\,\, (n=1,2, \cdots). 
	\label{Aargmin_tDMDpc} 
\end{equation}
This minimization problem is a variant of the Procrustes problem \citep{golub2013matrix}, which can be expressed in a recursive form as
\begin{equation} e^{z_n} = e^{z_{n-1}}\!\left( {\left\langle \boldsymbol{\varphi}_{f_k}(\boldsymbol{x},(n-1)\Delta T),\, \boldsymbol{\varphi}_{f_k}(\boldsymbol{x},n\Delta T)\right\rangle} \right). \label{eq:zn_log} \end{equation}
In this study, we denote $e^{z_n}\boldsymbol{\varphi}_{f_k}(\boldsymbol{x},n\Delta T)$ simply as $\boldsymbol{\varphi}_{f_k}(\boldsymbol{x},n\Delta T)$, where the rotation factor $e^{z_n}$ is implicitly applied.

The amplitude coefficients $a_{f_k}(t)$ can be computed from the base process $\boldsymbol{u}(\boldsymbol{x}, t, 0)$ and eigenmodes as follows
 \begin{equation}
  \begin{split}
\left[a_{f_1}(t), a^*_{f_1}(t), \cdots \right]& \stackrel{\mathrm{def}}{=} \boldsymbol{\Phi}^{\dagger}\left\{\boldsymbol{u}(\boldsymbol{x},t, 0)-\boldsymbol{u}_{t}(\boldsymbol{x},t)\right\},
  \label{calcoefficient}
  \end{split}
\end{equation}
where 
 \begin{equation}
  \begin{split}
\boldsymbol{\Phi}^{\dagger}=\left[\boldsymbol{\varphi}_{f_1}(\boldsymbol{x},t),\boldsymbol{\varphi}^*_{f_1}(\boldsymbol{x},t), \cdots\right].
  \end{split}
\end{equation}
In this paper, the Moore–Penrose pseudoinverse for computing amplitude coefficients is computed by preconditioning the QR decomposition. Here, we consider the average fields $\boldsymbol{u}_{t}(\boldsymbol{x}, t)$ over $\alpha$, defined as follows
\begin{eqnarray}
\boldsymbol{u}_{t}(\boldsymbol{x}, t) \stackrel{\mathrm{def}}{=} \frac{1}{2\pi} \int^{2\pi}_0 \boldsymbol{u}(\boldsymbol{x}, t, \alpha) d \alpha.
   \label{time-average}
\end{eqnarray}
We use the flow field averaged over $\alpha$ as a time-dependent base flow. Note that the inital base flow $\boldsymbol{u}_{t}(\boldsymbol{x}, t=0)$ coressponds to $\boldsymbol{u}_{b}(\boldsymbol{x})$. Strictly speaking, subtraction of $\boldsymbol{u}_{t}(\boldsymbol{x}, t)$ is not required when computing the amplitude coefficients. However, in this study, the base flow and the amplitude coefficients are separated to maintain consistency with the subsequent energy-transfer analysis. 

Moreover, when the phase-shift angle is assumed to be time-invariant, even from a general process $\boldsymbol{u}(\boldsymbol{x}, t, \alpha)$, the amplitude coefficients can be computed as  
\begin{equation}
  \begin{split}
  a_{f_l}(t) = e^{-\frac{f_l}{f_c}\alpha}b_{f_l}(t),
  \label{calcoefficient_general}
  \end{split}
\end{equation}
where
 \begin{equation}
  \begin{split}
\left[b_{f_1}(t), b^*_{f_1}(t), \cdots \right]& \stackrel{\mathrm{def}}{=} \boldsymbol{\Phi}^{\dagger}\left\{\boldsymbol{u}(\boldsymbol{x},t, \alpha)-\boldsymbol{u}_{t}(\boldsymbol{x},t)\right\}.\\
  \end{split}
\end{equation}
However, in data acquisition using the time-stepping method, the phase-shift angle $\alpha$ is not necessarily constant during the numerically integrated evolution. The temporal variation of the phase-shift angle $\alpha$ will be discussed in a later section.
 
\begin{figure}
	\centering\includegraphics[width=12cm,keepaspectratio]{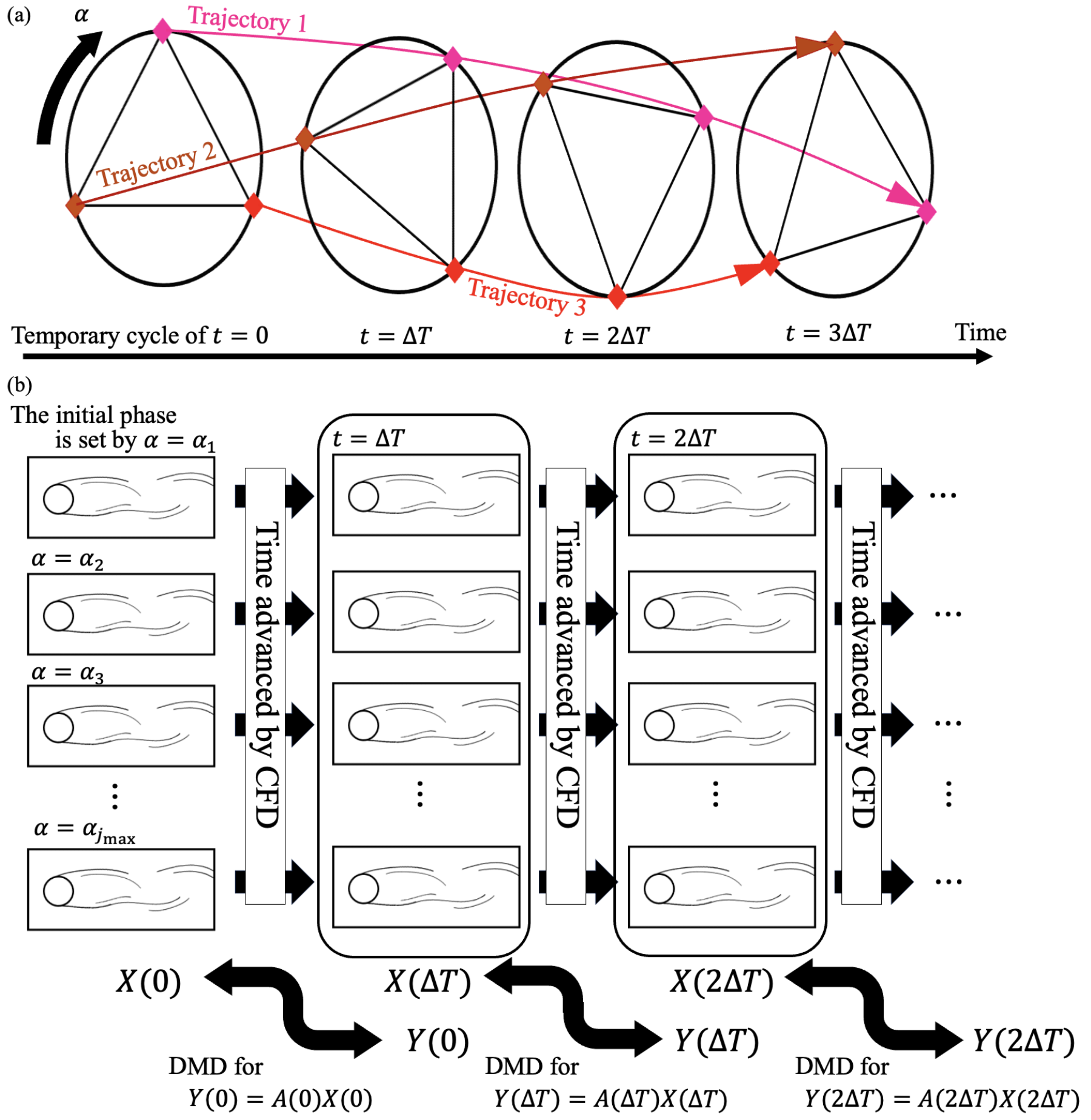}
  \captionsetup{justification=raggedright,singlelinecheck=false}
  \caption{Schematic of tDMDpc. By changing the temporal phase of the mode in the initial flow and time progressing independently by CFD, a transient flow with different phases can be obtained.}
 \label{fig_tDMDpc}
\end{figure}

The schematic of this modal decomposition procedure is shown in figure \ref{fig_tDMDpc}. In this method, the phase of the dominant frequency $f_1$ is controlled initially using a parameter $\alpha$, with the expectation that the phase of the solution trajectory will vary depending on $\alpha$ at all times. As a result, the snapshot matrices at each time correspond to flow fields that are phase-shifted along the same cycle. We refer to this method as time-varying DMD with phase controlling (tDMDpc).

\section{Mode extraction of tDMDpc for circular cylinder using LSA}\label{sec4}
In this section, we demonstrate the application of the tDMDpc method to the transient process in the cylinder wake, where a steady flow evolves into a periodic flow through nonlinear growth. During this process, the amplitudes of a pair of eigenmodes that initially grow linearly from the steady flow are progressively amplified, eventually saturating to form the periodic state. Accordingly, in this system, the initial condition for tDMDpc is constructed by superimposing a pair of eigenmodes onto the steady flow.

\subsection{LSA results for initial flow fields}
\begin{figure}
	\centering\includegraphics[width=10cm,keepaspectratio]{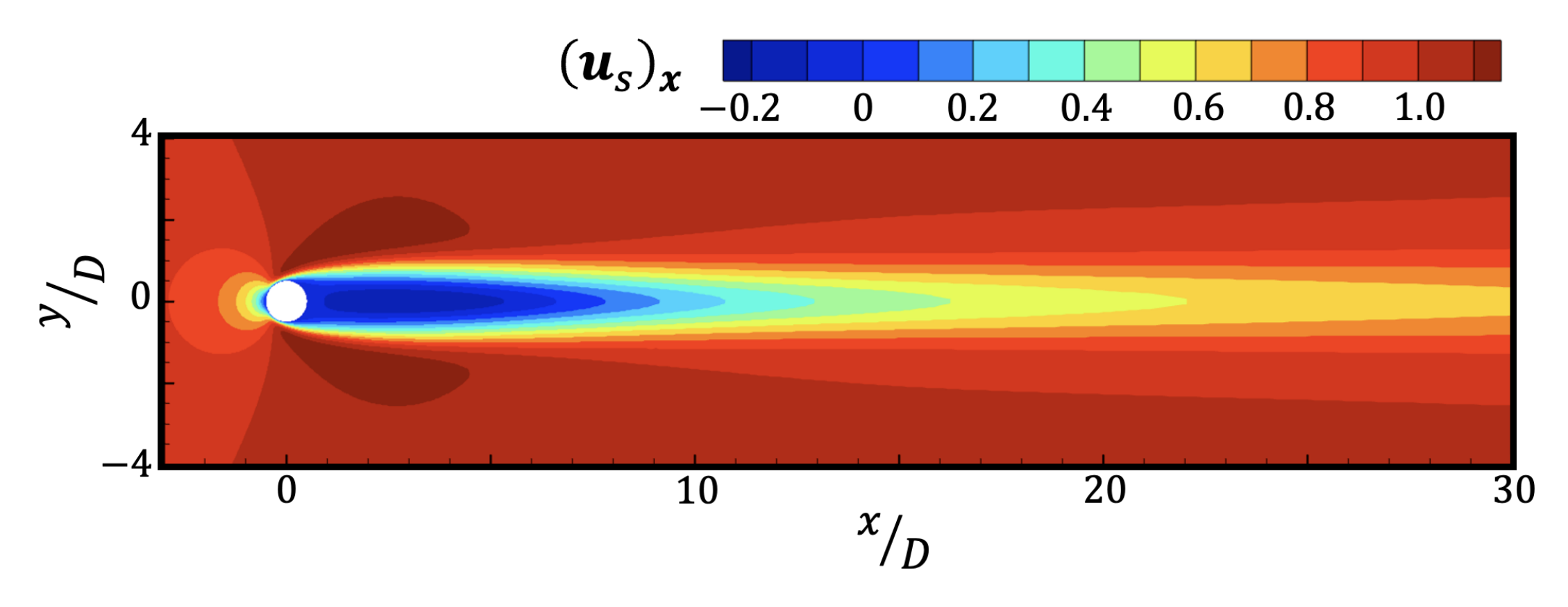}
  \captionsetup{justification=raggedright,singlelinecheck=false}
  \caption{$x$-direction velocity fields of the base flow at $\Rey=100$ obtained from the regular grid. The base flow is obtained by imposing symmetry on $y=0$.}
 \label{fig_steady_100}
\end{figure}

To construct the initial field for tDMDpc, LSA is performed around the steady flow to extract the unstable eigenmodes. The time-stepping LSA was performed for the flow around a cylinder. The base flow is computed by imposing symmetry on $y=0$. Figure \ref{fig_steady_100} shows the streamwise velocity fields of the base flow at $\Rey=100$ obtained from a regular grid. A recirculation region with negative streamwise velocity is formed in the cylinder wake. Many previous studies \citep{Sen,Fornberg_1980} have reported that the recirculation region of the wake expands with increasing $\Rey$.

\begin{figure}
	\centering\includegraphics[width=11cm,keepaspectratio]{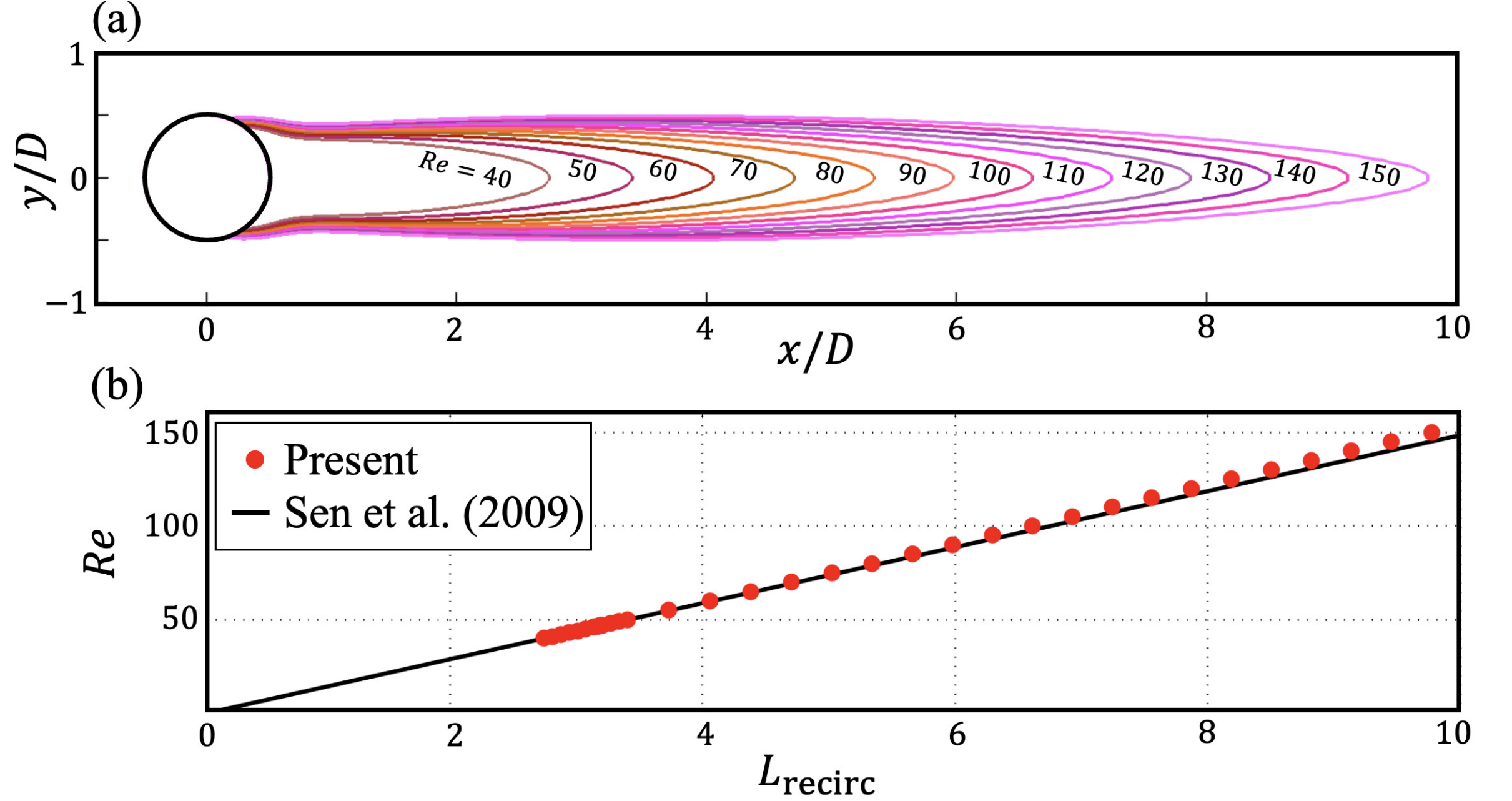}
  \captionsetup{justification=raggedright,singlelinecheck=false}
  \caption{Base flow for various $\Rey$ from the regular grid. (a) $0$-line of the $x$-direction velocity component of the base flow (steady flow) at various $\Rey$. (b) The $x$-direction length of the recirculation region, $L_\text{recirc}$. The $x$-direction length of the $0$ line expands linearly with increasing $Re$.}
 \label{steady_Re}
\end{figure}

Figure \ref{steady_Re} (a) shows the $0$-lines of $x$-direction  velocity for the base flow at various $\Rey$, which characterize the wake recirculation region. As $\Rey$ increases, the recirculation region becomes larger. Here, the length of the recirculation region, $L_{\text{recirc}}$, is defined as the $x$-position where the $0$-line and $y=0$ intersect, excluding the cylinder surface. Figure \ref{steady_Re} (b) shows the relationship between $L_{\text{recirc}}$ and $\Rey$. $L_{\text{recirc}}$ increases linearly with $\Rey$, in close agreement with the results of \citet{Sen}.

The LSA was performed using the obtained base flow $\boldsymbol{u}_{s}$. The details of the parameter choices in the time-stepping method are discussed in Appendix \ref{apenb}. Figure \ref{fig_faif} shows the most dominant eigenmode at $\Rey = 40, 60, 100,$ and $150$, obtained from the LSA, along with the absolute value of the modes. The $0$ line of the base flow for the same $\Rey$ values is shown in the figure as a black line. In the case of cylinder flow, the most dominant mode has the lowest frequency, $f_1$. Thus, the most unstable mode is referred to as $\boldsymbol{\varphi}_{f_1}$. The spatial distribution of $\boldsymbol{\varphi}_{f_1}$ is asymmetric with respect to $y = 0$ for all $\Rey$ cases. Focusing on the distribution around the $0$ line of the base flow, asymmetric fluctuations exist vertically above and below the recirculation region enclosed by the $0$ line. This represents the beginning of forming an asymmetric Karman vortex that arises from a symmetric recirculation region.

The distribution of absolute values of $\boldsymbol{\varphi}_{f_1}$ is in good agreement with \citet{GIANNETTI_2007, Mittal_2009}. For $\Rey > 60$, $\boldsymbol{\varphi}_{f_1}$ shows that the values become small sufficiently far from the cylinder. The peak value of $\sqrt{\boldsymbol{\varphi}^H_{f_1}\boldsymbol{\varphi}_{f_1}}$ occurs at approximately $x/D \approx 10$, except for $\Rey = 40$. However, this position does not follow a monotonous trend with respect to $\Rey$. 

\begin{figure}
	\centering\includegraphics[width=12cm,keepaspectratio]{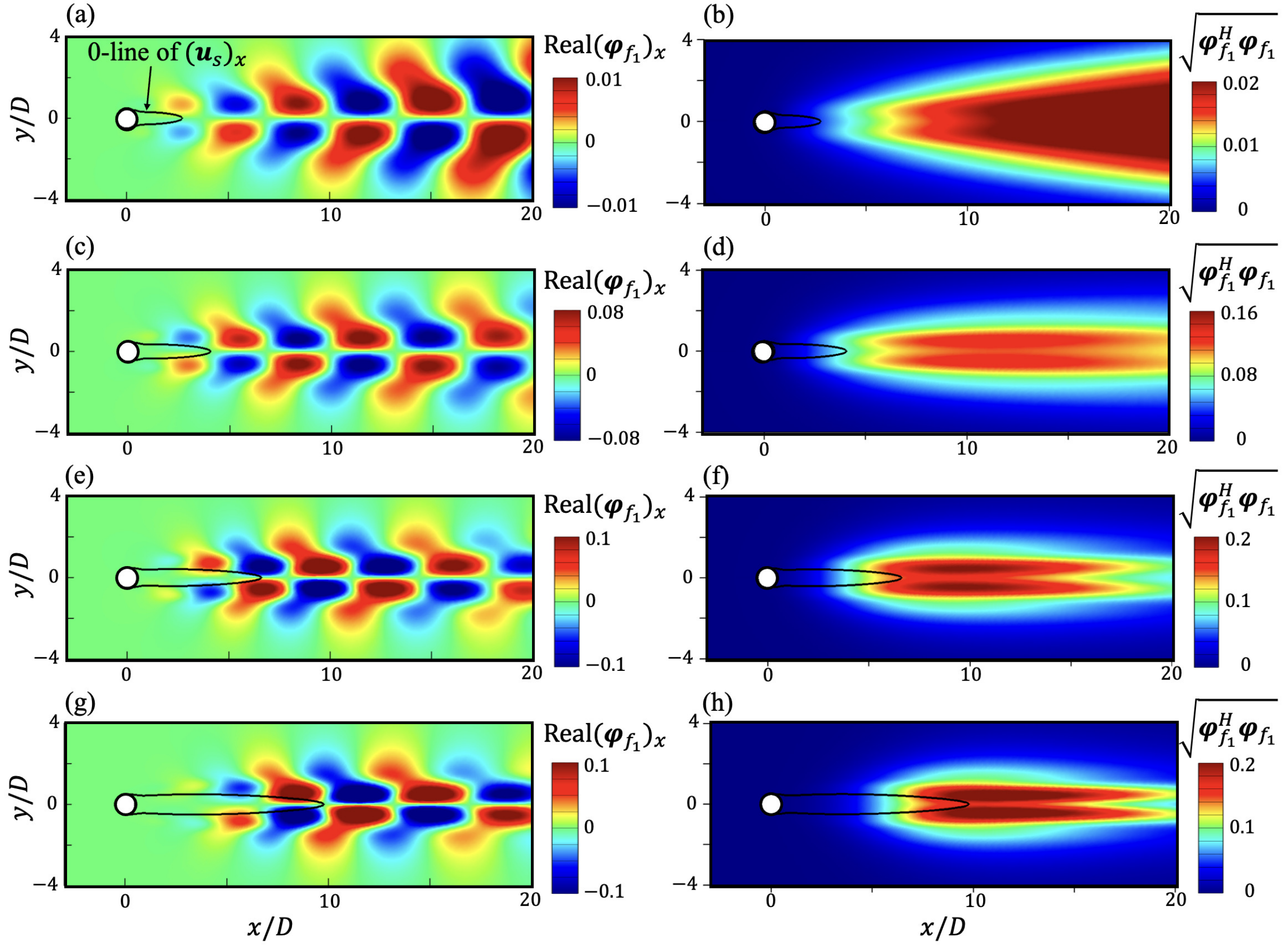}
  \captionsetup{justification=raggedright,singlelinecheck=false}
  \caption{Spatial distribution of $x$-direction component of eigenmode $(\boldsymbol{\varphi}_{f_1})_x$ and its absolute value: (a) and (b) $\Rey=40$, (c) and (d) $60$, (e) and (f) $100$, and (g) and (h) $150$. The black line indicates the zero line of $x$-direction velocity from the base flow. All modes were obtained from the regular grid.}
 \label{fig_faif}
\end{figure}

\begin{figure}
	\centering\includegraphics[width=14cm,keepaspectratio]{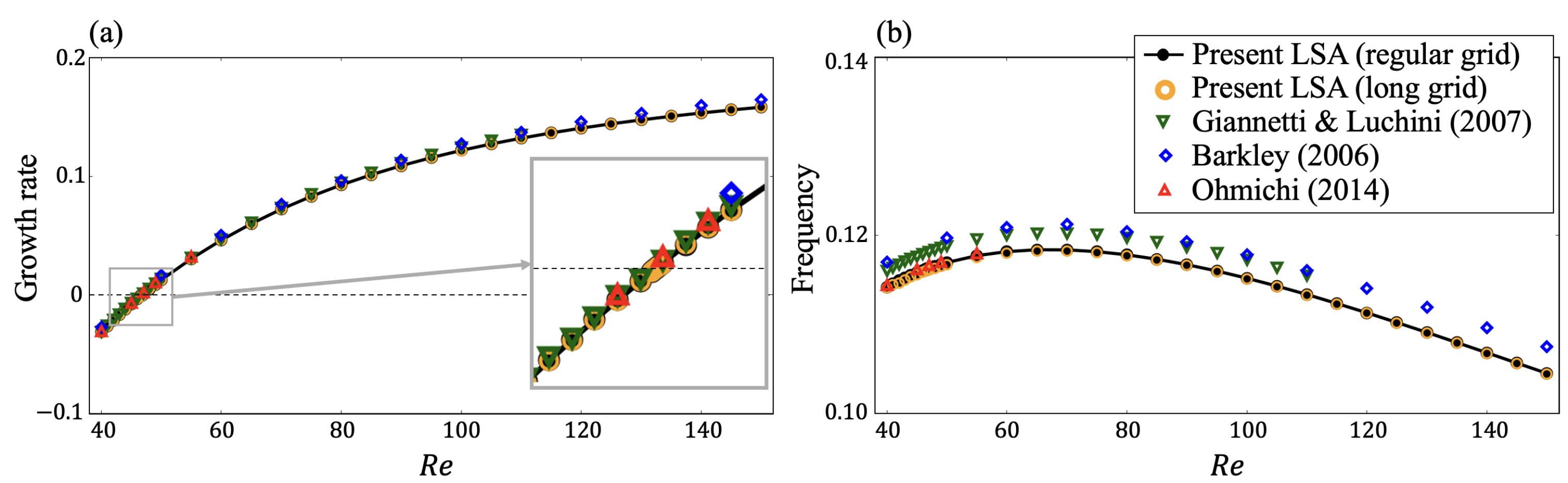}
  \captionsetup{justification=raggedright,singlelinecheck=false}
  \caption{Comparison of (a) growth rate and (b) frequency obtained by the time-stepping LSA as a function of $Re$ with previous works.}
 \label{fig_growth_LSA}
\end{figure}

Figure \ref{fig_growth_LSA} shows the growth rate of $\boldsymbol{\varphi}_{f_1}$ and the frequency $f_1$ at each $\Rey$. The growth rates and frequencies obtained with the regular and long grids are in close agreement across all $\Rey$. The growth rate monotonically increases with increasing $\Rey$, and the sign of the growth rate changes at $\Rey \approx 46.8$. The sign of the growth rate indicates whether the most unstable mode grows from the base flow, which implies the onset of a Hopf bifurcation. Thus, the critical $\Rey$ is $46.8$, which is quite similar to previous studies \citep{Ohmichi_D, cilinderreview, Kumar_2006, barkley2006linear, GIANNETTI_2007}. The frequency increases up to $\Rey = 65$ and then starts to decrease at that $\Rey$. This trend is consistent with previous studies. The frequency values are in close agreement with the results of \citet{Ohmichi_D} but are slightly smaller than those of \citet{barkley2006linear} and \citet{GIANNETTI_2007}. The difference is possibly due to the LSA methodology, as Ohmichi's result is based on LSA using the time-stepping method (matrix-free method), while the other two results are obtained using the matrix method. As shown in Appendix \ref{apenb}, the frequencies obtained by the time-stepping method are close to those obtained from numerical simulations. From this perspective, we conclude that the time-stepping method is reasonable.

\subsection{tDMDpc result for transient cylinder flow}
\begin{figure}
\centering
\centering\includegraphics[width=11cm,keepaspectratio]{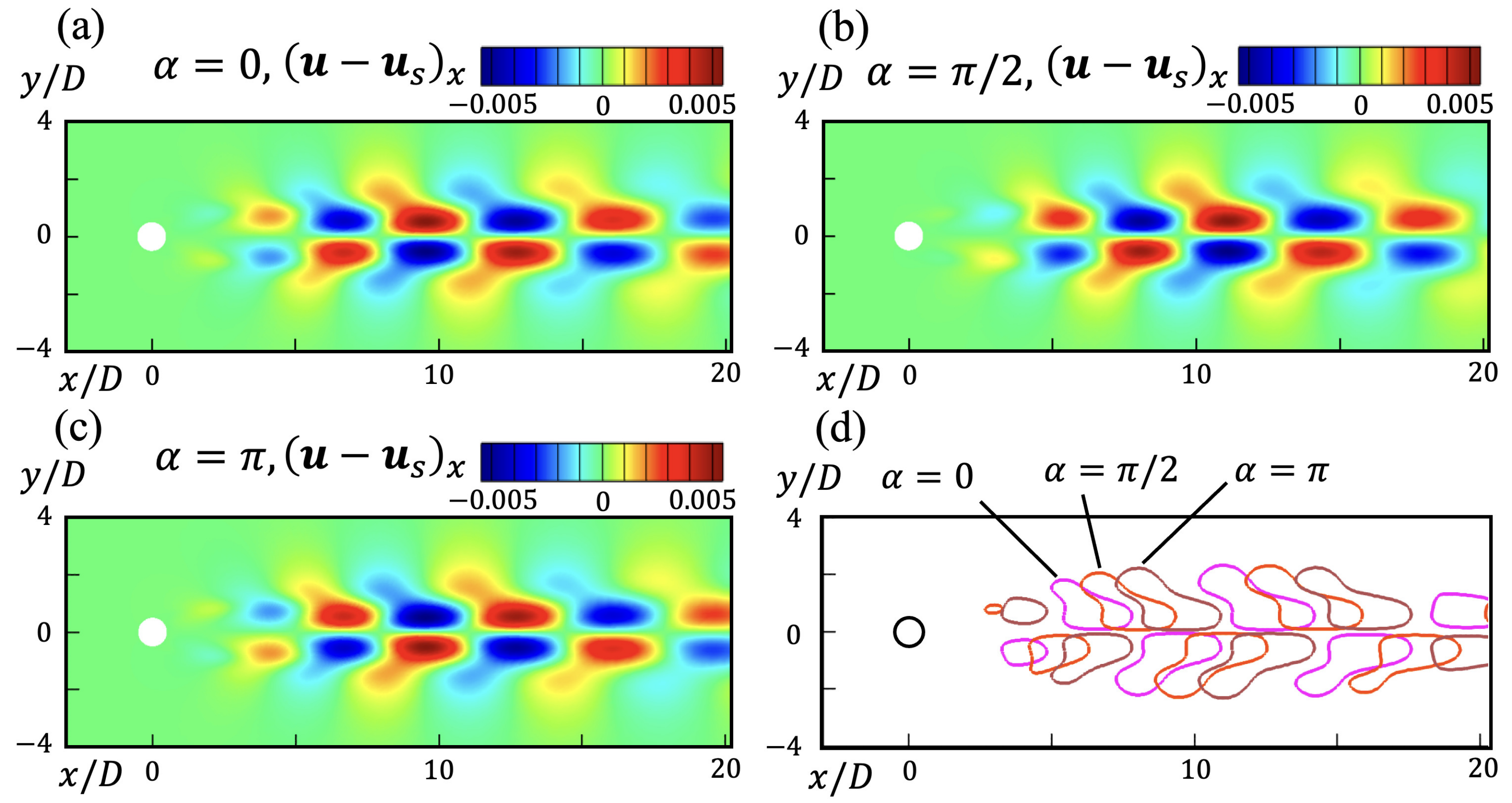}
\captionsetup{justification=raggedright,singlelinecheck=false}
\caption{Initial perturbation fields of tDMDpc at $\Rey=100$ (a) $\alpha=0$, (b) $\alpha=\pi/2$, (c) $\alpha=\pi$, and (d) contour lines of $0.001$ at $\alpha=0, \pi/2,$ and $\pi$. All perturbations are streamwise component.}
\label{fig_flowvariation}
\end{figure}

The transient flow process around a cylinder is decomposed using tDMDpc, with modes obtained from time-stepping LSA around a steady flow. Since LSA yields only one unstable frequency $f_1$, the initial flow field is constructed by assigning a sufficiently small amplitude $\epsilon_{f_1}$, thereby allowing the influence of the harmonic mode $f_n = nf_1$, where $n$ is natural number, to be neglected. In summary, the initial flow field for tDMDpc is given by
\begin{equation}
\boldsymbol{u}(\boldsymbol{x}, 0, \alpha) = \boldsymbol{u}_{s}+{\epsilon}_0|\boldsymbol{u}_{s}|_2\left\{e^{\alpha i}\boldsymbol{\varphi}_{f_1}(\boldsymbol{x})+e^{-\alpha i}\boldsymbol{\varphi}_{f_{-1}}(\boldsymbol{x})\right\},
   \label{initial_DMDpc}
\end{equation}
where $|\cdot|_2$ represents the $L2$ norm of the $N$-dimensional vector. In this context, the basic process $\boldsymbol{u}(\boldsymbol{x}, t, 0)$ is defined as the transient evolution initiated from the real part of the eigenmode initialy prepared by LSA. The parameters $\Delta T$ and $\epsilon_0$ are set to $0.1$ and $0.001$, respectively, which are the same values used in the time-stepping LSA. The initial perturbation fields at $\alpha = 0, \pi/2, \pi$ for $\Rey = 100$ are shown in figure \ref{fig_flowvariation}. According to equation (\ref{initial_DMDpc}), the initial perturbation field at $\alpha = 0$ corresponds to the real part of the eigenmode, with its absolute value scaled by $\epsilon_0$. Thus, the spatial distribution in figure \ref{fig_flowvariation} (a) is identical to that in figure \ref{fig_faif} (e). The initial perturbation field at $\alpha = \pi/2$ corresponds to the imaginary part of the eigenmode, also scaled by $\epsilon_0$. The initial perturbation field at $\alpha = \pi$ is equivalent to that at $\alpha = 0$ but with the sign inverted. Figure \ref{fig_flowvariation} (d) presents an overlay plot of the contours at $0.001$ for the three fields. The spatial distribution of the perturbation field changes depending on the value of $\alpha$, indicating that the phase angles of the oscillations differ.

\begin{figure}
	\centering\includegraphics[width=8cm,keepaspectratio]{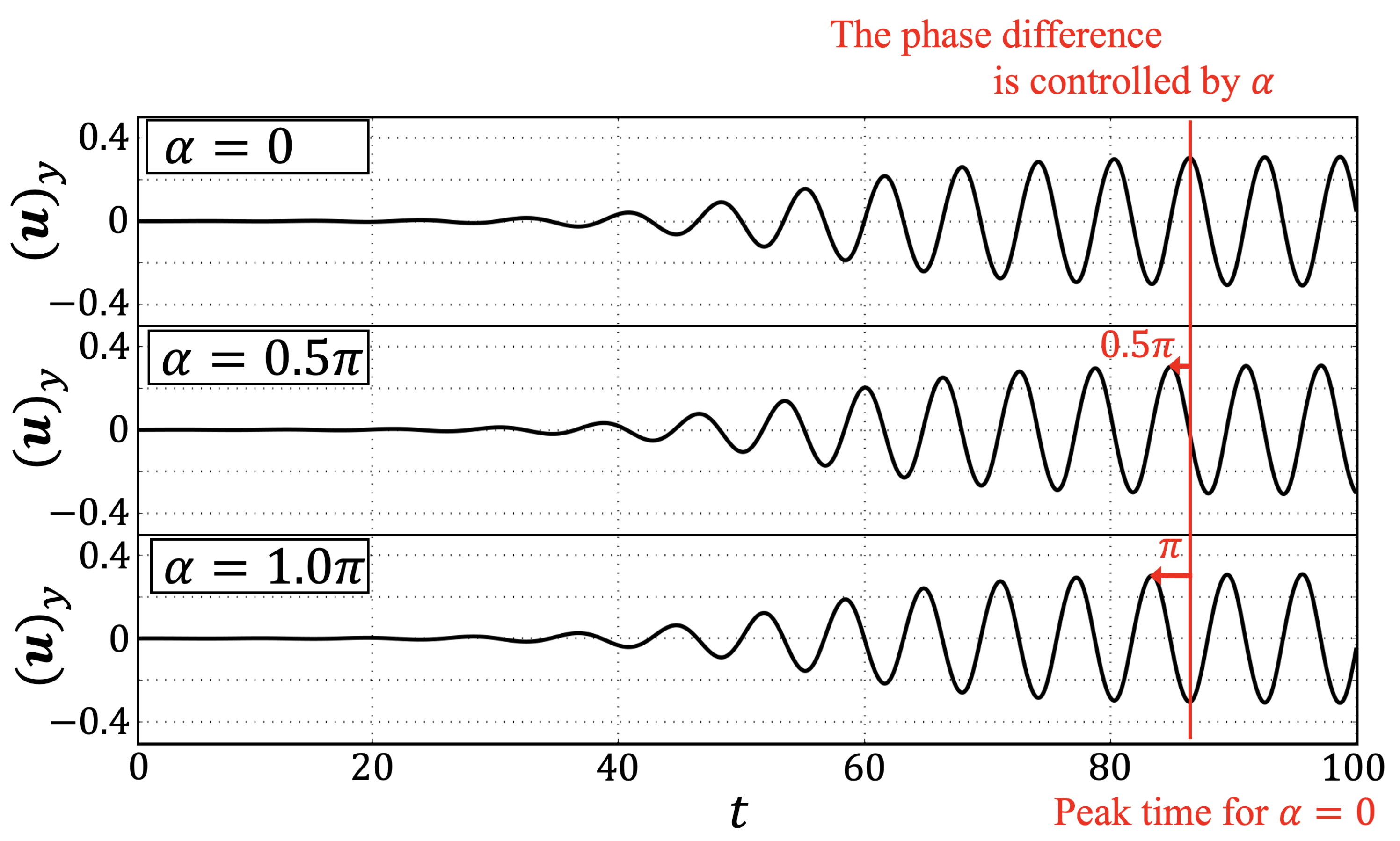}
  \captionsetup{justification=raggedright,singlelinecheck=false}
  \caption{Time variation of transverse velocity component at $(x/D,y/D) = (1,0)$ at $\Rey=100$, with different $\alpha$ in tDMDpc. The phase angle in the transient process can be controlled by $\alpha$.}
 \label{fig_velphase}
\end{figure}

The initial flow fields are prepared by varying the value of $\alpha$ under $j_\text{max} = 20$. Figure \ref{fig_velphase} shows the time variation of the transverse velocity at the wake position $(x/D, y/D) = (1, 0)$, computed from numerical simulations using the initial flow fields with phase-shifted perturbation presented in figure \ref{fig_flowvariation}. At $t=0$, the velocity variation is too small to be significant, but as time progresses, the velocity perturbation grows. Once the oscillation amplitude reaches a sufficiently large value, it stabilizes, and the flow field transitions to the post-transient state. A comparison of flow fields for different $\alpha$ values reveals that the oscillation phase shifts with respect to $\alpha$. Therefore, time-series data of flow fields evolving from initial flow fields with varying $\alpha$ provide the dataset of the development process with different phase angles.

Using the time-series data of the flow field developed from $j_\text{max} = 20$ different $\alpha$ values, we computed the flow field averaged over $\alpha$ by equation (\ref{time-average}) for snapshots at the same $t$. The number of snapshots, $j_\text{max}$, in tDMDpc, corresponds to the number of snapshots used for averaging and mode extraction. Therefore, a sufficiently large number of snapshots is required for accurate results. The convergence of the averaged fields and the modes extracted by tDMDpc with respect to $j_\text{max}$ is discussed in Appendix \ref{apenc}. Figure \ref{fig_0freq_transient} shows the $0$-contour line of the streamwise velocity in the flow field averaged over $\alpha$ at $t=30, 35, 40, 45, 55, 65,$ and $80$. The streamwise direction length of the recirculation region, bounded by the $0$-contour line, decreases as time progresses. The transient growth leading to the reduction of the recirculation region has been discussed in various studies \citep{shift, barkley2006linear, Manti_Lugo_2015}. According to \citet{Manti_Lugo_2015}, this reduction is caused by an increase in Reynolds stress induced by the growth of perturbations. Therefore, understanding the reduction of the recirculation region requires capturing the growth process of the perturbation components.

\begin{figure}
	\centering\includegraphics[width=10cm,keepaspectratio]{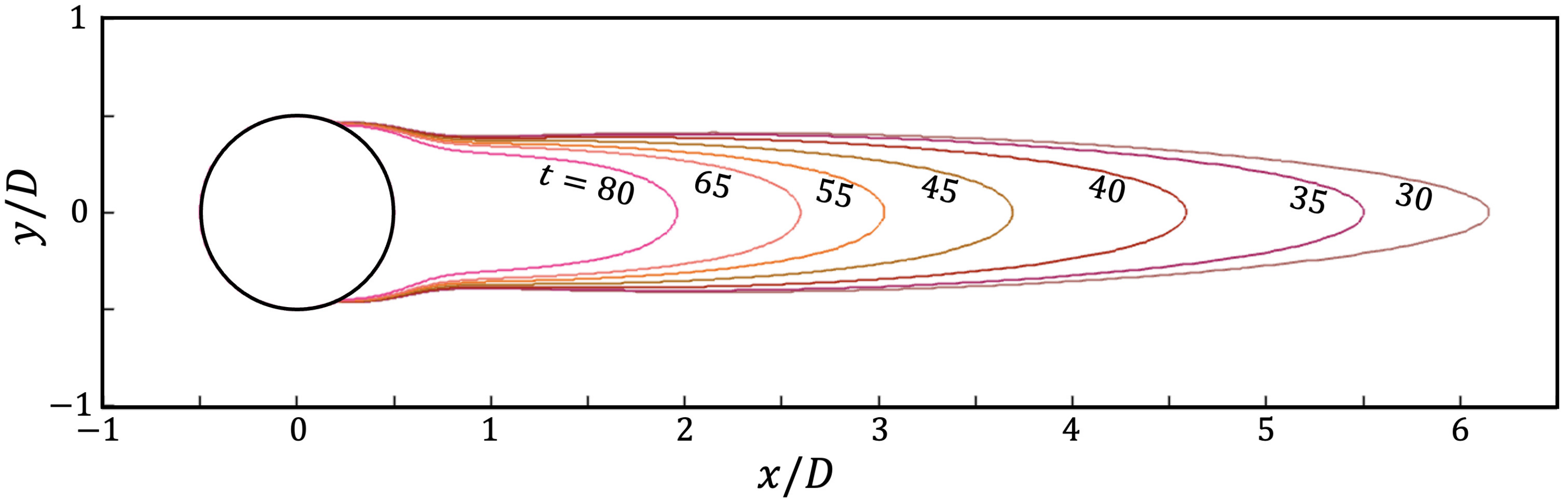}
  \captionsetup{justification=raggedright,singlelinecheck=false}
  \caption{Zero line for $x$-direction velocity fields (recirculation region) at various times averaged over $\alpha$ in the tDMDpc for $\Rey = 100$. As the perturbation grows, the recirculation region and length to the $x$-direction become small.}
 \label{fig_0freq_transient}
\end{figure}

\begin{figure}
	\centering\includegraphics[width=10cm,keepaspectratio]{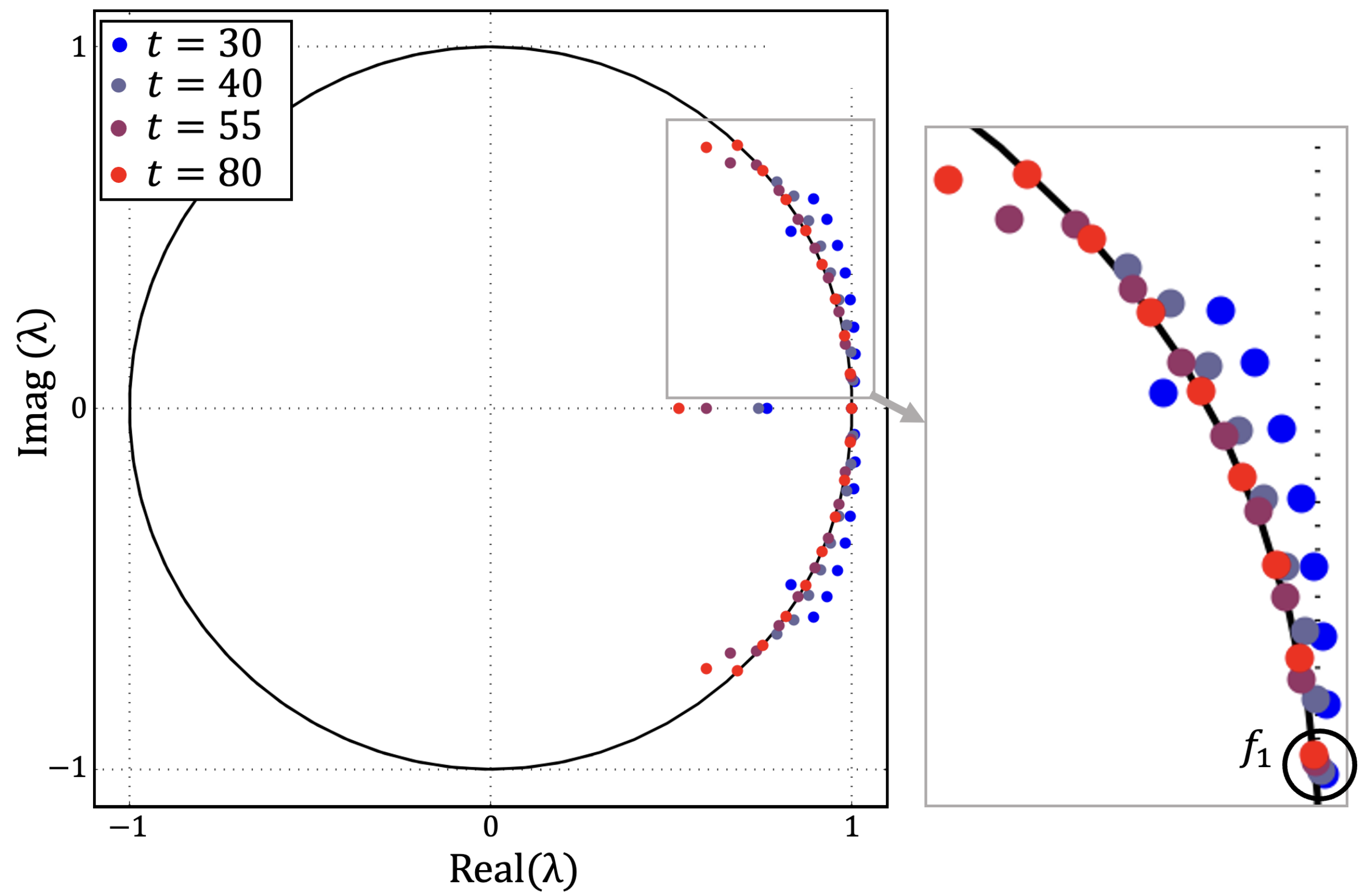}
  \captionsetup{justification=raggedright,singlelinecheck=false}
\caption{Ritz value  distribution of tDMDpc at $\Rey = 100$. The black line represents the unit circle, which indicates the zero-growth rate. Flow fields become a stable state because the Ritz value converges to the unit circle as time progresses.}
\label{fig_eigen_tDMDpc}
\end{figure}

Time-varying modes and eigenvalues (Ritz values) were extracted using tDMDpc from the time-series data $\boldsymbol{u}(\boldsymbol{x}, t, \alpha)$ obtained with $j_\text{max} = 20$, as illustrated in the conceptual diagram in figure \ref{fig_tDMDpc} (b). Figure \ref{fig_eigen_tDMDpc} shows the distribution of the  Ritz value of tDMDpc at times $t = 30, 40, 55, 80$. These times were chosen based on the variation in the $x$-direction length of the recirculation region (see figure \ref{fig_0freq_transient}). The  Ritz values that exist on the unit circle, indicated by the black line in the figure, correspond to a zero growth rate. At $t = 30$ and $40$, a large number of  Ritz values are found outside the unit circle, indicating that the flow fields are growing. As time progresses, the  Ritz values concentrate around the unit circle, suggesting that the flow field converges to a stable quasi-steady state. The few  Ritz values located inside the unit circle are likely numerical errors near the outflow boundary.

 The argument of each Ritz value, that is, its declination from the $x$-axis, represents the angular frequency. At $t=30$, the  Ritz values  with the lowest angular frequency, except for the zero-frequency, closely match the  Ritz values  of the most unstable mode $\boldsymbol{\varphi}_{f_1}(\boldsymbol{x})$ obtained from LSA. Therefore, the  Ritz value with the smallest angular frequency result from the nonlinear growing of the most unstable modes of LSA, which were used in the initial perturbation fields. Excluding numerical errors, it can be observed that the other frequencies correspond to harmonics of the $f_1$ frequency. This behavior arises from the nonlinear interaction between the $f_1$-frequency component and other finite-frequency components, leading to the appearance of modes with frequencies $f_n = f_{n-1} + f_1$ for $n = 2, 3, \cdots$.

\begin{figure}
	\centering\includegraphics[width=13cm,keepaspectratio]{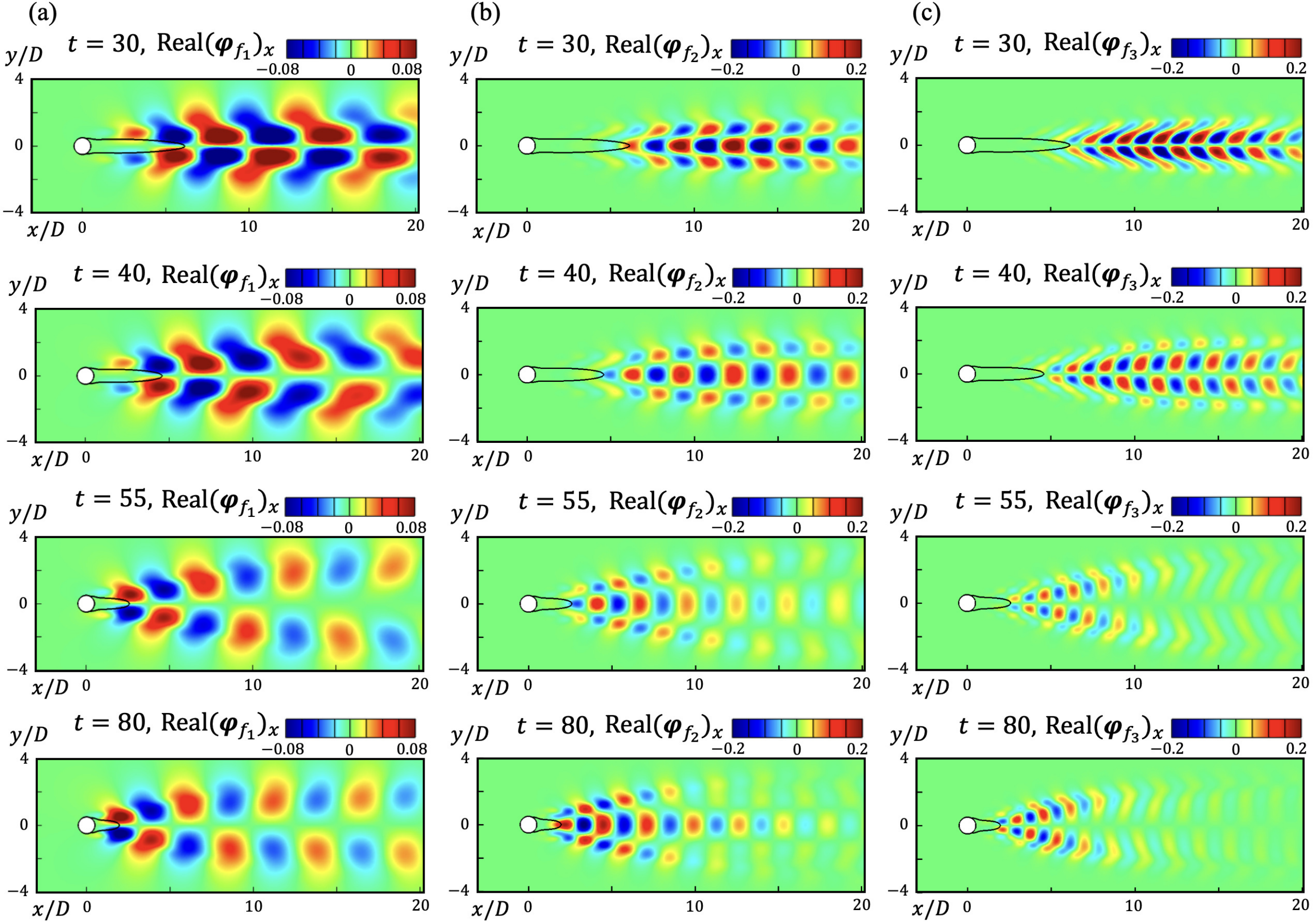}
  \captionsetup{justification=raggedright,singlelinecheck=false}
  \caption{Time variation of eigenmodes distribution from tDMDpc for (a) most lowest-frequency $f_1$, (b) $f_2 = 2 f_1$, and (c)  $f_3 = 3 f_1$. All distributions represent real parts of eigenmode, and $x$-direction components.}
 \label{fig_tDMDpc_fai}
\end{figure}

Figure \ref{fig_tDMDpc_fai} presents the modes obtained from tDMDpc at frequencies $0.1 < f_1 < 0.2$, $f_2 = 2 f_1$, and $f_3 = 3 f_1$, extracted at $t=30$, $40$, $55$, and $80$. The black line indicates the $0$-contour line of the averaged fields over $\alpha$. For all three frequency components, as the $x$-direction recirculation length decreases, the mode distribution gradually approaches a cylinder. At $t=80$, the well-known mode distributions for post-transient periodic flow around a cylinder, as reported in several studies \citep{Satomanifold_2024, AkhtarD, yeung_2024, shift, POD15}, are clearly visible. 
At $t = 30$, the fluctuation magnitude at a downstream location (e.g., $x/D \sim 10$) just beyond the recirculation region is comparable to that observed near the end of the recirculation region. However, as time advances, the fluctuations in the downstream region decay more rapidly, while those near the recirculation region persist or even intensify. These variations clearly depict the gradual evolution of the fluctuation field throughout the transient process, approaching the modes of the post-transient flow. Therefore, tDMDpc is an effective tool for extracting coherent structures during the transient process.

\begin{figure}
	\centering\includegraphics[width=13cm,keepaspectratio]{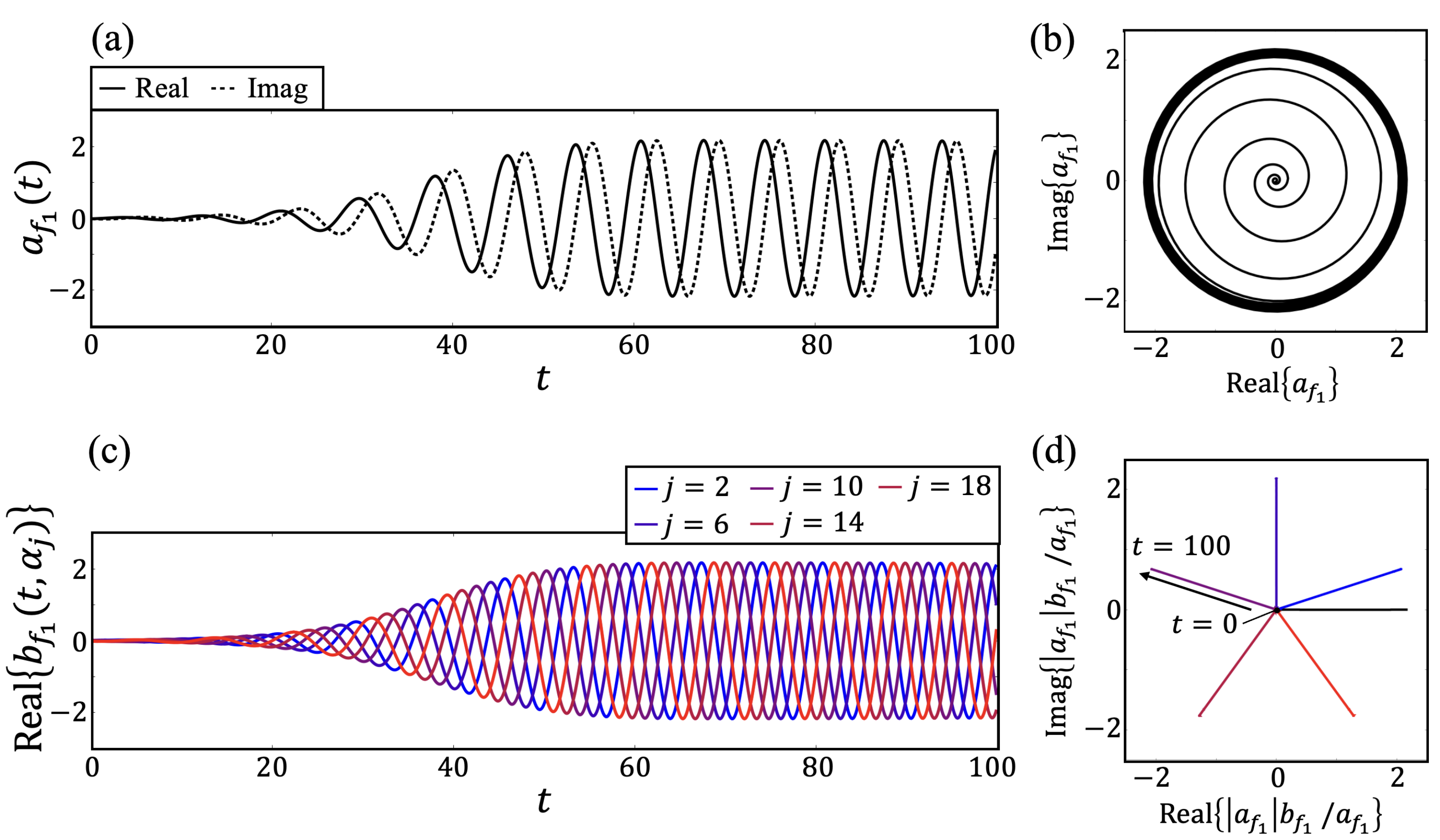}
  \captionsetup{justification=raggedright,singlelinecheck=false}
\caption{Temporal variation of coefficients for the tDMDpc modes. 
(a) Real and imaginary parts for $j=1$. 
(b) Phase portrait in the complex plane for $j=1$. 
(c) Real parts for $j=2,\,6,\,10,\,14,\,18$. 
(d) Phase relationships, with each coefficient normalized by that at $j=1$, showing constant phase offsets equal to the initial phase-shift angles $\alpha_j$.}
\label{fig_ampcoe}
\end{figure}

We focus on the transient quantities obtained from tDMDpc and their relationships. Figure \ref{fig_ampcoe}(a) shows the time evolution of the amplitude coefficient, computed from equation (\ref{calcoefficient}) using the tDMDpc modes and the time series data of the basic process $\boldsymbol{u}(\boldsymbol{x}, t, 0)$. As time progresses, the oscillation amplitude increases. Because the amplitude coefficient is complex, it can be plotted in the complex plane, as in figure \ref{fig_ampcoe}(b), to obtain a phase portrait. When the amplitude is small, the trajectory stays near the origin; at later times, it approaches a limit cycle. This behavior is consistent with phase-portrait models reported previously.

Figure \ref{fig_ampcoe}(c) presents the temporal evolution of the coefficients obtained from equation (\ref{calcoefficient_general}) using the tDMDpc modes and the time series $\boldsymbol{u}(\boldsymbol{x}, t, \alpha_j)$ with $j=2, 6, 10, 14, 18$. Because their phase-shift angles differ, these series exhibit distinct, constant phase offsets. To quantify this, we divided each coefficient by the normalized amplitude coefficient for $j=1$. The results, plotted in the complex plane in figure \ref{fig_ampcoe}(d), show that for $j=1$ the values are real (zero phase-shift angle) at all times, corresponding to the amplitude coefficient. For $j \neq 1$, the complex argument gives the phase-shift angle, and the modulus gives the oscillation amplitude. Notably, the phase-shift angles remain constant in time for all time-series data. Thus, during the nonlinear growth of the cylinder wake, the evolution for $j \neq 1$ can be represented by the basic process at $j=1$ shifted by an initial phase-shift angle $\alpha_j$; equivalently,
\begin{equation}
b_{f_1}(t,\alpha_j)=e^{i\alpha_j}\,a_{f_1}(t).
\end{equation}
We also confirmed that higher-frequency tDMDpc modes maintain constant phase angles, but those results are omitted.

Focusing on the distinction from other operator-based methods, our modes capture both the fundamental frequency component and its harmonics within a single mode. This is achieved by allowing the best-fit linear operator itself to vary in time, thereby representing the time evolution of eigenmodes. In contrast, other operator-based methods, such as recursive DMD \citep{noack2016recursive} and Koopman mode decomposition \citep{Bagheri_2013}, represent the entire time series using time-invariant modes. The frequencies and growth rates obtained from such operators correspond to representative temporal properties of the eigenmodes rather than instantaneous values. Consequently, although the tDMDpc approach is less efficient in data compression, it provides greater interpretability.

We perform tDMDpc for $\Rey$ values other than $100$ to investigate the dependence of the flow variation on $\Rey$. For $\Rey \approx 150$, the vortex street transitions to a secondary vortex in the post-transient flow \citep{taneda1959downstream, Jiang_2019}. In this case, the wake vortex is no longer composed solely of harmonics of $f_1$. However, since the scope of this study is the transient process from a steady flow, the secondary vortex lies outside the scope of this work. Therefore, we address the transient process only for $\Rey \leq 100$, which is well below $150$.

\begin{figure}
	\centering\includegraphics[width=13cm,keepaspectratio]{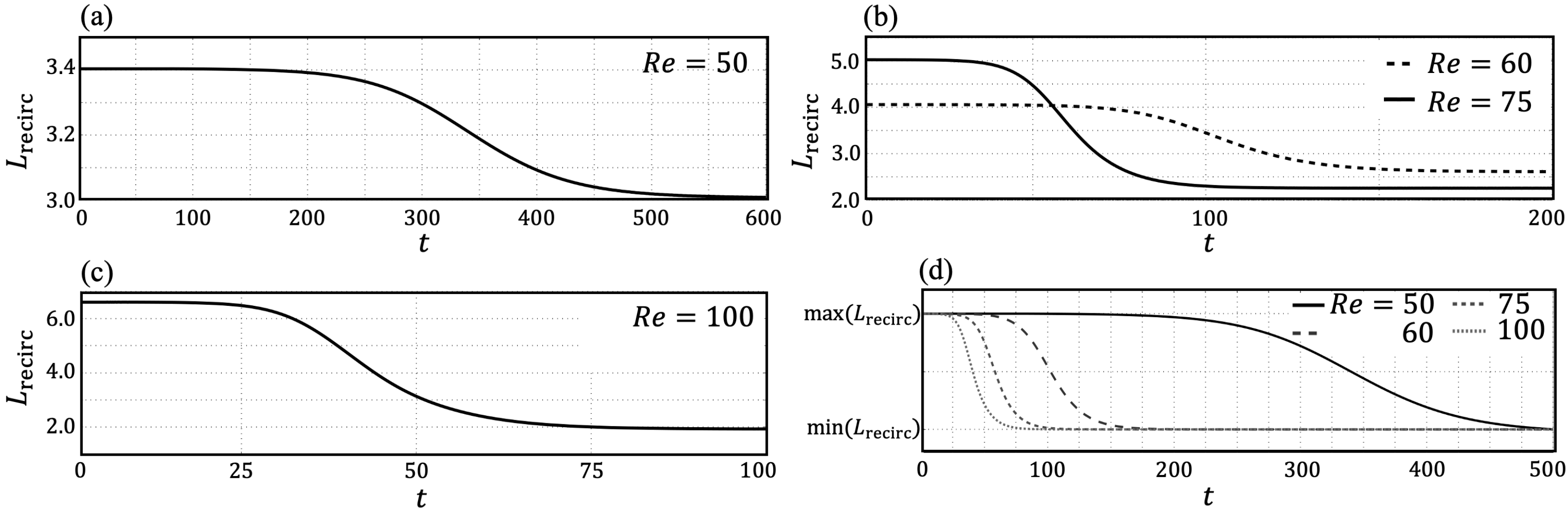}
  \captionsetup{justification=raggedright,singlelinecheck=false}
  \caption{Time variation of $L_{\text{recirc}}$ in the transient process: (a) $\Rey=50$, (b) $60$ and $75$, (c) $100$, and (d) $\Rey=50, 60, 75,$ and $100$, normalized by the maximum and minimum $L_{\text{recirc}}$.}
 \label{fig_Lrecirc_variousRe}
\end{figure}

For $\Rey = 50, 60,$ and $75$, initial fields were prepared using an eigenmode from LSA with $j_\text{max}=20$ for each $\Rey$, and time-series data for the transient process were obtained. From these transient data, we computed the flow field averaged over $\alpha$ as described in equation (\ref{time-average}). Figure \ref{fig_Lrecirc_variousRe} illustrates the temporal evolution of the streamwise recirculation length, $L_\text{recirc}$, in the averaged field for $\Rey = 50, 60, 75,$ and $100$. For all $\Rey$ values, $L_\text{recirc}$ decreases over time. After sufficient growth, the recirculation region stabilizes at a constant length, although the time required to reach this state is longer for smaller $\Rey$ values. Figure \ref{fig_Lrecirc_variousRe} (d) presents a normalized plot of $L_\text{recirc}$ for the four $\Rey$ cases, where the vertical axis is scaled by the maximum and minimum lengths of the respective recirculation regions. The temporal trends of $L_\text{recirc}$ are remarkably similar across all $\Rey$ cases. 

\begin{figure}
	\centering\includegraphics[width=13cm,keepaspectratio]{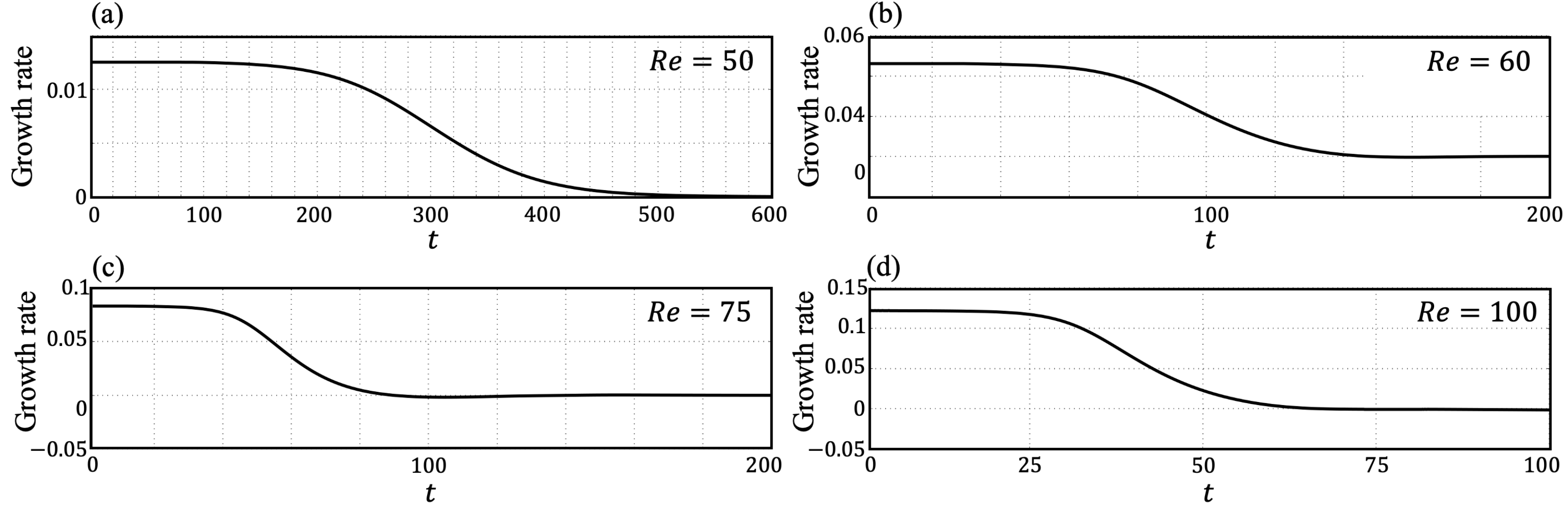}
  \captionsetup{justification=raggedright,singlelinecheck=false}
  \caption{Time variation of growth rate in the transient process: (a) $\Rey=50$, (b) $60$, (c) $75$, and (d) $100$.}
 \label{fig_growth_variousRe}
\end{figure}

We performed tDMDpc using time-series data of transient processes at four $\Rey$ values and extracted time-varying modes along with their growth rates. A gradual variation in the mode distributions over time was observed for all four $\Rey$ values; however, for simplicity, the mode distributions are omitted here. For a more quantitative analysis, figure \ref{fig_growth_variousRe} shows the time variation of the growth rates of the $f_1$-frequency mode at the four $\Rey$ values. The growth rate gradually decreases over time and eventually reaches zero. This indicates that the perturbation field, which grows rapidly about the steady flow, ultimately converges to a stable periodic flow. The trend in the time variation of the growth rates closely resembles the time variation of the recirculation region. This suggests a potential connection between the recirculation region and the growth of perturbation fields. \citet{barkley2006linear} demonstrated that the growth rate of the unstable mode decreases as the recirculation region of the base flow becomes shorter using LSA. A similar trend was also reported by \citet{Manti_Lugo_2014,Manti_Lugo_2015}, and the present results are consistent with these findings. Since the growth rate is directly linked to energy transfer, as derived from the global energy transfer equation, investigating energy transfer in transient processes could provide further insight into these potential relationships.


\section{Model reduction approach and derivation of energy transfer equation}\label{sec3}
In this section, we describe a method for analyzing energy transfer and the energy budget using the tDMDpc modes. Our goal is to derive the energy transfer between modes and the energy budget of each eigenmode using the eigenmodes $\boldsymbol{\varphi}_{f_k}(\boldsymbol{x}, t)$ and their corresponding amplitude coefficients $a_{f_k}(t)$. The amplitude coefficients are computed directly from the dataset and the eigenmodes, without requiring time integration or numerical prediction.

\subsection{Derivation of phase averaged reduced-order model}
When employing a time-dependent base flow, the flow field that is phase-shifted by $\alpha$ from the basic process $\boldsymbol{u}(\boldsymbol{x}, t, 0)$ can be represented by the tDMDpc modes as follows:
\begin{equation}
\boldsymbol{u}(\boldsymbol{x},t, \alpha) = \boldsymbol{{u}}_{t}(\boldsymbol{x},t)+\sum_{l=-\infty}^{\infty}a_{f_l}(t) e^{\alpha^{f_l}i}\boldsymbol{\varphi}_{f_l}(\boldsymbol{x},t),
   \label{DMDrepresent_new}
\end{equation}
where $\alpha^{f_l} = \frac{f_l}{f_c}\,\alpha$ represents the phase-shift angle from the basic process at $f_l$ frequency.
Strictly speaking, DMD extracts only a finite set of modes; consequently, the summation in equation (\ref{DMDrepresent_new}) is truncated to a finite number of terms, and the representation holds only approximately. In addition, the reconstruction of the instantaneous flow field using tDMDpc modes is performed around the base flow, and therefore the zero-frequency mode is not required in that representation. However, since the energy-budget equations are derived for each amplitude coefficient, the zero-frequency mode is included to obtain the corresponding budget equation for the mean flow. An example of such a zero-frequency energy budget can be found in the mean-field correction introduced by \citet{shift}, often referred to as the shift mode.

To derive the energy transfer and budget equation, equation (\ref{DMDrepresent_new}) is substituted into the governing equation (\ref{eqnavi}) as follows:
\begin{equation}
\begin{split}
\sum_{l=-\infty}^{\infty}&\left\{\frac{d}{d{t}}a_{f_l} e^{\alpha^{f_l}i}{\boldsymbol{\varphi}_{f_l}}\right\}=\\
&-\sum_{l=-\infty}^{\infty}\sum_{n=-\infty}^{\infty}{a_{f_l}a_{f_n}e^{(\alpha^{f_l} + \alpha^{f_n})i}}(\boldsymbol{\varphi}_{f_l}\cdot\nabla)\boldsymbol{\varphi}_{f_n}\\
\,&\,\,\,+\sum_{l=-\infty}^{\infty}a_{f_l} e^{\alpha^{f_l}i}\left\{-(\boldsymbol{{u}}_{t}\cdot\nabla)\boldsymbol{\varphi}_{f_l}-(\boldsymbol{\varphi}_{f_l}\cdot\nabla)\boldsymbol{{u}}_{t}
+\frac{1}{\Rey}{\nabla^2}\boldsymbol{\varphi}_{f_l}\right\}\\
\,&\,\,\,\,\,\,\,\,\,\,\,\,-\frac{d\boldsymbol{u}_{t}}{d{t}}
-(\boldsymbol{u}_{t}\cdot\nabla)\boldsymbol{{u}}_{t}
+\frac{1}{\Rey}{\nabla^2}\boldsymbol{u}_{t}
-\frac{1}{\rho}\nabla{p}.
   \label{LSAeq_GP_new}
   \end{split}
\end{equation}
Here, the time derivative on the left-hand side is replaced with the instantaneous linear operator of the continuous dynamical system, so that
\begin{eqnarray}
\frac{d}{dt}\left(a_{f_l} e^{i\alpha^{f_l}}\boldsymbol{\varphi}_{f_l}\right)&=& \mathcal{A}(t) a_{f_l} e^{i\alpha^{f_l}} \nonumber \\
&=& (\sigma_{f_l}+2\pi f_l i)\,a_{f_l} e^{i\alpha^{f_l}}\boldsymbol{\varphi}_{f_l}.
\label{LSAeq_timederive_new}
\end{eqnarray}
Strictly speaking, this deformation requires treating the dynamical system as fixed under the instantaneous linear operator. The time variation of the numerically computed eigenmodes and eigenvalues satisfies this requirement only when the dynamically orthogonal condition is fulfilled. Further details are provided in Appendix~\ref{dynamicalyorthognal}.

In equation (\ref{LSAeq_GP_new}), taking the inner product with the $f_k$-frequency component $a_{f_k} e^{i\alpha^{f_k}}\boldsymbol{\varphi}_{f_k}$, we obtain 
\begin{eqnarray}
\sum_{l=-\infty}^{\infty}&&(\sigma_{f_l}+2\pi f_l i)\,a_{f_l}a^*_{f_k} e^{(\alpha^{f_l}-\alpha^{f_k})i}D_{{f_l}{f_k}}\nonumber \\
&=&\sum_{l=-\infty}^{\infty}\sum_{n=-\infty}^{\infty}{a_{f_l}a_{f_n}a^*_{f_k}e^{(\alpha^{f_l} + \alpha^{f_n} - \alpha^{f_k})i}}F_{{f_l}{f_n}{f_k}}\nonumber\\
&&\,\,\,\,\,\,\,\,\,\,\,\,\,\,\,\,\,\,\,\,\,\,\,
+\sum_{l=-\infty}^{\infty}a_{f_l}a^*_{f_k} e^{(\alpha^{f_l}-\alpha^{f_k})i}G_{{f_l}{f_k}}
+(H_{f_k}+J_{f_k}+P_{f_k}) a^*_{f_k}e^{- \alpha^{f_k}i},
 \label{LSAeq_GP2_new}
\end{eqnarray}
where 
\begin{eqnarray}
&D_{{f_l}{f_k}}& \stackrel{\mathrm{def}}{=} \langle\boldsymbol{\varphi}_{f_l}, \boldsymbol{\varphi}_{f_k} \rangle, \label{LSAeq_D_new} \\
&F_{{f_l}{f_n}{f_k}}& \stackrel{\mathrm{def}}{=} -\langle(\boldsymbol{\varphi}_{f_l}\cdot\nabla)\boldsymbol{\varphi}_{f_n}, \boldsymbol{\varphi}_{f_k}\rangle, \label{LSAeq_F_new} \\
&G_{{f_l}{f_k}}& \stackrel{\mathrm{def}}{=} {\frac{1}{\Rey}}\langle{\nabla^2}\boldsymbol{\varphi}_{f_l},{\boldsymbol{\varphi}}_{f_k}\rangle-{\langle(\boldsymbol{{u}}_{t}\cdot\nabla)\boldsymbol{\varphi}_{f_l},{\boldsymbol{\varphi}}_{f_k}\rangle}-{\langle(\boldsymbol{\varphi}_{f_l}\cdot\nabla)\boldsymbol{{u}}_{t},{\boldsymbol{\varphi}}_{f_k}\rangle}, \label{LSAeq_G_new} \\
&H_{{f_k}}& \stackrel{\mathrm{def}}{=} {{\frac{1}{\Rey}}\langle{\nabla^2}\boldsymbol{{u}}_{t},{\boldsymbol{\varphi}}_{f_k}\rangle}-{\langle(\boldsymbol{u}_{t}\cdot\nabla)\boldsymbol{u}_{t},{\boldsymbol{\varphi}}_{f_k}\rangle}, \label{LSAeq_H_new} \\
&J_{{f_k}}& \stackrel{\mathrm{def}}{=} -\langle\frac{d\boldsymbol{u}_{t}}{d{t}}, \boldsymbol{\varphi}_{f_k} \rangle, \label{LSAeq_J_new} \\
&P_{f_k}& \stackrel{\mathrm{def}}{=} -\langle \frac{1}{\rho}\nabla{p},\boldsymbol{\varphi}_{f_k} \rangle. \label{LSAeq_P_new} 
\end{eqnarray}
Note that the pressure term $P_{f_k}$ in equation (\ref{LSAeq_GP2_new}) is negligible for cylinder flow \citep{cylinder7,pressure2}.

However, since eigenmodes of the linear operator are not necessarily orthogonal, $D_{f_l f_k} \neq \delta_{lk}$, where $\delta_{lk}$ denotes the Kronecker delta. To eliminate the cross terms $e^{i(2\pi f_l t + \alpha^{f_l})} D_{f_l f_k}$ ($l \neq k$) appearing in the first term on the left-hand side of equation (\ref{LSAeq_GP2_new}), we introduce averaging in the phase angle $\theta$, denoted by
\begin{eqnarray}
\overline{f(\theta)}^{\theta} \stackrel{\mathrm{def}}{=} \lim_{\Theta \to \infty} \frac{1}{2\Theta} \int^{\Theta}_{-\Theta} f(\theta) d\theta,
   \label{infty_average_new}
\end{eqnarray}
where $\overline{\cdot}^{\theta}$ denotes the phase-averaging operation, and $f(\theta)$ is a real- or complex-valued function. For convenience, the phase average is defined as an average over the interval $[-\infty, \infty]$. In numerical computations, however, the averaging is truncated to a finite interval. Nevertheless, as shown later, when the phase-shift angle $\alpha$ and amplitude coefficients $a_{f_k}$ can be treated independently, the phase-averaged quantities can be obtained analytically, and numerical averaging is not required.
If the phase of the basic state is denoted by $\theta_0$, then the phase-shifted state by $\alpha$ is given by $\theta = \theta_0 + \alpha$. Hence, since $d\theta = d\alpha$, we obtain
\begin{eqnarray}
\overline{f(\theta)}^{\theta}=\overline{f(\alpha)}^{\alpha},
\label{infty_average_alpha_new}
\end{eqnarray}
where, for notational convenience, we set $f(\alpha) = f(\theta_0 + \alpha)$.

By averaging equation (\ref{LSAeq_GP2_new}) with respect to the phase-shift angle $\alpha$, we obtain 
\begin{eqnarray}
\sum_{l=-\infty}^{\infty}&&\overline{(\sigma_{f_l}+2\pi f_l i)e^{(\alpha^{f_l}-\alpha^{f_k})i}a_{f_l}a^*_{f_k}D_{{f_l}{f_k}}}^{\alpha}\nonumber \\
&=&\sum_{l=-\infty}^{\infty}\sum_{n=-\infty}^{\infty}\overline{{a_{f_l}a_{f_n}a^*_{f_k}e^{(\alpha^{f_l} + \alpha^{f_n} - \alpha^{f_k})i}}F_{{f_l}{f_n}{f_k}}}^{\alpha} \nonumber\\
&&\,\,\,\,\,\,\,\,\,\,\,\,\,\,\,\,\,\,\,\,\,\,\,\,\,\,
+\sum_{l=-\infty}^{\infty}\overline{a_{f_l}a^*_{f_k} e^{(\alpha^{f_l}-\alpha^{f_k})i}G_{{f_l}{f_k}}}^{\alpha} +\overline{(H_{f_k}+J_{f_k}) a^*_{f_k}e^{- \alpha^{f_k}i}}^{\alpha}.
 \label{LSAeq_average1_new}
\end{eqnarray}
In the context of tDMDpc, since the base flow $\boldsymbol{u}_{t}$ and eigenmodes are computed from datasets using the phase-shift angle $\alpha$ as a statistical parameter, $\boldsymbol{u}_{t}$ and $\boldsymbol{\varphi}_{f_l}$ are not functions of the phase-shift angle $\alpha$. Thus, all the coefficients defined in equations  (\ref{LSAeq_D_new}–\ref{LSAeq_J_new}) are independent of $\alpha$. 
Strictly speaking, the amplitude $a_{f_l}$ may vary with $\alpha$. However, if we assume that all eigenmodes have the same growth rate and frequency for any phase-shift angle, then $a_{f_l}$, which represents the time-integrated value of growth rate, becomes independent of $\alpha$. Consequently, the phase-averaged equation yields
\begin{eqnarray}
\sum_{l=-\infty}^{\infty}&&(\sigma_{f_l}+2\pi f_l i)a_{f_l}a^*_{f_k}\overline{e^{(\alpha^{f_l}-\alpha^{f_k})i}}^{\alpha}D_{{f_l}{f_k}} \nonumber \\
&=&\sum_{l=-\infty}^{\infty}\sum_{n=-\infty}^{\infty}{a_{f_l}a_{f_n}a^*_{f_k}\overline{e^{(\alpha^{f_l} + \alpha^{f_n} - \alpha^{f_k})i}}^{\alpha}F_{{f_l}{f_n}{f_k}}} \nonumber\\
&&\,\,\,\,\,\,\,\,\,\,\,\,\,\,\,\,\,\,\,\,\,
+\sum_{l=-\infty}^{\infty}a_{f_l}a^*_{f_k}\overline{e^{(\alpha^{f_l}-\alpha^{f_k})i}}^{\alpha}G_{{f_l}{f_k}}
+(H_{f_k}+J_{f_k}) a^*_{f_k}\overline{e^{-\alpha^{f_k}i}}^{\alpha}.
 \label{LSAeq_average2_new}
\end{eqnarray}

Here, since the general results
\begin{eqnarray}
\overline{e^{- c \alpha i}}^{\alpha}=\lim_{\Theta \to \infty} \frac{1}{2\Theta}
\int_{-\Theta}^{\Theta} e^{- c\alpha i} d\alpha =     
\begin{cases}
        1, & c = 0, \\
        0, & c \neq 0,
    \end{cases}
\label{infty_average_alpha_new}
\end{eqnarray}
holds, it follows that 
\begin{eqnarray}
&\overline{e^{\alpha^{f_l}-\alpha^{f_k}i}}^{\alpha}& = \overline{e^{\frac{f_l-f_k}{f_c}\alpha i}}^{\alpha} = \delta_{lk},\\
&\overline{e^{-\alpha^{f_k}i}}^{\alpha}& = \overline{e^{-\frac{f_k}{f_c}\alpha i}}^{\alpha} = 0,\,\,\, \text{when} \,\,\, f_k \neq 0,\\
&\overline{e^{-\alpha^{f_k}i}}^{\alpha}& = \overline{e^{-\frac{f_k}{f_c}\alpha i}}^{\alpha} = 1,\,\,\, \text{when} \,\,\, f_k = 0,\\
&\overline{e^{(\alpha^{f_l} + \alpha^{f_n} - \alpha^{f_k})i}}^{\alpha}& = \overline{e^{\frac{f_l+f_n-f_k}{f_c}\alpha i}}^{\alpha} = 0 \,\,\, \text{when} \,\,\, f_l+f_n-f_k \neq 0,\\
&\overline{e^{(\alpha^{f_l} + \alpha^{f_n} - \alpha^{f_k})i}}^{\alpha}& = \overline{e^{\frac{f_l+f_n-f_k}{f_c}\alpha i}}^{\alpha} = 1 \,\,\,  \text{when} \,\,\, f_l+f_n-f_k = 0,
\label{infty_average_alpha_new}
\end{eqnarray}
where $\alpha^{f_k} = \frac{f_k}{f_c}\alpha$.
Thus, equation (\ref{LSAeq_average2_new}) become
\begin{eqnarray}
(\sigma_{f_k}+2\pi f_k i)a_{f_k}a^*_{f_k}D_{{f_k}{f_k}}
=\sum_{l=-\infty}^{\infty}{a_{f_l}a_{f_{k-l}}a^*_{f_k}F_{{f_l}{f_{k-l}}{f_k}}}+a_{f_k}a^*_{f_k}G_{{f_k}{f_k}}.
 \label{LSAeq_average3_new}
\end{eqnarray}
Since we chose the absolute value of tDMDpc mode as unity, thus, $D_{{f_k}{f_k}}=1$, the real and imaginary parts of the equations are
\begin{eqnarray}
\sigma_{f_k}|a_{f_k}|^2
&=&\sum_{l=-\infty}^{\infty}\text{Real}\{{a_{f_l}a_{f_{k-l}}a^*_{f_k}F_{{f_l}{f_{k-l}}{f_k}}}\}+|a_{f_k}|^2\text{Real}\{G_{{f_k}{f_k}}\}, \label{globalenergytransfer_transient}\\
2\pi f_k|a_{f_k}|^2
&=&\sum_{l=-\infty}^{\infty}\text{Imag}\{{a_{f_l}a_{f_{k-l}}a^*_{f_k}F_{{f_l}{f_{k-l}}{f_k}}}\}+|a_{f_k}|^2\text{Imag}\{G_{{f_k}{f_k}}\}.
 \label{LSAeq_average4_new}
\end{eqnarray}
As a result, all time-derivative terms except those associated with the frequency $f_k$ vanish, yielding an instantaneous-frequency-domain representation of the governing equation. When a finite number of modes is employed, this formulation can be regarded as a reduction obtained by projecting onto an instantaneous frequency space. Therefore, we refer to this model as the phase-averaged ROM.
The real part of the equation represents a time-varying energy budget that considers the time variation of energy. Here, the first term on the left-hand side represents the triadic energy transfer among different frequency eigenmodes. The term triad refers to the dynamically triadic relationship at arbitrary times,
\begin{equation}
	f_k(t) \mp f_l(t) \pm f_{k-l}(t) = 0.
	\label{triadic_dynamic}
\end{equation}

\subsection{In the case of time-independent eigenmodes}
To assess the validity of the derivation based on tDMDpc modes, we compare it with the frequency-domain Navier–Stokes equations for a conventional stationary flow field. 
When the eigenmode remains unchanged in time, varying only with a constant frequency and zero growth rate, the amplitude coefficient is determined by the initial amplitude $a_{f_k}(t=0)$ and the frequency as follows
\begin{eqnarray}
a_{f_k}(t) = a_{f_k}(0)e^{2\pi f_k t i}.
\end{eqnarray}
Thus, equations (\ref{globalenergytransfer_transient}) and (\ref{LSAeq_average4_new}) becomes
\begin{eqnarray}
2\pi f_k i|a_{f_k}(0)|^2
=&&\sum_{l=-\infty}^{\infty}{e^{2\pi  (f_l+f_{k-l}-f_k )ti}a_{f_l}(0)a_{f_{k-l}}(0)a^*_{f_k}(0)F_{{f_l}{f_{k-l}}{f_k}}}\nonumber \\
&&\,\,\,\,\,\,\,\,\,\,\,\,\,\,\,\,\,\,\,\,\,\,\,\,\,\,\,\,\,\,\,\,\,\,\,\,\,\,\,\,\,\,\,\,\,\,\,\,\,\,\,\,\,\,\,\,\,\,\,\,\,\,\,\,\,\,\,\,\,\,\,\,\,\,\,\,\,\,\,\,\,\,\,\,\,\,\,\,\,\,\,\,\,\,\,\,\,
+|a_{f_k}(0)|^2G_{{f_k}{f_k}}.
 \label{LSAeq_average_static_new}
\end{eqnarray}
This has the equal form of the momentum equation\citep{freeman_2024}.
The real part of the derived equation is
\begin{eqnarray}
0=&&\text{Real}\left(\sum_{l=-\infty}^{\infty}{a_{f_l}(0)a_{f_{k-l}}(0)a^*_{f_k}(0)e^{2\pi (f_l + f_{k-l} - f_k) ti}F_{{f_l}{f_{k-l}}{f_k}}}\right)\nonumber \\
&&\,\,\,\,\,\,\,\,\,\,\,\,\,\,\,\,\,\,\,\,\,\,\,\,\,\,\,\,\,\,\,\,\,\,\,\,\,\,\,\,\,\,\,\,\,\,\,\,\,\,\,\,\,\,\,\,\,\,\,\,\,\,\,\,\,\,\,\,\,\,\,\,\,\,\,\,\,\,\,\,\,\,\,\,\,\,\,\,\,\,\,\,\,\,\,\,\,
+\text{Real}\left(|a_{f_k}(0)|^2G_{{f_k}{f_k}}\right).
 \label{LSAeq_average_static_budget_new}
\end{eqnarray}
This equation represents modal energy budget at frequency $f_k$, and which is equal to equation (\ref{ballance1}) with $\hat{\boldsymbol{u}}_{f_k} \rightarrow a_{f_k}(0){\boldsymbol{\varphi}}_{f_k}$. The first term of the right-hand side is a triadic term representing energy transfer between different frequency components, and the second term represents the viscous diffusion effect.

Since the set of DMD modes does not form an orthogonal basis, the sum of squared modal amplitudes does not necessarily correspond to the kinetic energy, and thus amplitude and energy are not always directly related. However, as shown in Appendix \ref{energyorthognal}, the sum of squared amplitudes equals the phase-averaged energy. Accordingly, equation (\ref{LSAeq_average_static_budget_new}) can be interpreted as representing a phase-averaged energy budget. In a stationary flow, temporal variations are not accompanied by amplitude growth and can therefore be regarded as equivalent to phase variations. As a result, the energy budget expressed in temporal statistics coincides with that expressed in phase statistics. Note that the phase-averaged energy budget assumes that the amplitude and phase are uncorrelated. Consequently, in turbulent flows where phase variations can exhibit intermittency, this assumption may not hold, and the applicability of the method should be carefully assessed.


\subsection{Linear growth from base flow}
We linearize the equation derived from the phase-averaged ROM around the base flow and compare it with the conventional energy budget equation for a pair of eigenmodes. In this case, the eigenmode is comparable to the LSA mode. When linearized about the base flow, the equation becomes
\begin{eqnarray}
(\sigma_{f_k}+2\pi f_k i)|a_{f_k}|^2 =|a_{f_k}|^2G_{{f_k}{f_k}}.
 \label{eq:ROMgrowth}
\end{eqnarray}
Considering that the real part of equation (\ref{LSAeq_average_static_budget_new}) for the stationary flow represents the modal energy budget equation, the real part of equation (\ref{eq:ROMgrowth}) can be regarded as the energy budget equation for eigenmodes with a finite growth rate.
The real part is given by
\begin{eqnarray}
\sigma_{f_k}|a_{f_k}|^2=\text{Real}(G_{f_k f_k})|a_{f_k}|^2.
\label{globalenergytransfer_steady}
\end{eqnarray}
We refer to the equation for the energy budget, which includes the linear growth of the modal energy, as the global energy transfer equation, since it represents the overall transfer of energy to the eigenmode of interest. Futhermore, growth rate and frequency of $f_k$-frequency eigenmode are
\begin{eqnarray}
\sigma_{f_k}&=& \text{Real}({G_{{f_k}{f_k}}}),  \label{eq:ROMgrowth2}\\
f_k&=& \frac{\text{Imag}{(G_{{f_k}{f_k}})}}{2\pi}. \label{eq:ROMgrowth_freq}
\end{eqnarray}
It has been noted by \citet{Mittal_2009} that the growth rate of an eigenmode can be represented as the sum of three terms, which constitute the quantity $G_{f_k f_k}$. Our formulation of the growth rate is in complete agreement with their result, in which the growth rate is represented in terms of the eigenmode and the base flow.

The coefficients of the second-order terms leads to the simple conclusion that the growth rate is equal to $\text{Real}(G_{{f_k}{f_k}})$. When kinetic energy is defined as $\boldsymbol{u}^T\boldsymbol{u}$, energy transfer represents $2\sigma\boldsymbol{u}^T\boldsymbol{u}$. That is, once $\boldsymbol{u}$ is excited, the dynamical system acts as an amplifier of $\boldsymbol{u}$'s oscillation. The same result has also been reported by \citet{shift}. Hence, the change in sign of the growth rate across the pre- and post-critical $\Rey$ indicates a shift in the energy transfer direction. This simple conclusion connects dynamical systems and fluid dynamics. Note that for obtaining the structure that first transfers energy to $\boldsymbol{u}$, an approach based on the adjoint LSA \citep{GIANNETTI_2007, Luchini_2014, Ohmichi_2021} or resolvent analysis is a reasonable choice.

\section{Energy transfer analysis for  cylinder flow}\label{sec5}
\subsection{Linear growth regime from steady base flow}
\begin{figure}
	\centering\includegraphics[width=13cm,keepaspectratio]{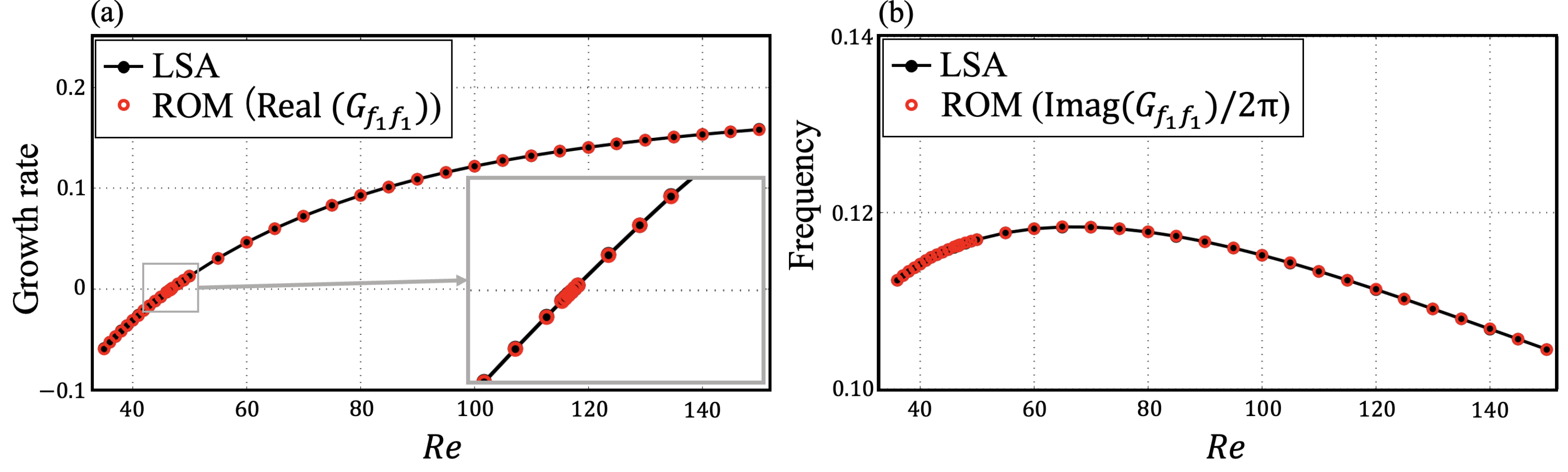}
  \captionsetup{justification=raggedright,singlelinecheck=false}
  \caption{Comparison of growth rate and frequency between LSA and ROM, where growth rate and frequency of ROM are given by $\text{Real}(G_{{f_1}{f_1}})$ and $\text{Imag}(G_{{f_1}{f_1}})/(2\pi)$, respectively. ROM closely approximates the growth rate and frequency of the eigenmode from LSA.}
 \label{ROMgrowth}
\end{figure}

Based on equation (\ref{globalenergytransfer_steady}), which provides expressions for the growth rate and frequency in the linear-growth regime of the phase-averaged ROM, we compute growth rates from the LSA modes and the steady flow, and discuss the resulting energy-transfer relationships. Figure \ref{ROMgrowth} shows the growth rates and those obtained from LSA (indicated by the black line). The growth rates obtained from the ROM are in close agreement with the LSA results. Therefore, the ROM with steady flow and unstable modes successfully reproduces the dynamical system around the steady flow.

Because the modal growth rate and frequency are given by the real and imaginary parts of $G_{f_1 f_1}$, respectively, examining the relationship between $G_{f_1 f_1}$, the LSA modes, and the steady base flow provides deeper insight. The treatment of the imaginary part is discussed in \citet{freeman_2024}: while it does not contribute to amplitude growth, it represents conservative energy transfer over one oscillation period. Here, we focus on the growth rate since it plays a significant role in the global energy transfer equation. From equation (\ref{eq:ROMgrowth2}), the growth rate of the ROM is given by $\text{Real}(G_{{f_1}{f_1}})$. The $\text{Real}(G_{{f_1}{f_1}})$ is expressed as the summation of three terms shown in equation (\ref{LSAeq_G_new}). 
The term $\frac{1}{\Rey} \langle \nabla^2 \boldsymbol{\varphi}_{f_k}, \boldsymbol{\varphi}_{f_k} \rangle$ clearly represents the viscous diffusion term. 
In addition, the two terms, $-\langle(\boldsymbol{u}_{b} \cdot \nabla)\boldsymbol{\varphi}_{f_k}, \boldsymbol{\varphi}_{f_k}\rangle$ and $\langle (\boldsymbol{\varphi}_{f_k} \cdot \nabla) \boldsymbol{u}_b, \boldsymbol{\varphi}_{f_k} \rangle$, also contribute to $G_{{f_k}{f_k}}$. 
From the momentum equation for the spectral energy budget (\ref{ballance1}), $\boldsymbol{\varphi}_{f_k}$ in the term $(\boldsymbol{\varphi}_{f_{n-k}} \cdot \nabla) \boldsymbol{\varphi}_{f_k}$ acts as the donor of energy to the recipient $\boldsymbol{\varphi}_{f_n}$ \citep{freeman_2024, yeung_2024}. Therefore, we now define the real parts of the three terms, considering their origins, as follows:
\begin{eqnarray}
\mathcal{D}_{f_k} & \stackrel{\mathrm{def}}{=} & \frac{1}{\Rey} \text{Real}(\langle \nabla^2 \boldsymbol{\varphi}_{f_k}, \boldsymbol{\varphi}_{f_k} \rangle), \\
\mathcal{T}_{f_k \rightleftarrows f_k} & \stackrel{\mathrm{def}}{=} & - \text{Real}\left\{ \langle (\boldsymbol{u}_b \cdot \nabla) \boldsymbol{\varphi}_{f_k}, \boldsymbol{\varphi}_{f_k} \rangle \right\}, \\
\mathcal{T}_{b \rightarrow f_k} & \stackrel{\mathrm{def}}{=} & - \text{Real}\left\{ \langle (\boldsymbol{\varphi}_{f_k} \cdot \nabla) \boldsymbol{u}_b, \boldsymbol{\varphi}_{f_k} \rangle \right\},
\end{eqnarray}
where $\mathcal{D}$ represents the diffusion term, $\mathcal{T}$ is the transfer term, and the direction of the arrow in the subscript indicates the direction of energy transfer. Here, the subscript $b$ on the transfer term $\mathcal{T}_{b \rightarrow f_k}$ matches the identifier used for the base flow: $b=s$ for a steady base flow and $b=t$ for a phase-averaged base flow in a time-dependent system.

\begin{figure}
	\centering\includegraphics[width=13cm,keepaspectratio]{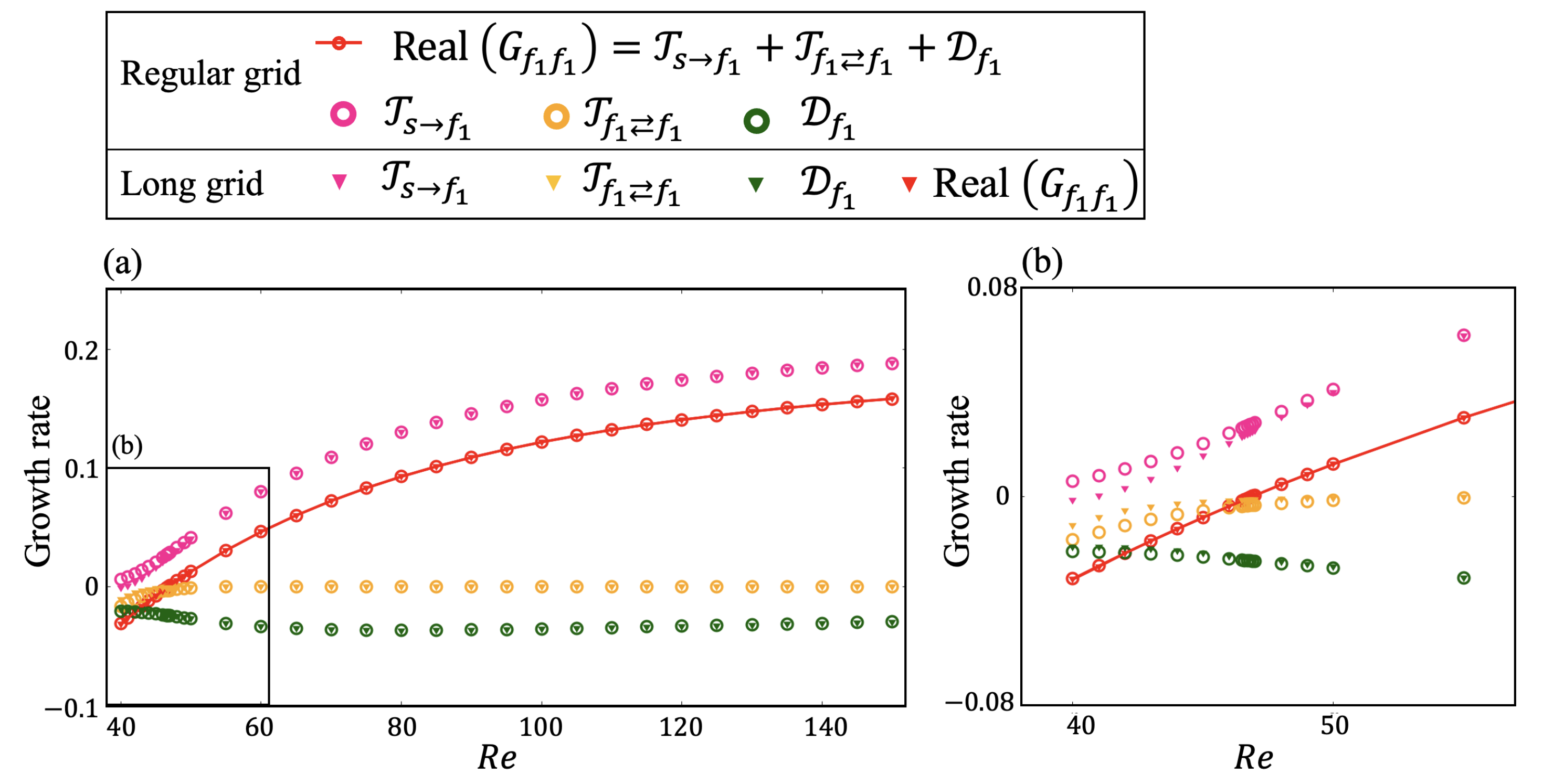}
  \captionsetup{justification=raggedright,singlelinecheck=false}
  \caption{Individual plots of the three terms that comprise the ROM growth rate ($\text{Real}(G_{{f_1}{f_1}})$): (a) overall view, (b) close-up view around bifurcation point.}
 \label{fig_growth_separate}
\end{figure}

Figure \ref{fig_growth_separate} shows the values of the three terms computed numerically at each $\Rey$. Note that the red symbols in the figure represent the growth rate of the ROM, which is the sum of the three terms, as shown by the red symbols in figure \ref{ROMgrowth}. For all $\Rey$ cases, only $\mathcal{T}_{s \rightarrow f_1}$ takes a positive value, indicating that $\mathcal{T}_{s \rightarrow f_1}$ is the term that amplifies the energy of $\boldsymbol{\varphi}_{f_1}$. The diffusion term $\mathcal{D}_{f_1}$ is always negative and remains nearly constant regardless of the $\Rey$ value. A closer examination of the diffusion effect reveals a slight variation for $Re < 70$. This variation is consistent with the results reported by \citet{Mittal_2009}.

 Since $\mathcal{D}_{f_1}$ has a coefficient of $1/\Rey$, $\text{Real}(\langle \nabla^2 \boldsymbol{\varphi}_{f_1}, \boldsymbol{\varphi}_{f_1} \rangle)$ increases with $\Rey$. If $\boldsymbol{\varphi}_{f_1}$ is assumed to have a characteristic wavenumber, then $\nabla^2 \boldsymbol{\varphi}_{f_1}$ can be considered proportional to the square of that wavenumber. Consequently, the squared wavenumber would scale with the $Re$. However, as shown by the spatial distribution of the eigenmodes in figure \ref{fig_faif}, the wavenumber of the eigenmode is not necessarily uniform across the entire domain. Therefore, the relationship between wavenumber and $Re$ remains speculative.

The term $\mathcal{T}_{f_1 \rightleftarrows f_1}$ is also negative for all $\Rey$ cases, reflecting the effect of convection by the base flow \citep{Mittal_2009}. This indicates that energy is lost through convection, whereby $\boldsymbol{\varphi}_{f_1}$ transfers energy to itself. This self-decaying nature from convection is also reported by \citet{yeung_2024} in the post-transient, periodic cylinder flow at $\Rey=100$. However, the slight negative values and resulting decay may indicate that energy is being advected out of the domain through the outflow boundaries.
When the $\Rey$ is smaller than the critical $\Rey$, this self-decaying property is more pronounced. Comparing the regular-grid and long-grid results shows that their discrepancy increases at low $\Rey$. With a wider computational domain, the two transfer terms approach zero. As noted by \citet{yeung_2024}, the boundary influence on the nonlinear term depends on fluxes through the domain boundary. This dependence is evaluated in Appendix~\ref{recipient-donor}. Despite the altered energy transfer with the exterior in the enlarged domain, the modal growth rates are unchanged, indicating that the essential amplification and attenuation mechanisms are already captured within the regular-grid domain.

Going back to the fact that the value of each of the three terms is computed by the spatial integration, the spatial distribution of the integrated function determines the value of each term. 
This motivates us to define the spatial fields that determine the values of the three terms as follows:
\begin{eqnarray}
\boldsymbol{\mathcal{D}}_{f_k}(\boldsymbol{x}) & \stackrel{\mathrm{def}}{=} & \frac{1}{\Rey} \text{Real}( \boldsymbol{\varphi}_{f_k}^H \nabla^2 \boldsymbol{\varphi}_{f_k}), \\
\boldsymbol{\mathcal{T}}_{f_k \rightleftarrows f_k}(\boldsymbol{x}) & \stackrel{\mathrm{def}}{=} & - \text{Real} \left\{\boldsymbol{\varphi}^H_{f_k} (\boldsymbol{u}_{b} \cdot \nabla) \boldsymbol{\varphi}_{f_k}\right\}, \\
\boldsymbol{\mathcal{T}}_{b \rightarrow f_k}(\boldsymbol{x}) & \stackrel{\mathrm{def}}{=} & - \text{Real}\left\{ \boldsymbol{\varphi}^H_{f_k}(\boldsymbol{\varphi}_{f_k} \cdot \nabla) \boldsymbol{u}_b\right\},
\end{eqnarray}
inspired by the interaction map of BMD \citep{schmidt_2020} and the transfer field of TOD \citep{yeung_2024}. 
 Specifically, we denote the spatial fields in bold, $\boldsymbol{\mathcal{D}}(\boldsymbol{x})$ and $\boldsymbol{\mathcal{T}}(\boldsymbol{x})$. To distinguish these fields from their spatially integrated counterparts $\mathcal{D}$ and $\mathcal{T}$, we retain the explicit dependence on $\boldsymbol{x}$ whenever referring to spatial distributions.
We refer to the spatial field $\boldsymbol{\mathcal{D}}_{f_k}(\boldsymbol{x})$ as the diffusion field, and $\boldsymbol{\mathcal{T}}_{f_k \rightleftarrows f_k}(\boldsymbol{x})$ and $\boldsymbol{\mathcal{T}}_{b \rightarrow f_k}(\boldsymbol{x})$ as transfer fields. 

\begin{figure}
	\centering\includegraphics[width=13cm,keepaspectratio]{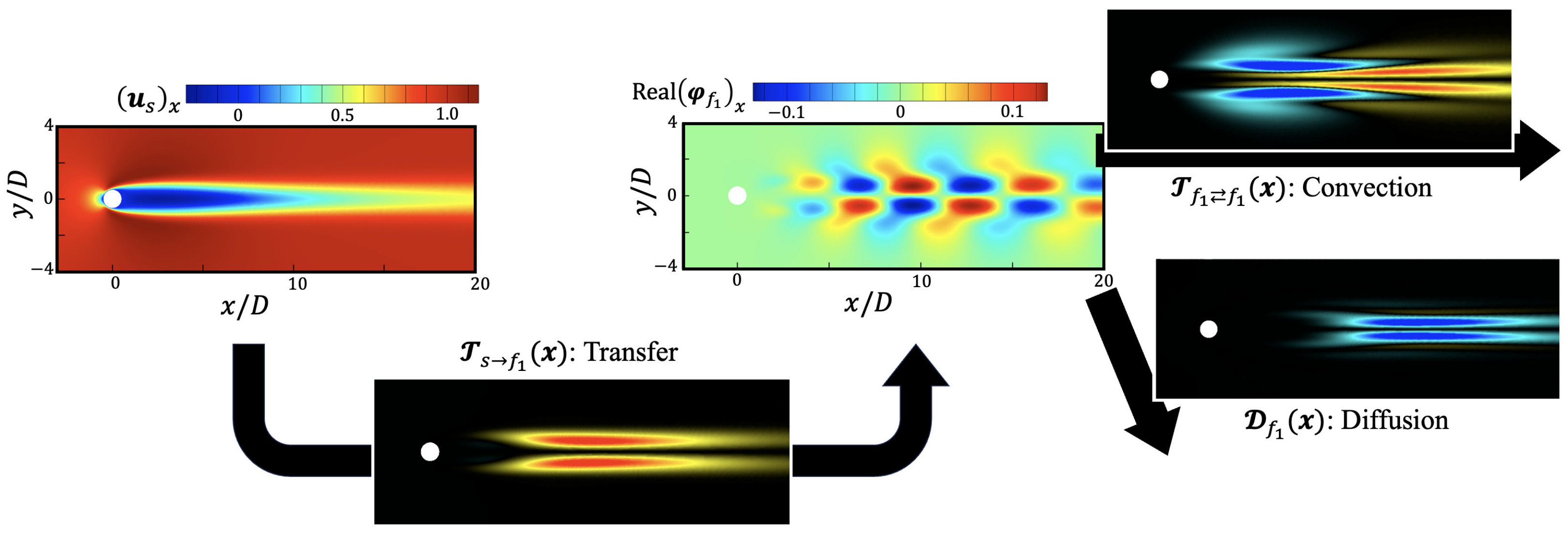}
  \captionsetup{justification=raggedright,singlelinecheck=false}
  \caption{Diagram of energy transfer relationship of steady base flow $\boldsymbol{u}_{s}$ and $\boldsymbol{\varphi}_{f_1}$ in the case of linearized system around the base flow. All contour map is the case of $\Rey=100$. The direction of the arrow indicates the signature of the integral value of the diffusion field and transfer field.}
 \label{fig_transfer_diagram}
\end{figure}

Figure \ref{fig_transfer_diagram} shows a conceptual diagram at $\Rey = 100$ with steady base flow, considering three spatial field distributions: $\boldsymbol{\mathcal{D}}_{f_1} (\boldsymbol{x})$, $\boldsymbol{\mathcal{T}}_{f_1 \rightleftarrows f_1} (\boldsymbol{x})$, and $\boldsymbol{\mathcal{T}}_{s \rightarrow f_1} (\boldsymbol{x})$, as well as the energy transfer direction determined by the sign of their spatial integrations: $\mathcal{D}_{f_1}$, $\mathcal{T}_{f_1 \rightleftarrows f_1}$, and $\mathcal{T}_{s \rightarrow f_1}$. These distributions are in close agreement with those presented by \citet{Mittal_2009}. Only the transfer term $\mathcal{T}_{s \rightarrow f_1}$ acts as an amplifier, supplying energy, while the other two act as energy dampers. Therefore, whether $\boldsymbol{\varphi}_{f_1}(\boldsymbol{x})$ develops or not depends on the sum of these three terms. The diffusion field $\boldsymbol{\mathcal{D}}_{f_1}(\boldsymbol{x})$ is entirely negative, whereas the transfer field from steady base flow $\boldsymbol{\mathcal{T}}_{s \rightarrow f_1}(\boldsymbol{x})$ shows an almost entirely positive distribution. In the field of $\boldsymbol{\mathcal{T}}_{f_1 \rightleftarrows f_1}(\boldsymbol{x})$, both positive and negative value regions exist.

Transfer fields and diffusion fields for $\Rey = 40, 60, 100,$ and $150$ are shown in figure \ref{fig_transfer}. The $0$-line of the streamwise velocity for the steady flow is shown for $\boldsymbol{\mathcal{T}}_{s \rightarrow f_j}(\boldsymbol{x})$, and the contour lines of $\sqrt{\boldsymbol{\varphi}^H_{f_1} \boldsymbol{\varphi}_{f_1}}$ (see figure \ref{fig_faif}) are shown for the other fields by the white line. The diffusion field has a similar distribution to $\sqrt{\boldsymbol{\varphi}^H_{f_1} \boldsymbol{\varphi}_{f_1}}$. This means that diffusion effects appear at all positions where fluctuations in the $f_1$ frequency component exist. The transfer fields $\boldsymbol{\mathcal{T}}_{f_1 \rightleftarrows f_1}(\boldsymbol{x})$ exhibit a negative distribution near the cylinder and a positive distribution farther away from the cylinder, except for $\Rey = 40$. Compared to the distribution of $\sqrt{\boldsymbol{\varphi}^H_{f_1} \boldsymbol{\varphi}_{f_1}}$, the distribution of $\boldsymbol{\mathcal{T}}_{f_1 \rightleftarrows f_1}(\boldsymbol{x})$ switches between positive and negative values around the peak-$x$ position of $\sqrt{\boldsymbol{\varphi}^H_{f_1} \boldsymbol{\varphi}_{f_1}}$. These positive and negative regions cancel each other out, resulting in a net small negative transfer, $\mathcal{T}_{f_1 \rightleftarrows f_1}$. Moreover, the presence of positive values in the far wake at $\Rey = 60$, $100$, and $150$ suggests that convection by the base flow contributes to the growth of the Karman vortices.

The transfer field $\boldsymbol{\mathcal{T}}_{s \rightarrow f_1}(\boldsymbol{x})$ has a distribution along the $x$-direction of the recirculation region, bounded by the $0$-line. This distribution is very similar to the sensitivity region where fluctuations are caused by the wake recirculation region, as reported by \citet{Ohmichi_2021}. Since $\boldsymbol{\mathcal{T}}_{s \rightarrow f_1}(\boldsymbol{x})$ represents the energy transfer from the base flow to $\boldsymbol{\varphi}_{f_1}(\boldsymbol{x})$, the formation of $\boldsymbol{\varphi}_{f_1}(\boldsymbol{x})$ originates from the recirculation region of the base flow. Therefore, the growth of $\boldsymbol{\varphi}_{f_1}(\boldsymbol{x})$ is driven by instability in the recirculation region of the wake. However, at $\Rey = 40$, this energy transfer is minimal, which is also evident from the integral value in figure \ref{fig_growth_separate}. As $\Rey$ increases, the length of the recirculation region also increases (see figure \ref{steady_Re}), making the wake recirculation more unstable and enhancing the energy transfer to $\boldsymbol{\varphi}_{f_1}(\boldsymbol{x})$.

\begin{figure}
	\centering\includegraphics[width=13.5cm,keepaspectratio]{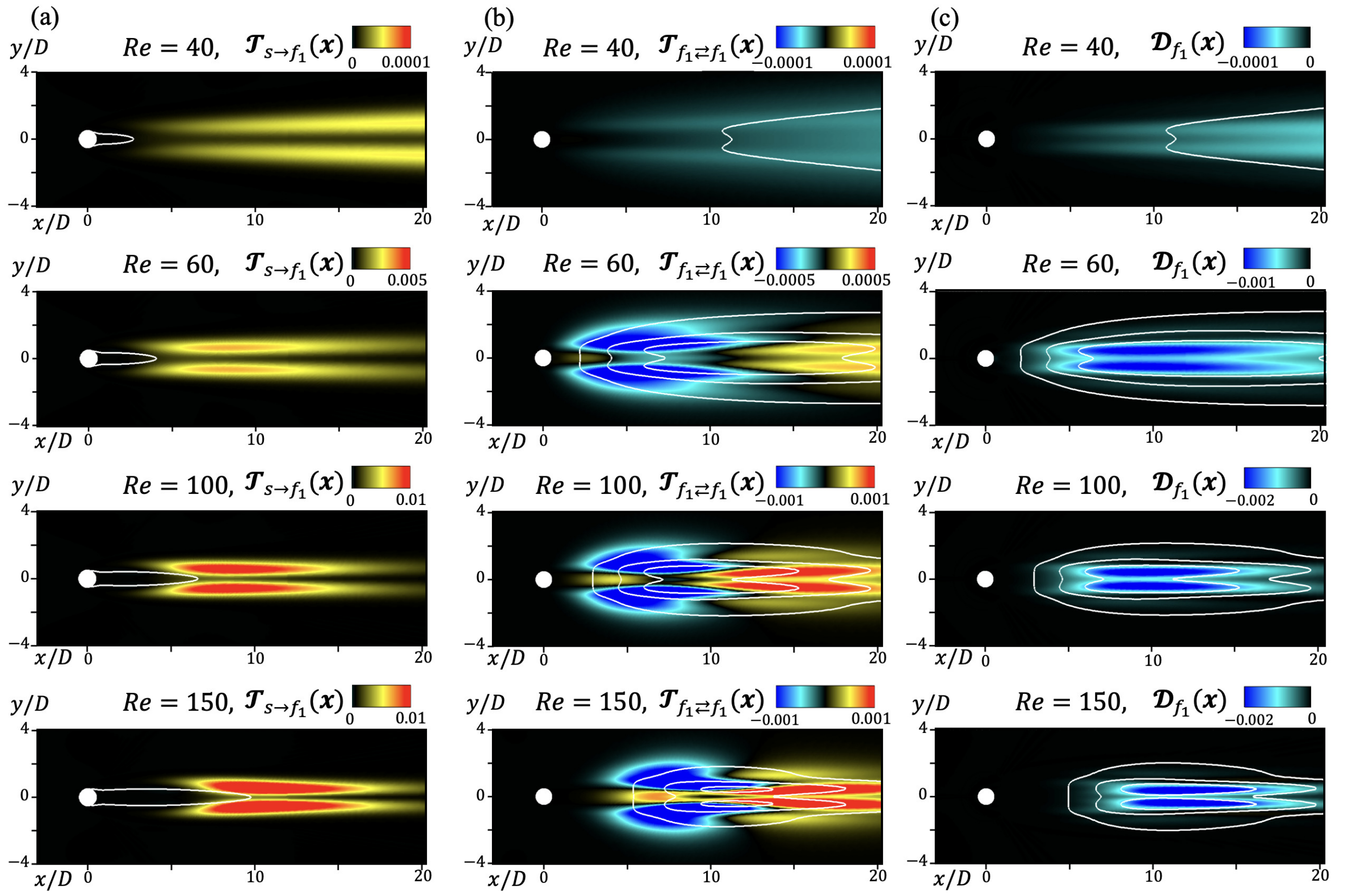}
  \captionsetup{justification=raggedright,singlelinecheck=false}
  \caption{Spatial distribution of transfer fields and diffusion fields at $\Rey = 40, 60, 100,$ and $150$.}
 \label{fig_transfer}
\end{figure}

To clarify the amplification mechanism of the eigenmode driven by energy transfer, we examine the energy source, $\mathcal{T}_{s \rightarrow f_1}$, with its transfer field $\boldsymbol{\mathcal{T}}_{s \rightarrow f_1}(\boldsymbol{x})$. In general, the transfer term can be decomposed into contributions from individual velocity components. Writing the components explicitly, we obtain
\begin{eqnarray}
\mathcal{T}_{b \rightarrow f_k} &=& - \text{Real}\left\{ \langle (\boldsymbol{\varphi}_{f_k} \cdot \nabla) \boldsymbol{u}_b, \boldsymbol{\varphi}_{f_k} \rangle \right\} \nonumber \\
&=& - \text{Real}\left\{ 
\int_{\Omega}(\boldsymbol{\varphi}_{f_k})_x \left(\frac{\partial \boldsymbol{u}_b}{\partial x}\right)_x (\boldsymbol{\varphi}_{f_k})_x d\boldsymbol{x}
 + \int_{\Omega}(\boldsymbol{\varphi}_{f_k})_x \left(\frac{\partial \boldsymbol{u}_b}{\partial x}\right)_y (\boldsymbol{\varphi}_{f_k})_y d\boldsymbol{x}\right\} \nonumber \\
&&
 - \text{Real}\left\{\int_{\Omega}(\boldsymbol{\varphi}_{f_k})_y \left(\frac{\partial \boldsymbol{u}_b}{\partial y}\right)_x (\boldsymbol{\varphi}_{f_k})_x d\boldsymbol{x}
    + \int_{\Omega}(\boldsymbol{\varphi}_{f_k})_y \left(\frac{\partial \boldsymbol{u}_b}{\partial y}\right)_y (\boldsymbol{\varphi}_{f_k})_y d\boldsymbol{x} \right\}.
\end{eqnarray}
Based on this decomposition, we introduce the following four contributions:
\begin{eqnarray}
\mathcal{T}_{b \rightarrow f_k}^{(x,x)} & \stackrel{\mathrm{def}}{=} & - \text{Real}\left\{ 
\int_{\Omega}(\boldsymbol{\varphi}_{f_k})_x \left(\frac{\partial \boldsymbol{u}_b}{\partial x}\right)_x (\boldsymbol{\varphi}_{f_k})_x d\boldsymbol{x}\right\},\label{fourterm_1}\\
\mathcal{T}_{b \rightarrow f_k}^{(x,y)} & \stackrel{\mathrm{def}}{=} & - \text{Real}\left\{ 
\int_{\Omega}(\boldsymbol{\varphi}_{f_k})_x \left(\frac{\partial \boldsymbol{u}_b}{\partial x}\right)_y (\boldsymbol{\varphi}_{f_k})_y d\boldsymbol{x}\right\},\label{fourterm_2}\\
\mathcal{T}_{b \rightarrow f_k}^{(y,x)} & \stackrel{\mathrm{def}}{=} & - \text{Real}\left\{ 
\int_{\Omega}(\boldsymbol{\varphi}_{f_k})_y \left(\frac{\partial \boldsymbol{u}_b}{\partial y}\right)_x (\boldsymbol{\varphi}_{f_k})_x d\boldsymbol{x}\right\},\label{fourterm_3}\\
\mathcal{T}_{b \rightarrow f_k}^{(y,y)} & \stackrel{\mathrm{def}}{=} & - \text{Real}\left\{ 
\int_{\Omega}(\boldsymbol{\varphi}_{f_k})_y \left(\frac{\partial \boldsymbol{u}_b}{\partial y}\right)_y (\boldsymbol{\varphi}_{f_k})_y d\boldsymbol{x}\right\},\label{fourterm_4}
\end{eqnarray}
where the first superscript corresponds to the catalyst and recipient components direction, and the second superscript denotes the donor component direction. We note that the sum of these four contributions is equivalent to the total transfer.

\begin{figure}
	\centering\includegraphics[width=13cm,keepaspectratio]{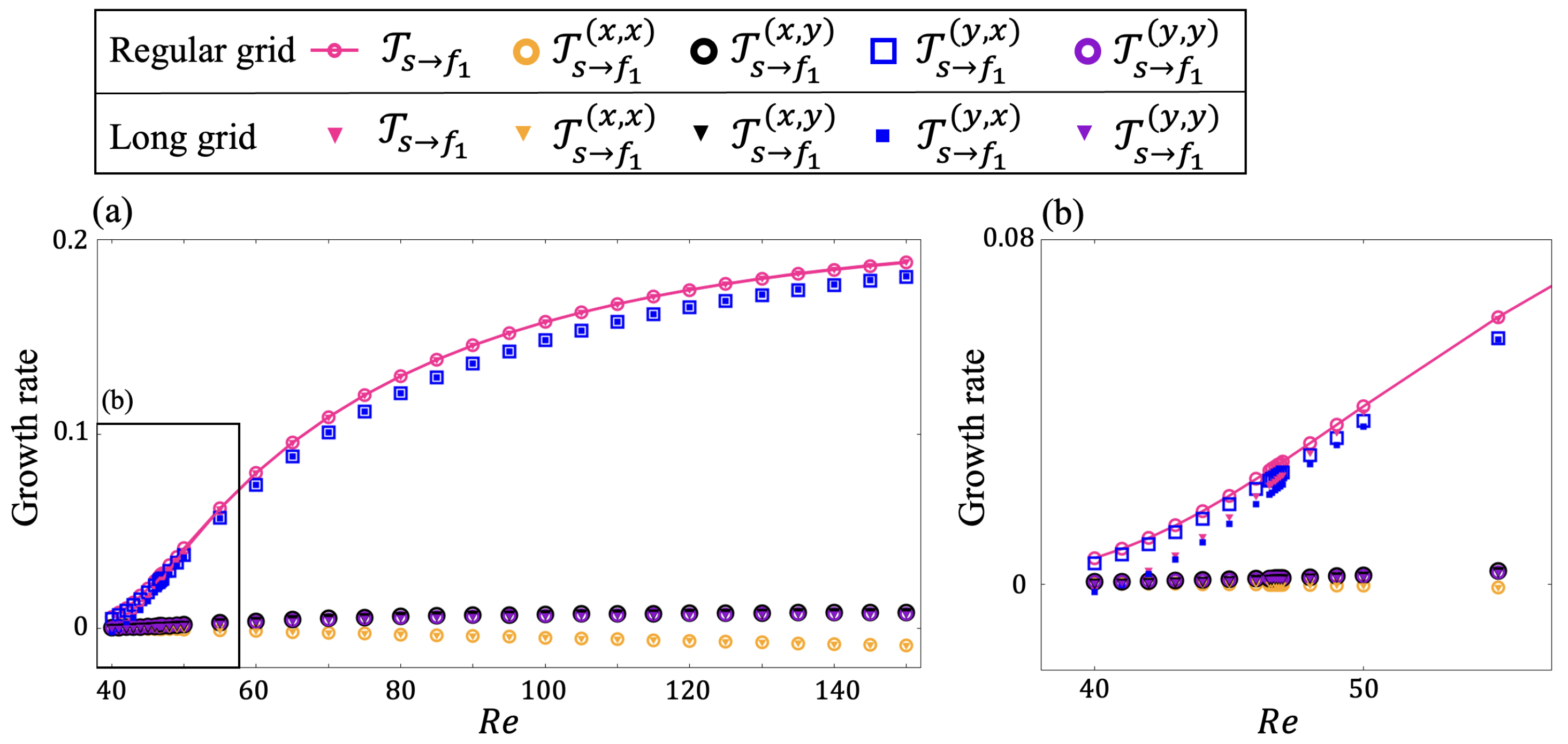}
  \captionsetup{justification=raggedright,singlelinecheck=false}
\caption{$Re$ dependence of the four terms contributing to the transfer term: $\mathcal{T}_{s \rightarrow f_1}^{(x,x)}$, $\mathcal{T}_{s \rightarrow f_1}^{(x,y)}$, $\mathcal{T}_{s \rightarrow f_j}^{(y,x)}$, and $\mathcal{T}_{s \rightarrow f_1}^{(y,y)}$. The net transfer $\mathcal{T}_{s \rightarrow f_1}$ is dominated by $\mathcal{T}_{s \rightarrow f_1}^{(y,x)}$.}
 \label{fig_transfer_separate}
\end{figure}

Figure~\ref{fig_transfer_separate} presents the $\Rey$ dependence of the four contributions, $\mathcal{T}_{s \rightarrow f_1}^{(x,x)}$, $\mathcal{T}_{s \rightarrow f_1}^{(x,y)}$, $\mathcal{T}_{s \rightarrow f_1}^{(y,x)}$, and $\mathcal{T}_{s \rightarrow f_1}^{(y,y)}$. Most of the contribution to the energy-transfer term $\mathcal{T}_{s \rightarrow f_1}$ comes from $\mathcal{T}_{s \rightarrow f_1}^{(y,x)}$. Because each contribution is defined by a spatial integral, its spatial distribution can be visualized in the same manner as $\mathcal{T}_{s \rightarrow f_1}(\boldsymbol{x})$; however, since the distribution is nearly identical to $\mathcal{T}_{s \rightarrow f_1}(\boldsymbol{x})$, we omit it here.

Focusing on \(\mathcal{T}_{s \rightarrow f_1}^{(y,x)}\), in addition to the eigenmode, the contribution from donor-component arises from the $x$-component of the $y$-gradient of the base flow $\boldsymbol{u}_s$. Figure \ref{fig_transfer} shows the distribution of $\left(\partial \boldsymbol{u}_s/\partial y\right)_x$ at $\Rey=40$ and $100$. Interestingly, although the value of $\mathcal{T}_{s \rightarrow f_1}^{(y,x)}$ differs markedly between these cases, the magnitude of the donor gradient is comparable. As seen in figure \ref{fig_transfer}(a), the difference stems from the downstream extent associated with the size of the recirculation region.

Figure~\ref{fig_transfer}(b) shows overlay plots for contours of the complex absolute value of catalyst $\sqrt{(\boldsymbol{\varphi}_{f_1})_y^{*}(\boldsymbol{\varphi}_{f_1})_y}$ at $10$, $30$, and $50$\% of its maximum. At lower $\Rey$, the perturbations represented by eigenmodes are distributed farther downstream; as $\Rey$ increases, they become more localized near the cylinder. Consequently, the recirculation region extends downstream, and the donor gradient extends in the same direction, increasing its overlap with the near-cylinder distribution of the perturbation components. This increased spatial overlap between the donor and the catalyst strengthens the transfer contributions of $\mathcal{T}_{s \rightarrow f_1}^{(y,x)}$ and, in turn, raises the growth rate.

\begin{figure}
	\centering\includegraphics[width=10cm,keepaspectratio]{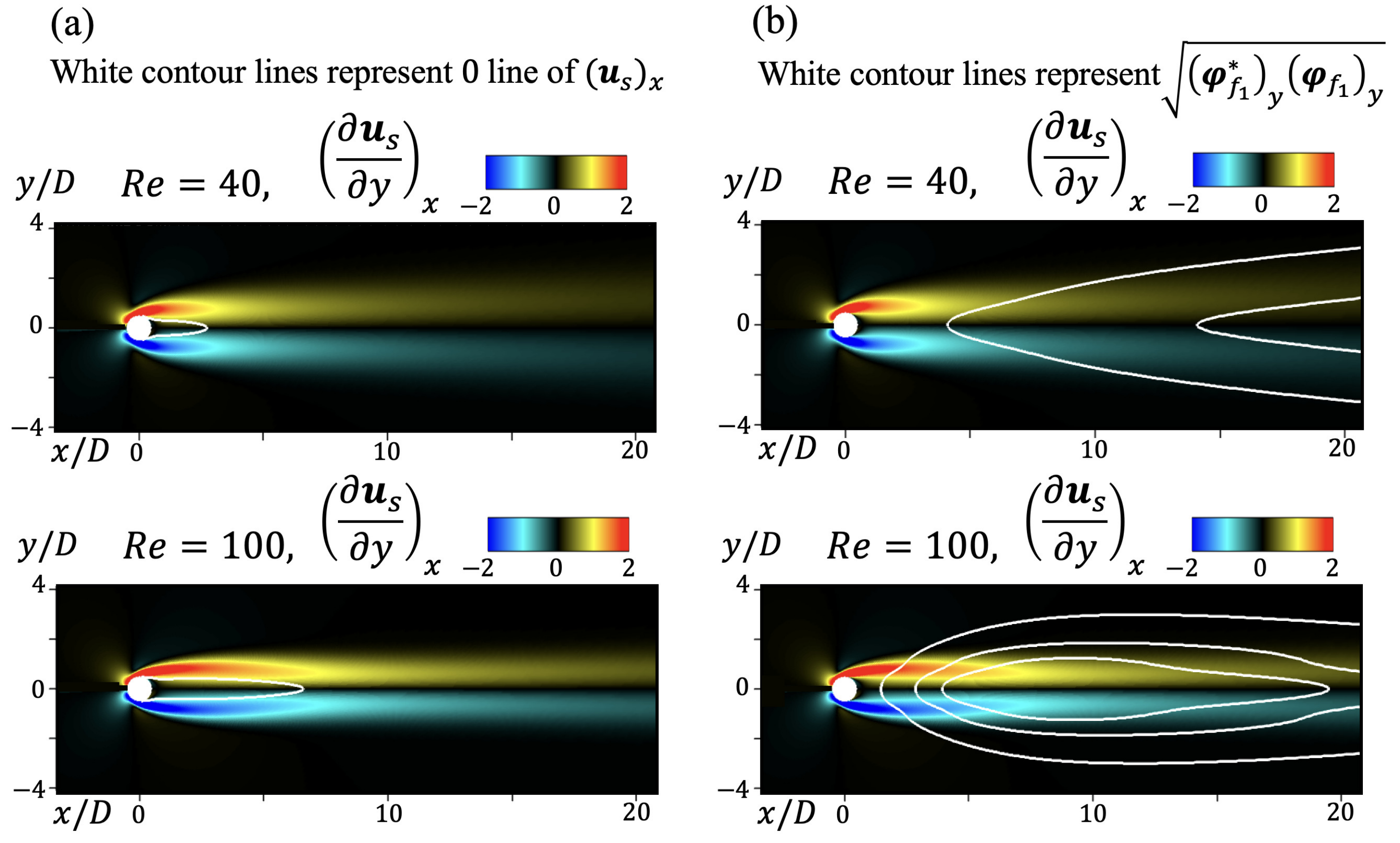}
  \captionsetup{justification=raggedright,singlelinecheck=false}
  \caption{Spatial distribution of the $y$-direction gradient of $\boldsymbol{u}_s$ at $\Rey = 40$ and $100$: (a) zero contour of $(\boldsymbol{u}_s)_x$, and (b) contours of $\sqrt{(\boldsymbol{\varphi}^*_{f_1})_y(\boldsymbol{\varphi}_{f_1})_y}$ at $10$, $30$, and $50\%$ of the maximum value.}
 \label{fig_transfer}
\end{figure}

The equation obtained from phase-averaged ROM, linearized system around the steady flow, successfully captures the system's dynamics near the steady flow obtained from LSA. This section extends the transfer analysis to a transient nonlinear development process beyond the linear growth from the steady field. In such cases, the linearized energy transfer equation (\ref{globalenergytransfer_steady}) alone cannot fully describe the evolution of the solution \citep{stankiewicz2017need}. Therefore, we investigate the energy-transfer dynamics of the transient process using the tDMDpc modes. We aim to provide a more comprehensive understanding of the energy transfer among different frequency components during the evolution of the flow from the steady state to the post-transient regime.

\begin{figure}
	\centering\includegraphics[width=13cm,keepaspectratio]{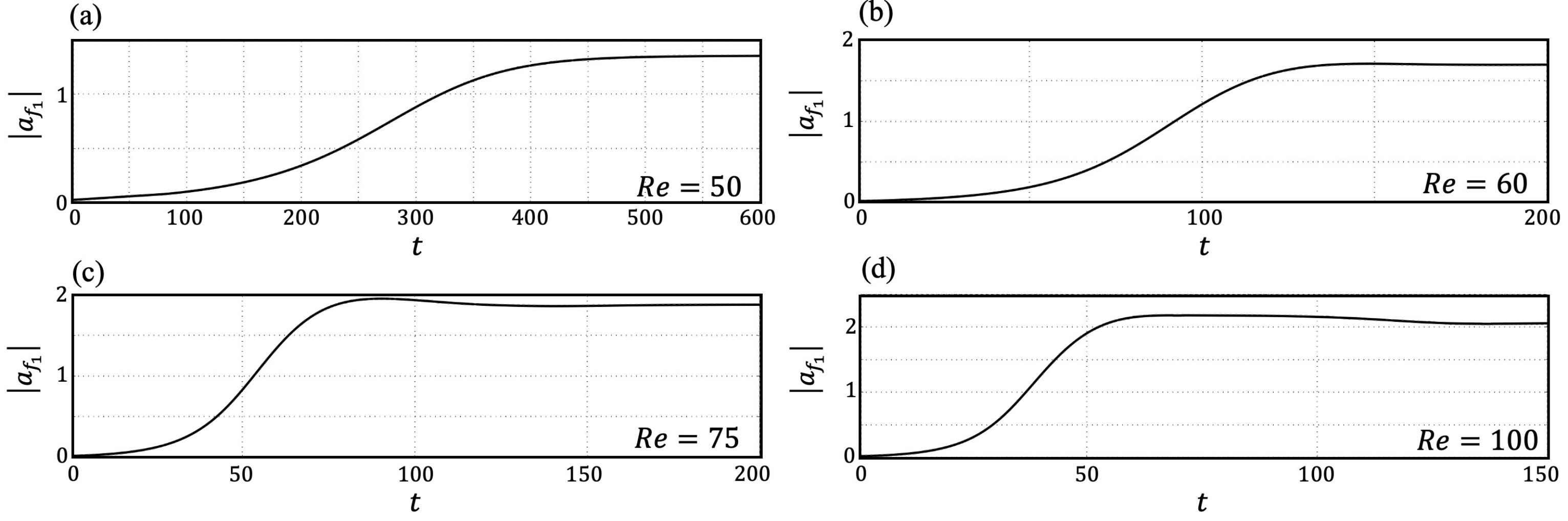}
  \captionsetup{justification=raggedright,singlelinecheck=false}
  \caption{Time variation of mode amplitude $|a_{f_1}(t)|$: (a) $\Rey=50$, (b) $60$, (c) $75$, and (d) $100$.}
 \label{fig_amplitude}
\end{figure}

To investigate the time variation of the energy at arbitrary times in the transient process, we computed the amplitudes coefficients of tDMDpc modes. Figure \ref{fig_amplitude} shows the time variation of absolute value of amplitude coefficients $|a_{f_1}(t)|$ at $\Rey=50, 60, 75,$ and $100$ computed from equation (\ref{calcoefficient}) using tDMDpc mode. The amplitude coefficients absolute value $|a_{f_1}(t)|$ increases over time during the transient process and eventually converges to a stable constant. This is consistent with the convergence of the growth rate to zero in tDMDpc. The growth process is monotonically increasing for $\Rey = 50$ and $60$, whereas a temporary maximum is observed at  $\Rey = 75$ and $100$. 
This transient amplification has also been reported by \citet{Manti_Lugo_2015} in their DNS of flow past a circular cylinder at $\Rey = 100$, where they noted that such amplification cannot be captured by linearized equations. In contrast, modeling with tDMDpc provides a linear operator that best fits the instantaneous evolution of the flow field, thereby successfully capturing the transient amplification, a fundamentally nonlinear feature.

\begin{figure}
	\centering\includegraphics[width=13cm,keepaspectratio]{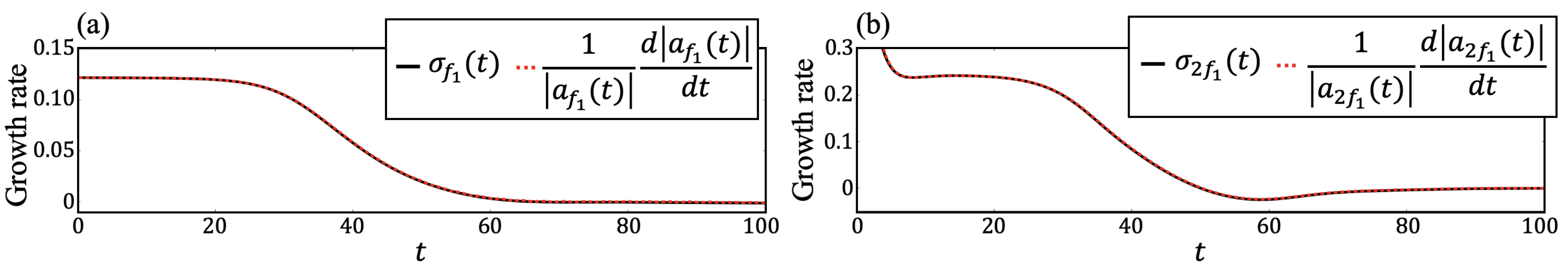}
  \captionsetup{justification=raggedright,singlelinecheck=false}
  \caption{Time variation of growth rate in the transient process comparing with growth rate computed from amplitude coefficient at $Re=100$: (a) fundamental frequency $f_1$, and (b) second-harmonic frequency $2f_1$.}
 \label{fig_growth_compare100}
\end{figure}

In the global energy transfer equation, the energy budget is related to the growth rate. Ideally, the growth rate obtained from the tDMDpc mode should coincide with that computed from the time-dependent amplitude coefficients. The instantaneous growth rate of the amplitude coefficient, $a_{f_k}(t)$, can be expressed as
\begin{equation}
  \begin{split}
  \text{\{Growth rate of $a_{f_k}(t)$\}} = \frac{1}{|a_{f_k}(t)|}\frac{d|a_{f_k}(t)|}{dt} \approx \frac{|a_{f_k}(t+\Delta T)| - |a_{f_k}(t-\Delta T)|}{|a_{f_k}(t)|2\Delta T}.
  \end{split}
\end{equation}
Figure \ref{fig_growth_compare100} shows the temporal evolution of the growth rates obtained from tDMDpc and those computed from the amplitude coefficients for the fundamental frequency $f_1$ and its harmonics at $\Rey=100$. In all cases, the two growth rates agree remarkably well. This demonstrates that the global energy transfer equation based on the amplitude coefficients can reproduce the energy budget governing the growth rates in tDMDpc. However, the growth rate of the second-harmonic frequency mode shows an unusually large value at early times. This behavior can be attributed to the fact that only the LSA mode of the fundamental frequency was introduced as the initial disturbance. Consequently, the second-harmonic mode rapidly amplifies through nonlinear energy transfer from the fundamental mode during the initial transient period.

In the nonlinear case, the global energy transfer equation (\ref{globalenergytransfer_transient}) requires the computation of the nonlinear term $F_{{f_l}{f_{k-l}}{f_k}}$. This term represents the energy transfer between different frequencies, and the energy budget of a mode at $f_k$ frequency requires contributions from other frequency modes. In practice, however, only a finite number of modes can be obtained from DMD, so the nonlinear energy transfer cannot be fully reproduced. Nevertheless, since the DMD algorithm extracts eigenmodes within a subspace that represents the energetically dominant structures, the nonlinear terms can be approximated provided that a sufficiently large but finite set of modes is retained.
Here, we define the nonlinear term using $r$ number of modes as follows:
\begin{equation}
  \begin{split}
  \mathcal{N}_{f_k}(t,r) \stackrel{\mathrm{def}}{=} \sum_{l=-\infty}^{\infty} \text{Real}\{{c_{f_{l}}(t)}{c_{f_{k-l}}(t)}c^*_{f_k}(t)F_{{f_l}{f_{k-l}}{f_k}}(t)\},
  \end{split}
\end{equation}
where
\begin{eqnarray}
c_{f_{j}}(t) = 
\begin{cases}
        a_{f_{j}}(t), & f_j = -rf_1, \cdots, -2f_1, -f_1, f_1, 2f_1, \cdots rf_1, \\
        0, &            f_j \neq -rf_1, \cdots, -2f_1, -f_1, f_1, 2f_1, \cdots rf_1.
    \end{cases}
\end{eqnarray}
Strictly speaking, ${c_{f_{l}}}(t){c_{f_{k-l}}}(t)c^*_{f_k}(t)$ should be averaged over transient processes with different phases. However, in the present case, where the amplitude coefficients are either independent of phase or the phase dependence is negligible, analytical averaging has already been performed in deriving the phase-averaged ROM, and additional numerical averaging is not required.

\begin{figure}
\centering\includegraphics[width=13cm,keepaspectratio]{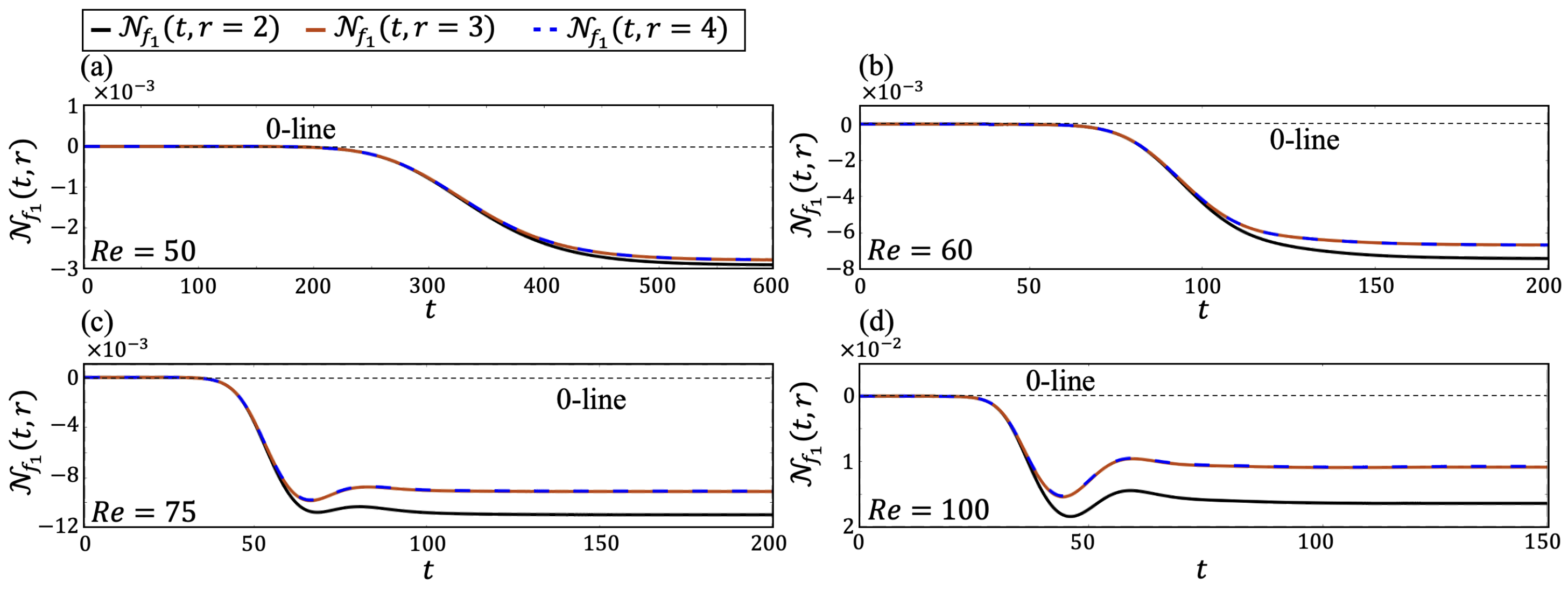}
  \captionsetup{justification=raggedright,singlelinecheck=false}
  \caption{Temporal evolution of the nonlinear term $\mathcal{N}_{f_k}(t,r)$ with $r = 2, 3,$ and $4$, (a) $\Rey=50$, (b) $60$, (c) $75$, (d) $100$.}
 \label{fig_triadicenergy}
\end{figure}

Figure~\ref{fig_triadicenergy} shows the temporal evolution of the nonlinear term $\mathcal{N}_{f_k}(t,r)$ computed at $\Rey = 50, 60, 75,$ and $100$ with $r = 2, 3,$ and $4$. For all $\Rey$, the case $r=2$ exhibits a different temporal behavior of $\mathcal{N}_{f_k}(t,r)$ from those of $r=3$ and $r=4$, while the results for $\mathcal{N}_{f_k}(t,r=3)$ and $\mathcal{N}_{f_k}(t,r=3)$ are nearly identical. This indicates that energy transfer between the fundamental frequency $f_1$ and the $4f_1$ component is much weaker than that with the $3f_1$ component. Previous studies \citep{shift, yeung_2024} have reported that, in periodic cylinder wakes, the energy transfer with the fundamental frequency decreases as the frequency increases. The present results suggest that this trend also holds during transient processes when viewed from the perspective of phase-averaged energy.

In all $\Rey$ cases, the energy transfer remains negative across all times, indicating that energy is consistently transferred to higher-frequency modes. Since $ f_1 $ is the lowest frequency obtained from tDMDpc, $\mathcal{N}_{f_1}(t,r)$ represents energy transfer to higher-frequency modes. Thus, the negative value confirms a totally forward cascade process, where energy is transferred from $f_1$ to its harmonics. 
Focusing on the time variation, for $\Rey = 50$ and $60$, $\mathcal{N}_{f_1}(t,r)$ decreases monotonically throughout the transient process. In contrast, for $\Rey = 75$ and $100$, $\mathcal{N}_{f_1}(t,r)$ reaches a temporary minimum before increasing and stabilizing in the steady state. These results indicate that nonlinearity becomes more pronounced as $\Rey$ increases.

\begin{figure}
	\centering\includegraphics[width=13cm,keepaspectratio]{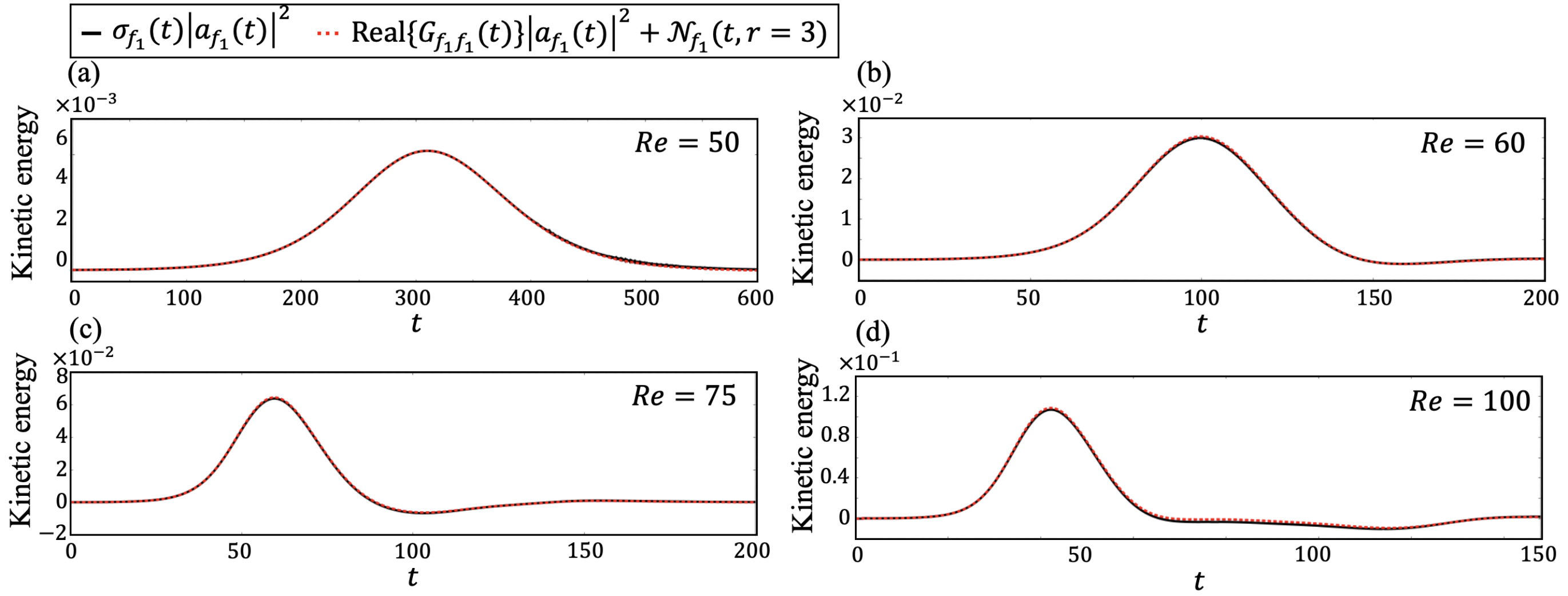}
  \captionsetup{justification=raggedright,singlelinecheck=false }
  \caption{Temporal evolution of the left- and right-hand sides of the global energy transfer equation (\ref{globalenergytransfer_transient}) computed from the nonlinear term $\mathcal{N}_{f_1}(t,r=3)$ and $\text{Real}\{G_{f_1f_1}(t)\}|a_{f_1}(t)|^2$ using the tDMDpc modes at  (a) $\Rey=50$, (b) $60$, (c) $75$, (d) $100$.}
 \label{fig_energybudget_new}
\end{figure}

To assess whether the energy budget can be evaluated from the nonlinear term $\mathcal{N}_{f_1}(t,r)$ and $\text{Real}\{G_{f_1f_1}(t)\}$, we computed $\text{Real}\{G_{f_1f_1}(t)\}$ using the tDMDpc modes. Figure~\ref{fig_energybudget_new} shows the temporal evolution of the left- and right-hand sides of the global energy transfer equation (\ref{globalenergytransfer_transient}) obtained with these terms. Across all $\Rey$, the two sides exhibit close agreement, indicating that the right-hand side of the equation provides a reliable representation of the energy budget during transient processes.

To investigate the variation of energy transfer in transient processes, we decompose $\text{Real}\{G_{{f_1}{f_1}}(t)\}$ into three terms: $\mathcal{D}_{f_1}(t)$, $\mathcal{T}_{t \rightarrow f_1}(t)$, and $\mathcal{T}_{f_1 \rightleftarrows f_1}(t)$, following the energy transfer analysis around a steady flow. These terms are computed from $\boldsymbol{\varphi}_{f_1}(\boldsymbol{x}, t)$ at each time step, and the corresponding values $\mathcal{D}_{f_1}|a_{f_1}|^2$, $\mathcal{T}_{t \rightarrow f_1}|a_{f_1}|^2$, and $\mathcal{T}_{f_1 \rightleftarrows f_1}|a_{f_1}|^2$ are obtained using time-varying $\mathcal{D}_{f_1}(t)$, $\mathcal{T}_{t \rightarrow f_1}(t)$, $\mathcal{T}_{f_1 \rightleftarrows f_1}(t)$, and $a_{f_1}(t)$. Here, it should be noted that the factor $|a_{f_1}(t)|^2$ is introduced to match the magnitude of the phase-averaged energy based on Appendix \ref{energyorthognal}.

\begin{figure}
	\centering\includegraphics[width=13cm,keepaspectratio]{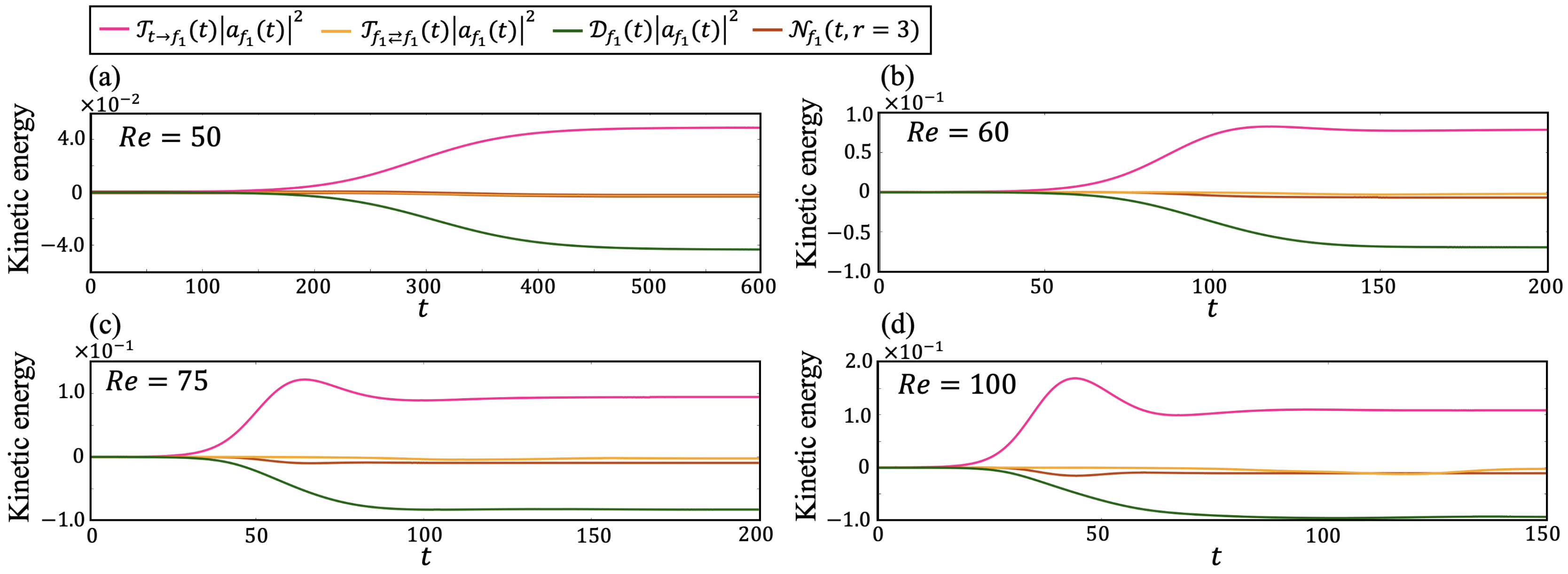}
  \captionsetup{justification=raggedright,singlelinecheck=false}
  \caption{Time variation of four energy transfer term $\mathcal{D}_{f_1}(t)|a_{f_1}(t)|^2$, $\mathcal{T}_{t \rightarrow f_1}(t)|a_{f_1}(t)|^2$, $\mathcal{T}_{f_1 \rightleftarrows f_1}(t)|a_{f_1}|^2$, and $\mathcal{N}_{f_1}(t,r=3)$: (a) $\Rey=50$, (b) $60$, (c) $75$, and (d) $100$. }
 \label{fig_triadicenergy_separate}
\end{figure}

 Figure \ref{fig_triadicenergy_separate} illustrates the time variation of the terms on the right-hand side of the global energy transfer equation (\ref{globalenergytransfer_transient}): $\mathcal{D}_{f_1}(t)|a_{f_1}(t)|^2$, $\mathcal{T}_{t \rightarrow f_1}(t)|a_{f_1}(t)|^2$, $\mathcal{T}_{f_1 \rightleftarrows f_1}(t)|a_{f_1}(t)|^2$, and $\mathcal{N}_{f_1}(t, r=3)$. Since $\mathcal{T}_{t \rightarrow f_1}(t)|a_{f_1}(t)|^2$ is positive, while $\mathcal{T}_{f_1 \rightleftarrows f_1}|a_{f_1}|^2$ and $\mathcal{D}_{f_1}(t)|a_{f_1}(t)|^2$ are negative, the energy transfer relationships observed in the steady field $\boldsymbol{u}_{s}(\boldsymbol{x})$ and the $f_1$-frequency LSA  mode also hold in transient processes.

Focusing on the temporal variation of each term, the diffusion term decreases monotonically over time for all $\Rey$, suggesting that diffusion effects become increasingly dominant as the $f_1$-frequency component evolves.
Similarly, $\mathcal{T}_{f_1 \rightleftarrows f_1}(t)|a_{f_1}(t)|^2$ remains slightly negative. As seen in the time variation of $\mathcal{N}_{f_1}(t,r)$ in figure \ref{fig_triadicenergy} (also shown as the brown line in figure \ref{fig_triadicenergy_separate}), all terms exhibit monotonic behavior at $\Rey = 50$. At $\Rey = 60$, $\mathcal{T}_{t \rightarrow f_1}(t)|a_{f_1}(t)|^2$ reaches a maximum before slightly decreasing to a steady state. Since this trend is absent in $\mathcal{N}_{f_1}(t,r)$ at $\Rey = 60$, this suggests that $\Rey = 60$ is close to the transition point where the monotonic development observed at $\Rey = 50$ shifts to the nonlinear development seen at $\Rey = 75$.  For $\Rey = 75$ and $100$, the decline after the peak of $\mathcal{T}_{t \rightarrow f_1}(t)|a_{f_1}(t)|^2$ becomes more pronounced. Notably, the peak of $\mathcal{T}_{t \rightarrow f_1}(t)|a_{f_1}(t)|^2$ nearly coincides with the minimum of $\mathcal{N}_{f_1}(t,r)$. This suggests that as energy transfer from the steady field to the $f_1$-frequency component increases, energy transfer to higher-frequency components also intensifies. That is, a hierarchical energy transfer structure can be inferred, where energy flows from the steady field into the $f_1$-frequency component and subsequently cascades from the $f_1$-frequency component to $f_1$-harmonic components.

\begin{figure}
\centering\includegraphics[width=13cm,keepaspectratio]{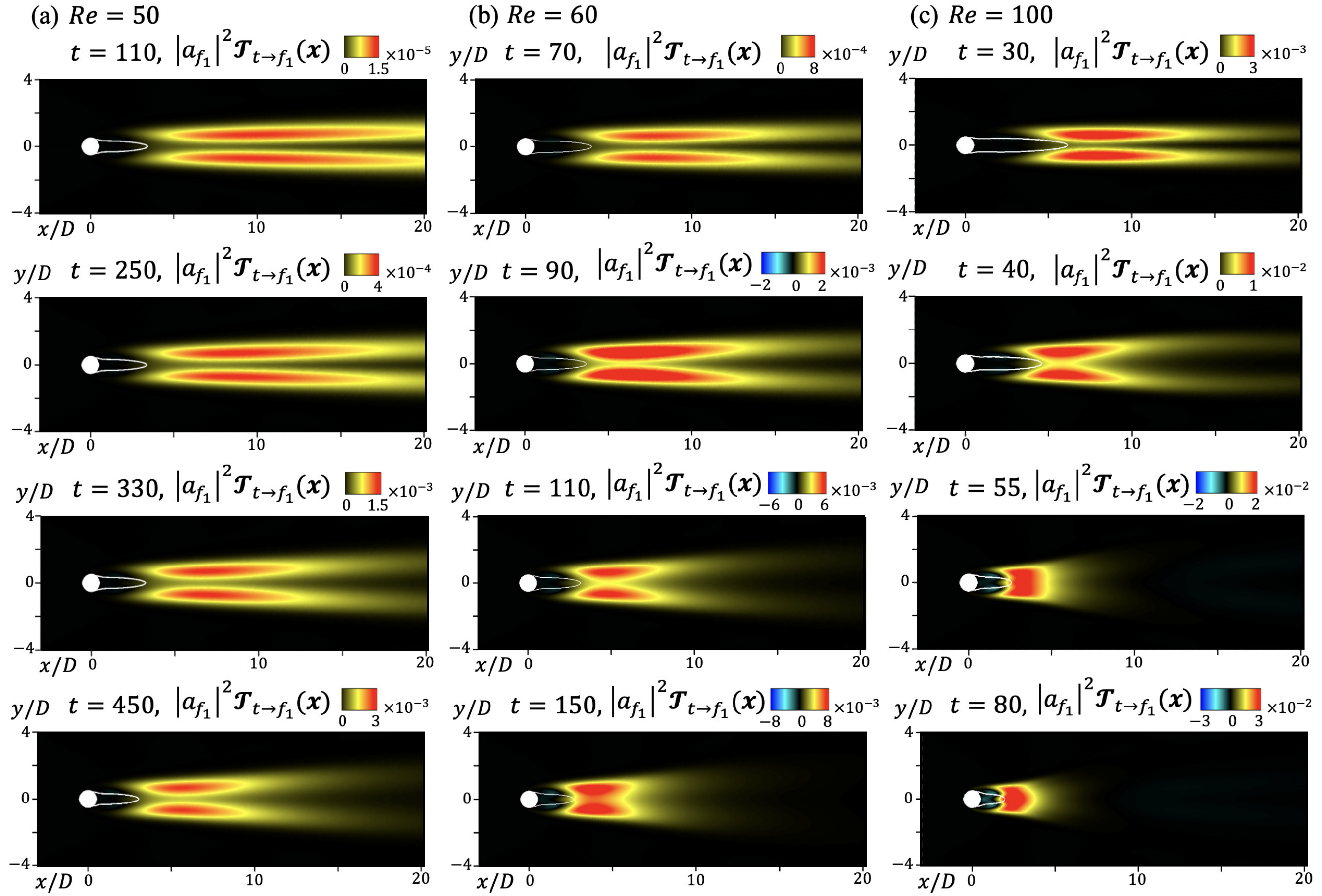}
  \captionsetup{justification=raggedright,singlelinecheck=false}
  \caption{Spatial distribution of transfer fields $\mathcal{T}_{t \rightarrow f_1}(\boldsymbol{x},t)|a_{f_1}(t)|^2$ obtained from eigenmodes of tDMDpc (a) $\Rey=50$, (b) $60$, and (c) $100$.}
 \label{fig_transfer_tDMDpc}
\end{figure}

From the preceding discussion, it is evident that the primary energy supply responsible for the amplification of the $f_1$-frequency component is provided by the transfer term $\boldsymbol{\mathcal{T}}_{t \rightarrow f_1}(\boldsymbol{x}, t)|a_{f_1}(t)|^2$. 
This motivates us to investigate the energy transfer fields $\boldsymbol{\mathcal{T}}_{t \rightarrow f_1}(\boldsymbol{x}, t)|a_{f_1}(t)|^2$.
Figure \ref{fig_transfer_tDMDpc} shows the time-variation of energy transfer fields $\boldsymbol{\mathcal{T}}_{t \rightarrow f_1}(\boldsymbol{x}, t)|a_{f_1}(t)|^2$ at $\Rey = 50, 60,$ and $100$. The $0$-line of $\boldsymbol{u}_{t}(\boldsymbol{x}, t)$ at the same $t$ is plotted by the white line. The positive energy transfer region moves to the cylinders with time variation. At $t=80$ with a $\Rey = 100$, particularly strong energy transfer is observed near the recirculation region of the cylinder wake and on the $y = 0$. This distribution of energy transfer fields is very similar to transfer fields at post-transient flow fields at $\Rey = 100$ reported in \citet{yeung_2024}. The distributions at $t=100$ and $t=30$ for $\Rey = 100$ are quite different, but we can see how they gradually change continuously with time evolution. Continuous changes can be seen even at $\Rey = 50$ and $60$.

We focus on the relationship between the location of the recirculation region indicated by the white line and the energy transfer distribution. At $\Rey = 100$, the $0$-line and energy transfer distributions maintain almost the same $x$ position at all times. Conversely, for $t=300$ at $\Rey = 60$ and $\Rey = 50$, the transfer distribution exists at a distance from the $0$ line. It is possible that the difference in the development process between $\Rey=50$ and $100$ (see figure \ref{fig_triadicenergy_separate}) is due to this difference in the transfer fields. Even under conditions that eventually lead to a periodic flow, the system undergoes different transient processes. 

\begin{figure}
	\centering\includegraphics[width=13cm,keepaspectratio]{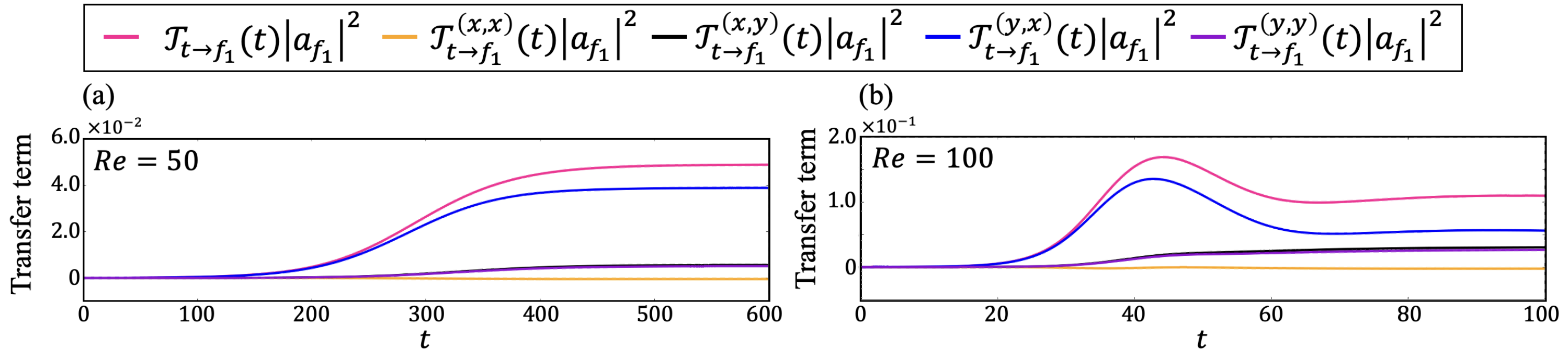}
  \captionsetup{justification=raggedright,singlelinecheck=false }
  \caption{Temporal evolution of the four energy transfer terms in equation~(\ref{fourterm_1}--\ref{fourterm_4}) at (a) $\Rey=50$ and (b) $100$. The contribution of ${\mathcal{T}}^{(y,x)}_{t \rightarrow f_1}(t)|a_{f_1}(t)|^2$ remains dominant at all times. At $\Rey=100$, the transient amplification of the total energy transfer is primarily attributed to ${\mathcal{T}}^{(y,x)}_{t \rightarrow f_1}(t)|a_{f_1}(t)|^2$.}
 \label{fig21_rev1}
\end{figure}

To examine in more detail the dynamics leading to a temporary peak in energy transfer during the transient process, we decompose the time-dependent energy transfer term at $\Rey=50$ and $100$ into the four terms of equation~(\ref{fourterm_1}--\ref{fourterm_4}). Figure~\ref{fig21_rev1} presents the temporal evolution of the contributions from these four terms. Similar to the contributions obtained from the LSA modes in the linear growth regime, the term $\mathcal{T}^{(y,x)}_{t \rightarrow f_1}(t)|a_{f_1}(t)|^2$ remains dominant at all times. At $\Rey=100$, the transient enhancement of the total energy transfer is found to arise primarily from the contribution of ${\mathcal{T}}^{(y,x)}_{t \rightarrow f_1}(t)|a_{f_1}(t)|^2$.  However, in the nonlinear growth process, slight increases in the contributions of ${\mathcal{T}}^{(y,y)}_{t \rightarrow f_1}(t)|a_{f_1}(t)|^2$ and ${\mathcal{T}}^{(x,y)}_{t \rightarrow f_1}(t)|a_{f_1}(t)|^2$ are also observed.

\begin{figure}
	\centering\includegraphics[width=11cm,keepaspectratio]{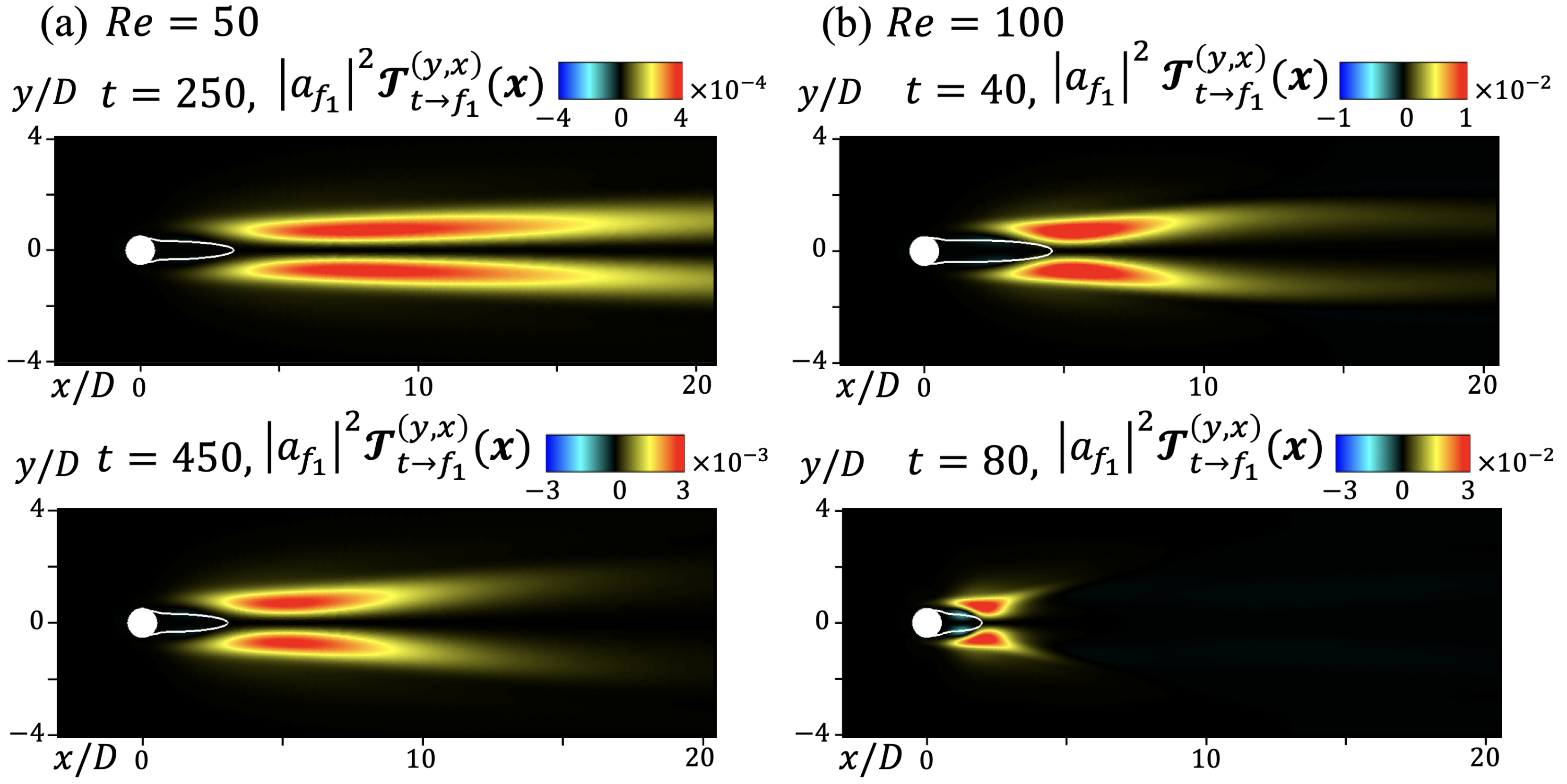}
  \captionsetup{justification=raggedright,singlelinecheck=false}
  \caption{Spatial distribution of $|a_{f_1}(t)|^2\boldsymbol{\mathcal{T}}^{(y,x)}_{t \rightarrow f_1}(\boldsymbol{x},t)$ at (a) $\Rey=50$ ($t=250, 450$) and (b) $100$ ($t=40, 80$).}
 \label{fig21_rev2}
\end{figure}

Figure~\ref{fig21_rev2} shows the spatial distribution of the transfer field $\boldsymbol{\mathcal{T}}^{(y,x)}_{t \rightarrow f_1}(\boldsymbol{x},t)|a_{f_1}(t)|^2$, which corresponds to the integrand of $\mathcal{T}^{(y,x)}_{t \rightarrow f_1}(t)|a_{f_1}(t)|^2$. At $\Rey = 50$, the distribution closely resembles that of the total transfer field $\boldsymbol{\mathcal{T}}_{t \rightarrow f_1}(\boldsymbol{x},t)|a_{f_1}(t)|^2$, obtained from the sum of all four terms, and shows little variation with time. In contrast, at $\Rey = 100$, the distribution differs in that transfer is absent along $y=0$, unlike in the total transfer field. This discrepancy reflects the influence of other terms. Based on the relative magnitudes shown in figure~\ref{fig21_rev1}, the difference can be attributed to the contributions of ${\mathcal{T}}^{(y,y)}_{t \rightarrow f_1}(t)|a_{f_1}(t)|^2$ and ${\mathcal{T}}^{(x,y)}_{t \rightarrow f_1}(t)|a_{f_1}(t)|^2$.

\begin{figure}
	\centering\includegraphics[width=11cm,keepaspectratio]{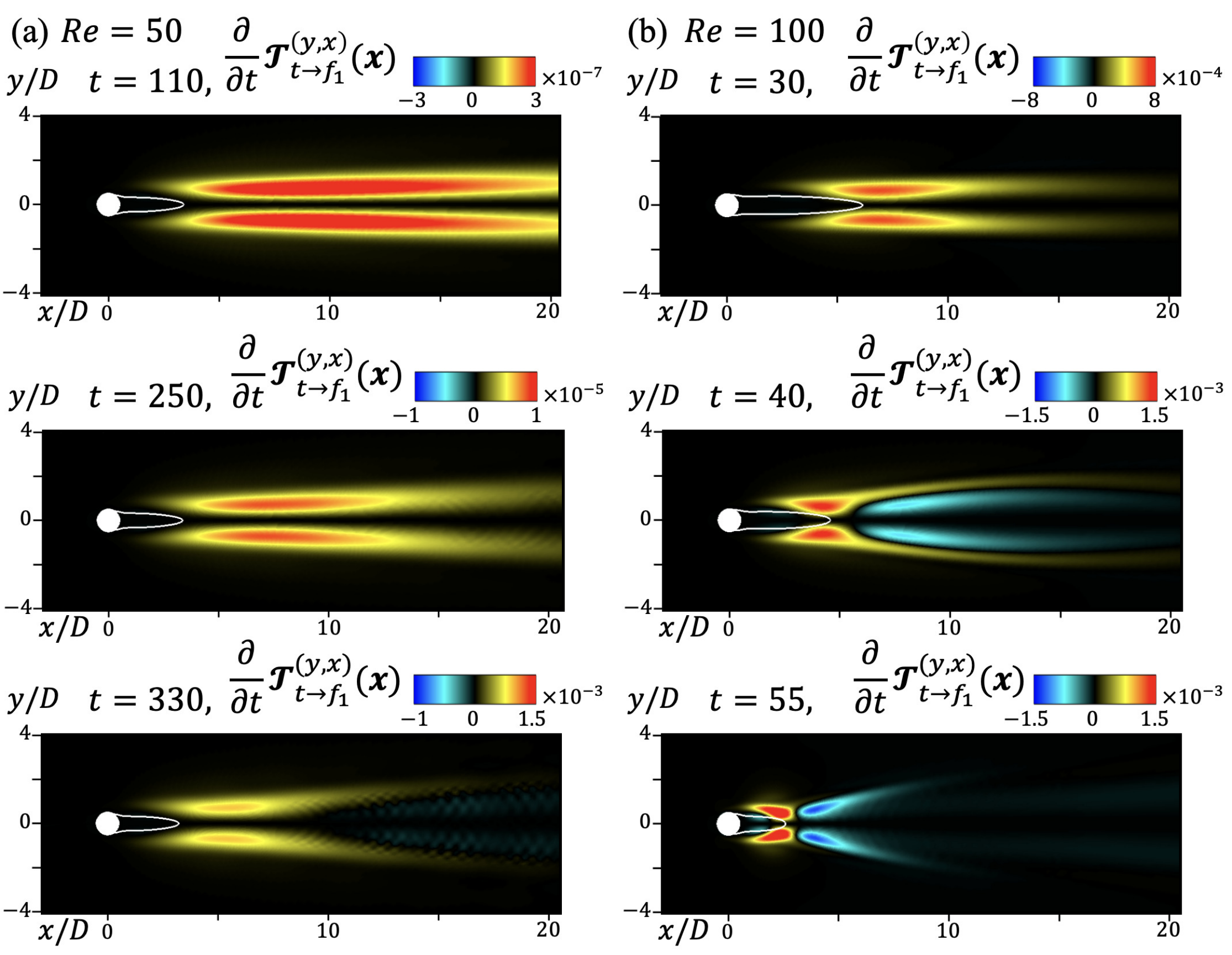}
  \captionsetup{justification=raggedright,singlelinecheck=false }
  \caption{Spatial distribution of the temporal derivative of 
$\boldsymbol{\mathcal{T}}^{(y,x)}_{t \rightarrow f_1}(\boldsymbol{x},t)|a_{f_1}(t)|^2$ 
at (a) $\Rey=50$ and (b) $100$. Temporal derivative is evaluated using a forward-difference scheme. Negative regions correspond to a decrease in the amount of energy transfer, 
observed for $t=40$ and $55$ at $Re=100$.}
 \label{fig21_rev4}
\end{figure}

The subsequent decrease following temporal amplification is clearly observed in the temporal variation of $\boldsymbol{\mathcal{T}}^{(y,x)}_{t \rightarrow f_1}(\boldsymbol{x},t)|a_{f_1}(t)|^2$. To identify the locations of energy-transfer decrease, we examine the temporal derivative of $\boldsymbol{\mathcal{T}}^{(y,x)}_{t \rightarrow f_1}(\boldsymbol{x},t)|a_{f_1}(t)|^2$. 
Figure~\ref{fig21_rev4} shows the spatial distribution of the temporal derivative, evaluated using a forward-difference scheme. 
Regions of negative values, corresponding to a decrease in the amount of energy transfer, are observed for $t=40$ and $55$ at $\Rey=100$. These times coincide closely with the decrease following the temporal peak seen in figure~\ref{fig21_rev1}(b). Such negative regions typically appear downstream of the positive regions, implying that the location of energy transfer shifts upstream: amplification occurs at the front while decrease develops behind. This forward progression of the transfer distribution is consistent with figures~\ref{fig_transfer_tDMDpc} and \ref{fig21_rev2}. However, because the overall amount of transfer decreases, the distribution becomes more locally concentrated even as it shifts forward. As a result, in the post-transient periodic state, the energy reaches a state of balance.

\begin{figure}
	\centering\includegraphics[width=11cm,keepaspectratio]{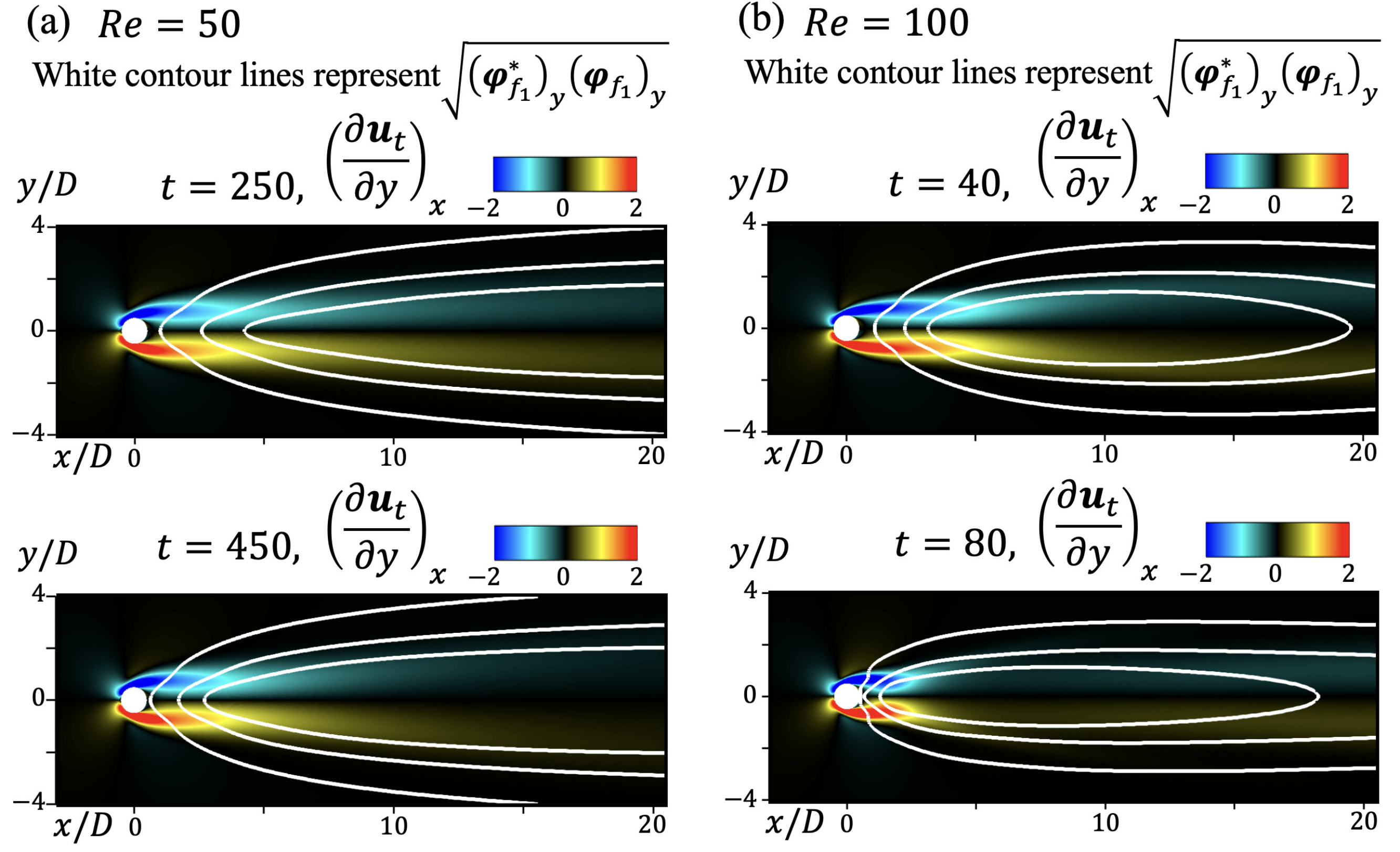}
  \captionsetup{justification=raggedright,singlelinecheck=false }
\caption{Spatial distribution of the $x$-component of the $y$-gradient of the donor component $\boldsymbol{u}_t(\boldsymbol{x}, t)$ in $\boldsymbol{\mathcal{T}}^{(y,x)}_{t \rightarrow f_1}(\boldsymbol{x},t)|a_{f_1}(t)|^2$ with contour lines of the catalyst component $(\boldsymbol{\varphi}_{f_1})_y$ at $10$, $30$, and $50$\% of its maximum value. (a) $Re=50$ and (b) $100$.}
 \label{fig21_rev3}
\end{figure}

To access more closely the localization of the energy-transfer distribution, we computed the $x$-direction component of the $y$-gradient of the donor component $\boldsymbol{u}_t(\boldsymbol{x}, t)$ in $\boldsymbol{\mathcal{T}}^{(y,x)}_{t \rightarrow f_1}(\boldsymbol{x},t)|a_{f_1}(t)|^2$. Figure~\ref{fig21_rev3} shows the resulting spatial distribution with contour lines of the catalyst component $(\boldsymbol{\varphi}_{f_1})_y$, plotted at $10$, $30$, and $50$\% of the maximum value. As explained in figure~\ref{fig_transfer} for the LSA-based energy-transfer analysis, the $x$-direction component of the $y$-gradient of base flow is distributed around the recirculation region. Therefore, as the recirculation region becomes small, the spatial distribution of the donor component becomes concentrated near the cylinder. Unlike the linear case at $\Rey=40$ in figure~\ref{fig_transfer}(b), however, the catalyst component extends further upstream even when the recirculation region is small, allowing energy transfer to remain active in the post-transient regime. This suggests that, throughout the transient process, the catalyst component continues to amplify energy from the early growth stage, adapting to the reduction of the recirculation region. At $\Rey=100$, the spatial localization originates from the donor component itself: as the recirculation region becomes small, its distribution becomes more localized. These results suggest that the amplification mechanisms underlying the post-transient periodic flow are closely linked to the size of the recirculation region.  
Based on these insights, energy transfer and budget analysis during the transient process is effective for investigating the underlying dynamics.

\section{Conclusion}\label{sec6}
This paper presents an energy transfer and budget analysis method for flow fields with growing or decaying modes. Furthermore, our approach extends to nonlinear development scenarios.  The energy transfer analysis is achieved using a projection-based ROM for linear operator eigenmodes obtained from LSA and DMD. The projected governing equations are organized using a phase-averaging procedure for linear operator eigenmodes. The introduction of phase averaging for transient processes is justified by the results obtained from tDMDpc.  
Based on the simplified equations, we derived the global energy transfer equations for the entire flow field at the frequencies of interest. When the growth rate of the eigenmode is zero, the global energy transfer equations coincide with the conventional spectral energy budget for the Fourier mode or the Navier–Stokes equation in the frequency domain. 
The global energy transfer equations were developed into a simplified form when the equation is linearized around a steady field and into an advanced form when accounting for the time-varying linear operators and modes. 

The transfer analysis results using LSA eigenmodes around a steady flow show that the growth rate obtained from LSA is expressed by the sum of energy transfer from the steady flow to the eigenmode, self-decay derived from the convection of the eigenmodes, and viscous diffusion. 
This indicates that the growth or decay of eigenmodes is governed by their energy budget.
For the flow around a cylinder, the only source of eigenmode energy is the steady flow, and this supply is nearly nonexistent for $\Rey$ values smaller than the bifurcation point. 
The decrease in the growth rate at $\Rey$ lower than the bifurcation point is caused by the self-decay from the convection of the eigenmode. Although the magnitude of this self-decay from convection depends on the computational domain, the resulting growth rate is only weakly affected by the boundary.
The spatial distributions causing these transfers and diffusion were computed from the eigenmodes and the steady field inspired by other modal energy analysis methods. 

The energy transfer and budget analysis were extended to the transient development process by considering the time variation of eigenmodes due to time-varying linear operators. 
To extract time-varying linear operator eigenmodes, we introduced tDMDpc, which regulates the initial phase and applies DMD to data obtained from multiple numerical simulation cases. 
Using tDMDpc for two-dimensional flow around a cylinder, we obtained time-varying eigenmodes that grow from a steady field to a periodic unsteady scenario. The base flow is obtained by averaging the numerical results across different initial conditions.
For flow around a cylinder, the recirculation region of the wake becomes smaller as the unstable eigenmode grows.
The amplitude coefficients and phase-shift angle obtained from tDMDpc reveal that the phase-shift angle remains constant in time during the nonlinear transient process of the cylinder wake. This result enables the application of phase-averaging operations to the transient process.

Using the time-varying eigenmodes obtained by tDMDpc, we computed the energy budget for the global energy transfer equation. 
In the flow around a cylinder, energy transfer from the base flow to the lowest-frequency eigenmode drives transient development. Simultaneously, nonlinear interactions induce an energy cascade into higher-frequency modes, which peaks alongside the energy transfer to the lowest-frequency eigenmode.
The spatial distribution of transfer fields in the transient process indicates that a substantial amount of energy transfer from the base flow is concentrated around the recirculation region. 

For future research, the proposed approach can also be used to compute energy budgets for frequencies other than the fundamental one, including the zero frequency. Furthermore, tDMDpc captures transient variations, such as temporary increases in amplitude and amount of energy transfer, that cannot be described by conventional theories based on linearized equations around the base flow. Even for the flow around a circular cylinder, examining how this transient amplification influences eigenmodes at other frequencies may lead to new insights into fluid physics.
In addition, high-frequency modes that do not appear when the system is linearized around the base flow cannot be addressed by conventional linearized equations. Therefore, tDMDpc and the associated energy budget analysis provide an effective framework for evaluating the influence of high-frequency modes on low-frequency ones during transient processes. 
While the present study focused only on a simple transient process, namely the two-dimensional flow around a circular cylinder, the proposed framework can be extended to more complex flow scenarios. In particular, applications involving multiple frequency components in the initial condition or phase-dependent variations in amplitude during the transient process could open new possibilities for analyzing transient flow dynamics.

For the data preparation strategy based on the modal-phase perspective in tDMDpc, this approach could also serve as a data-acquisition strategy for other modal decomposition techniques, such as operator-based methods (e.g. DMD variants or Koopman mode decomposition) and those based on Fourier or POD families. When combined with these methods, it may offer new possibilities for analyzing more complex transient flows.

Regarding the phase-averaged ROM, constructing a framework in which phase and amplitude (i.e. growth rate and frequency) are treated independently offers a new approach to analyzing the dynamics of transient flows with time-varying energy transfer fields.
Since this energy transfer field varies throughout the transient process, it serves as a powerful tool for observing the time-varying energy transfer distribution and discovering new flow physics. 

\appendix
\section{\label{apena}Grid convergence}
\begin{figure}
	\centering\includegraphics[width=10cm,keepaspectratio]{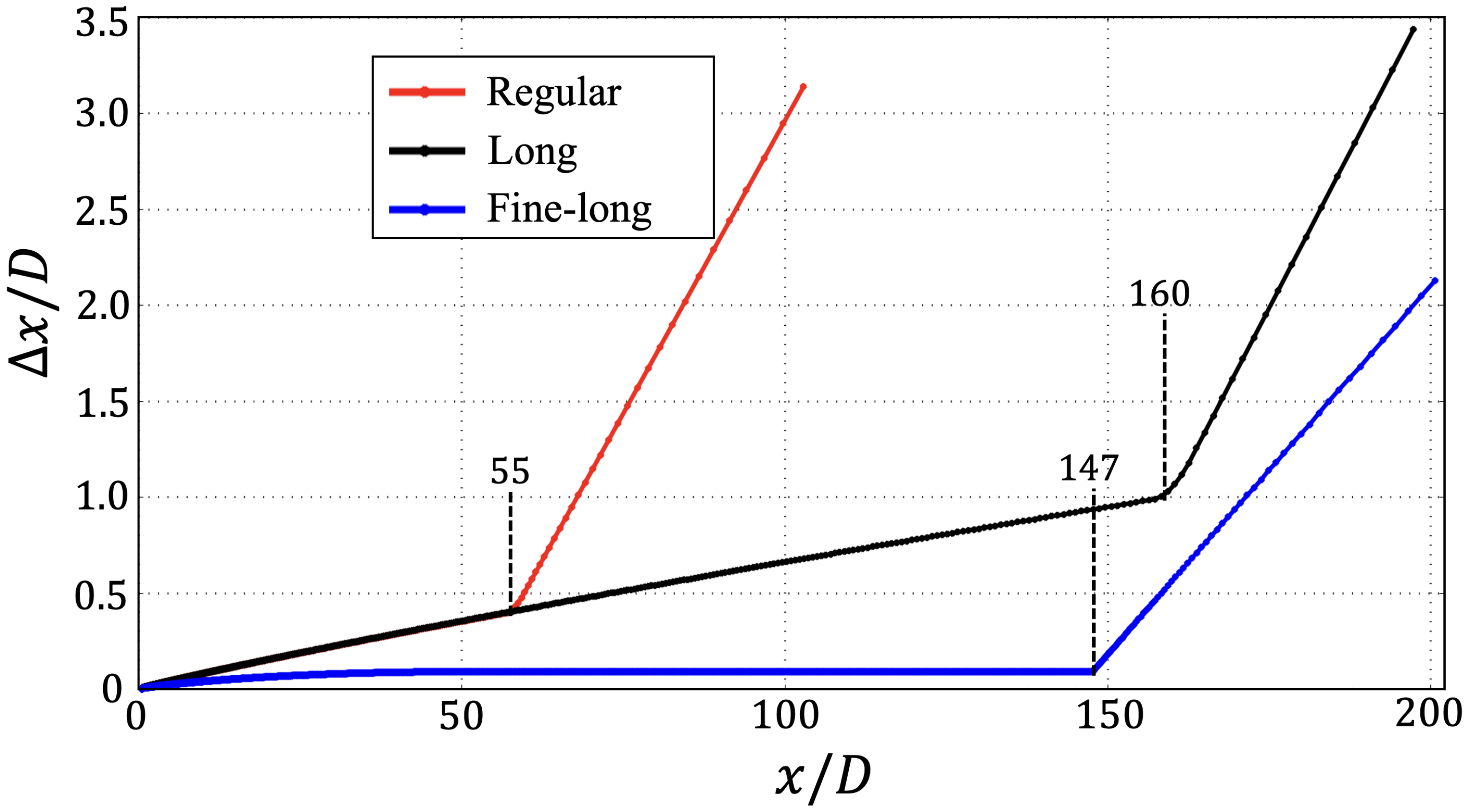}
  \captionsetup{justification=raggedright,singlelinecheck=false}
  \caption{Cell width for regular, long, and fine-long grid.}
 \label{figa1}
\end{figure}

The computational grid used in this study is validated based on the parameter dependence check. Regular grids and fine-long grids were prepared to test the effect on numerical results of the grid width and the far-field boundary size. Figure \ref{figa1} shows the grid widths of the grids located on the $x$-axis in the two computational grids. The regular and fine-long grids accelerate the expansion of the grid width when the radius is greater than $55$, $160$ and $147$, respectively. The expansion ratio of the neighboring cell is less than or equal to $1.1$. The number of cells for the fine-long grid is $580$ in the wall-parallel direction and $2260$ in the wall-normal direction.

Numerical simulations were performed for a $\Rey = 150$ using two computational grids. CFL number is set at less than or equal to $1.0$ based on the validation of our previous work \citep{12ICCFD}. For quantitative comparison, the time-averaged velocity field $\bar{\boldsymbol{u}}(\boldsymbol{x})$ and the time-averaged kinetic energy  $\overline{\tilde{u}\tilde{u}}(\boldsymbol{x})$ and $\overline{\tilde{v}\tilde{v}}(\boldsymbol{x})$ are computed as 
\begin{eqnarray}
\bar{\boldsymbol{u}}(\boldsymbol{x})=\frac{1}{T}\int \boldsymbol{u}(\boldsymbol{x}, t) dt,
\end{eqnarray}
\begin{eqnarray}
\overline{\tilde{u}\tilde{u}}(\boldsymbol{x})=\frac{1}{T}\int \tilde{u}(\boldsymbol{x}, t)\tilde{u}(\boldsymbol{x}, t) dt = \frac{1}{T}\int u((\boldsymbol{x}, t)u(\boldsymbol{x}, t) dt - \bar{u}(\boldsymbol{x})\bar{u}(\boldsymbol{x}),
\end{eqnarray}
\begin{eqnarray}
\overline{\tilde{v}\tilde{v}}(\boldsymbol{x})=\frac{1}{T}\int \tilde{v}(\boldsymbol{x}, t)\tilde{v}((\boldsymbol{x}, t) dt = \frac{1}{T}\int v((\boldsymbol{x}, t)v(\boldsymbol{x}, t) dt - \bar{v}(\boldsymbol{x})\bar{v}(\boldsymbol{x}),
\end{eqnarray}
based on \citet{Asada_master}. Here, $T$ is set to $300$, which is long enough since the primary dimensionless period of the vortex shedding is about $5$.

\begin{figure}
	\centering\includegraphics[width=13cm,keepaspectratio]{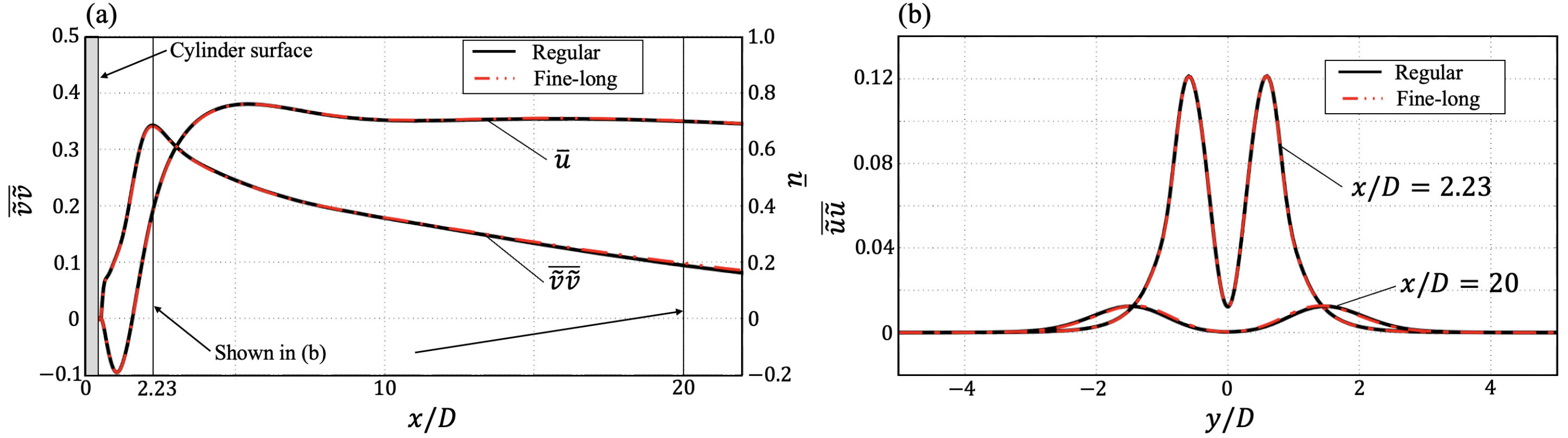}
  \captionsetup{justification=raggedright,singlelinecheck=false}
  \caption{Comparison of the regular grid and fine-long grid with the average value of kinetic energy distribution $\overline{\tilde{u}\tilde{u}}$, $\overline{\tilde{v}\tilde{v}}$, and streamwise velocity fields $\bar{u}$: (a) distribution on the $x$-axis of $\overline{\tilde{v}\tilde{v}}$ and $\bar{u}$, (b) $y$-direction distribution of $\overline{\tilde{u}\tilde{u}}$ and $\bar{u}$ at $x/D=20$, $2.23$. $x/D=2.23$ is the peak value position of $\overline{\tilde{v}\tilde{v}}$.}
 \label{fig_kineticenergy}
\end{figure}

Figure \ref{fig_kineticenergy} shows the distribution of $\bar{\boldsymbol{u}}(\boldsymbol{x})$, $\overline{\tilde{u}\tilde{u}}$, and $\overline{\tilde{v}\tilde{v}}$ in the $x$ and $y$ directions. The distributions in the $x$ direction for $x/D<20$ shown in figure \ref{fig_kineticenergy} (a) are in better agreement for the regular and fine-long grids. The range of the $x$-direction was chosen because most of the energy transfer occurs at $x/D<20$. Since the deviation of the kinetic energy distribution is relatively larger than that of the average field, the kinetic energy distribution is focused in the $y$ direction, distribution shown in figure \ref{fig_kineticenergy} (b). The $x$-section coordinates position was chosen to be $x/D=2.23$ and $20$. $x/D=2.23$ is selected since the peak value position of $\overline{\tilde{v}\tilde{v}}$ on the $x$-axis. The $y$-axis distribution of kinetic energy in the streamwise direction at $x/D=2.23$ and $20$ is in close agreement with the regular and fine-long grids. Therefore, the resolution of numerical results with the regular grid is enough for this work. Moreover, since the cell width for the regular and long grids is identical in the region $x/D \leq 55$, the local values from the long grid at $x/D<20$ are expected to exhibit a similar distribution with regular and fine-long grids.

\begin{figure}
	\centering\includegraphics[width=13cm,keepaspectratio]{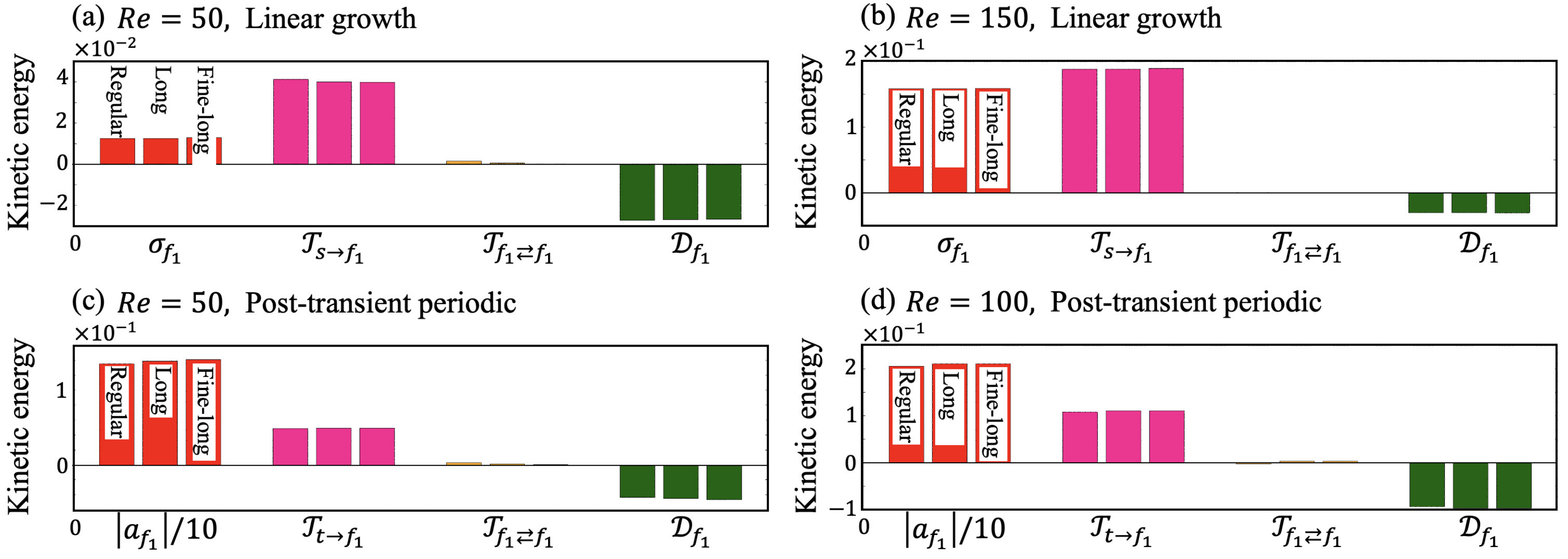}
  \captionsetup{justification=raggedright,singlelinecheck=false }
  \caption{Comparison of representative global parameters for different grids (a) linear regime at $Re=50$, (b) linear regime at  $Re=150$, (c) post-transient periodic state at $Re=50$, (d) post-transient periodic state at  $Re=100$. The parameters show only minor differences, indicating that the effects of grid resolution and domain extent are negligible.}
 \label{fig_globalconv}
\end{figure}

The terms associated with energy transfer are global parameters that involve spatial integration over the entire computational domain. Therefore, it is important to verify the effects of grid resolution and domain size on the integrated quantities. While the effect of domain size in the region where the growth rate becomes negative is discussed in the main text, this section focuses on the dependence on grid resolution and domain size during the growth phase. Representative global parameters, two transfer terms, and a diffusion term, are examined, along with the growth rate $\sigma_{f_1}$ in the linear-growth regime and the amplitude coefficient $|a_{f_1}|$ in the periodic regime. Figure~\ref{fig_globalconv} compares these representative global parameters obtained by LSA in the linear-growth and periodic regimes. First, comparing the long-grid and fine-long-grid results reveals that the differences in all global parameters are negligible. Because these two grids have nearly identical domain sizes and differ only in resolution, the grid resolution is considered sufficient for evaluating the global parameters. In contrast, a comparison between the regular and long grids shows a slight difference in the periodic regime, which can be attributed to the downstream convection of flow fluctuations being more pronounced in the periodic regime than in the linear-growth regime. Nevertheless, this variation is much smaller than the temporal variations in energy discussed in figure~\ref{fig_triadicenergy_separate}, indicating that the influence of grid and domain size is limited. 
 
\section{\label{apenb}Time-stepping LSA and validation}
We check the validity of time-stepping LSA and the resulting eigenvalues (Ritz values)  and eigenmodes.
The validity is divided into two parts: checking the dependence of parameters in time-stepping LSA and checking $\Rey$ dependence. 

\subsection{\label{apenb_1} Time-stepping LSA methodology}
This study uses matrix-free LSA, which is referred to as a time-stepping approach \citep{Chiba, Ohmichi_D, Ranjan_2020}. 
This method approximates the matrix $A'$ from the time evolution of the flow field with disturbances added to the base flow using numerical simulation. Early developments of matrix-free stability analysis based on the time-stepping approach were initiated by \citet{Chiba} using the Arnoldi method \citep{arnoldi1951principle} and were subsequently extended to tri-global stability analysis \citep{chiba2001three}. Independently, a similar CFD-based approach was conducted by \citet{BAGHERI_2009}. Among various formulations of time-stepping LSA, the Arnoldi-based approach remains the most established. In the present study, however, we employ a DMD-based time-stepping formulation \citep{Ranjan_2020}, which is closely related to the treatment of time-dependent operators. As discussed by \citet{DMD2}, the Arnoldi method and DMD are equivalent in their projection-based formulation.


A disturbance $\boldsymbol{u}'^{(n=0)}(t=0)$, where $n$ is the iteration number for collecting snapshots, is initially prepared by 
\begin{equation}
\boldsymbol{u}'^{(0)}(0) = {\epsilon}_0|\boldsymbol{u}_b|_2\frac{\boldsymbol{r}_0}{|\boldsymbol{r}_0|_2},
   \label{initialdist}
\end{equation}
where $|\cdot|_2$ represents the $L2$ norm of the $N$-dimensional vector, $\boldsymbol{u}_b$ is the base flow, $\boldsymbol{r}_0$ is the random disturbance, and $\epsilon_0$ is a parameter that can be set to any value, representing the ratio of the base flow and disturbance norms. The flow field at $\Delta T$ after a time progression by the numerical simulation is presented below:
\begin{equation}
\boldsymbol{u}_b + \boldsymbol{u}'^{(0)}(\Delta T)=A^{\text{CFD}}\{\boldsymbol{u}_b+\boldsymbol{u}'^{(0)}(0)\},
   \label{evoLSA}
\end{equation}
where, $\boldsymbol{u}'^{(0)}(\Delta T)$ is the flow field obtained by evolving $\boldsymbol{u}'^{(0)}(t=0)$ forward in time by $\Delta T$. Here, $A^{\text{CFD}}$ represents the linear operator that advances the flow field by a time interval $\Delta T$ in the CFD simulation. In this computation, the influence of the nonlinear term $-(\boldsymbol{u}' \cdot \nabla)\boldsymbol{u}'$ is excluded so that the second-order nonlinearity induced by the disturbance is removed. The disturbance field $\boldsymbol{u}'$ is obtained by subtracting the base flow from the instantaneous flow field at each time step, limited by CFL. Strictly speaking, the time advancement in CFD allows a slight temporal variation of the base flow. However, when a steady field is used as the base flow, this influence remains negligible if the disturbance norm is appropriately set. 

To collect data that approximates linearized operator $A'$, we iteratively compute the time progression of the flow field with disturbances added to the base flow. For the $n$th iteration ($n > 1$), the disturbance $\boldsymbol{u}'_n (t=0)$ to the base flow is computed from the flow field of ($n-1$)th iteration presented below:
\begin{equation}
\boldsymbol{u}'^{(n)}(0) = {\epsilon}_0|\boldsymbol{u}_{b}|_2\frac{\boldsymbol{u}'^{(n-1)}(\Delta T)}{|\boldsymbol{u}'^{(n-1)}(\Delta T)|_2}.
   \label{redist}
\end{equation}
And time progressing by numerical simuration linear linearized around $\boldsymbol{u}_b$ as follows
\begin{equation}
\boldsymbol{u}_b + \boldsymbol{u}'^{(n)}(\Delta T)=A^{\text{CFD}}\{\boldsymbol{u}_b + \boldsymbol{u}'^{(n)}(0)\}.
   \label{evoLSA}
\end{equation}

\begin{figure}
	\centering\includegraphics[width=14cm,keepaspectratio]{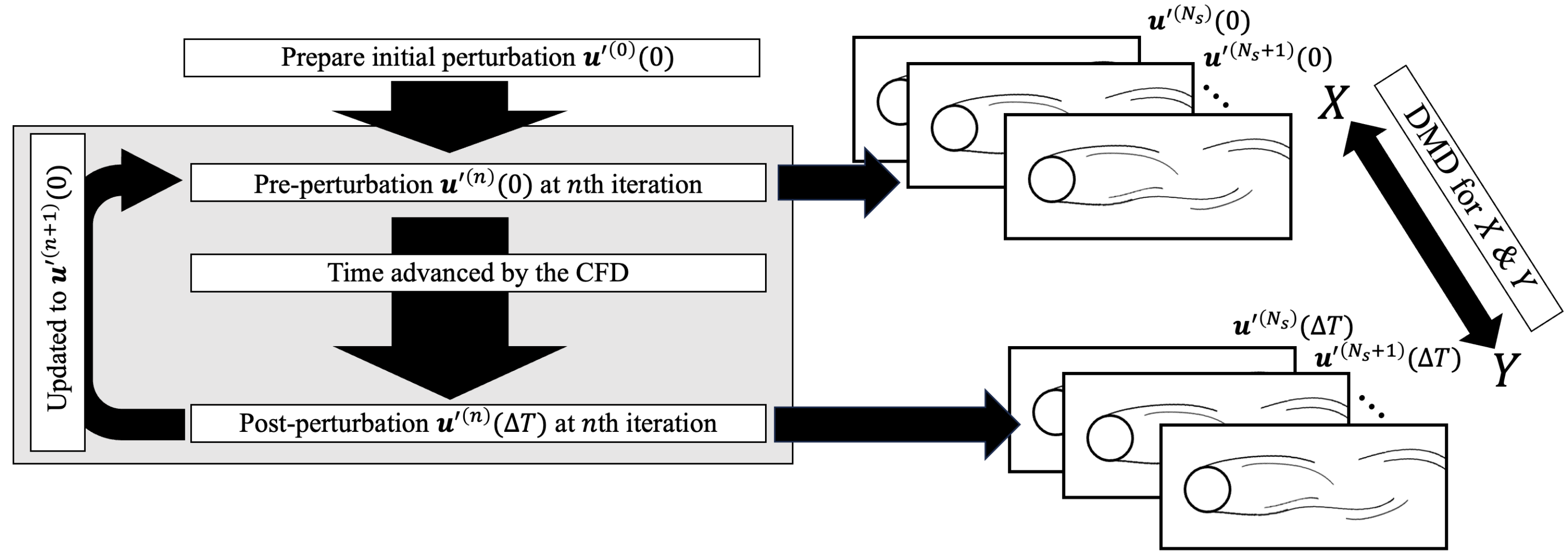}
  \captionsetup{justification=raggedright,singlelinecheck=false}
  \caption{Schematic of time-stepping LSA using DMD.}
 \label{fig1}
\end{figure}

The eigenvectors can be computed by applying DMD to the following matrices
\begin{eqnarray}
X&=&[\boldsymbol{u}'^{(N_s)}(0),\,\,\,\,\,\,\boldsymbol{u}'^{(N_s+1)}(0),\,\,\,\,\,\,\cdots,\,\,\,\,\,\boldsymbol{u}'^{(M+N_s-1)}(0)] \in \mathbb{R}^{N \times M},\\
Y&=&[\boldsymbol{u}'^{(N_s)}(\Delta T),\,\boldsymbol{u}'^{(N_s+1)}(\Delta T),\,\cdots,\,\boldsymbol{u}'^{(M+N_s-1)}(\Delta T)]  \in \mathbb{R}^{N \times M},
   \label{matrixQ}
\end{eqnarray}
where $N_s$ is the starting position of the DMD data and is set to remove the effect of snapshots with initial disturbance, and $M$ is the number of snapshots. 
In DMD, the linear operator $A$ does not necessarily represent the linearized operator $A'$ around the base flow but rather the linear operator that best fits the data. However, in time-stepping LSA, the snapshots obtained from disturbances that grow linearly from the base flow allow the DMD to approximate the linearized Navier–Stokes operator $A'$. The DMD-based mode extraction in the time-stepping LSA is proposed by \citet{Ranjan_2020}. In this study, $N_s$ is set to $150D/(U_{\infty}\Delta T)$. $M$ is set such that $N_sM-1 = 50D/(U_{\infty}\Delta T)$ based on \citet{mypaper2}. A conceptual diagram of the mode extraction process using the time-stepping LSA is shown in figure~\ref{fig1}.

\subsection{\label{apenb_1}Time-stepping LSA parameter dependance}
Time-stepping LSA has two parameters, $\Delta T$ and $\epsilon_0$. In this paper, $\Delta T$ is fixed at $0.1$, and the parameter dependence is tested by varying the magnitude of $\epsilon_0$. 

\begin{figure}
	\centering\includegraphics[width=8cm,keepaspectratio]{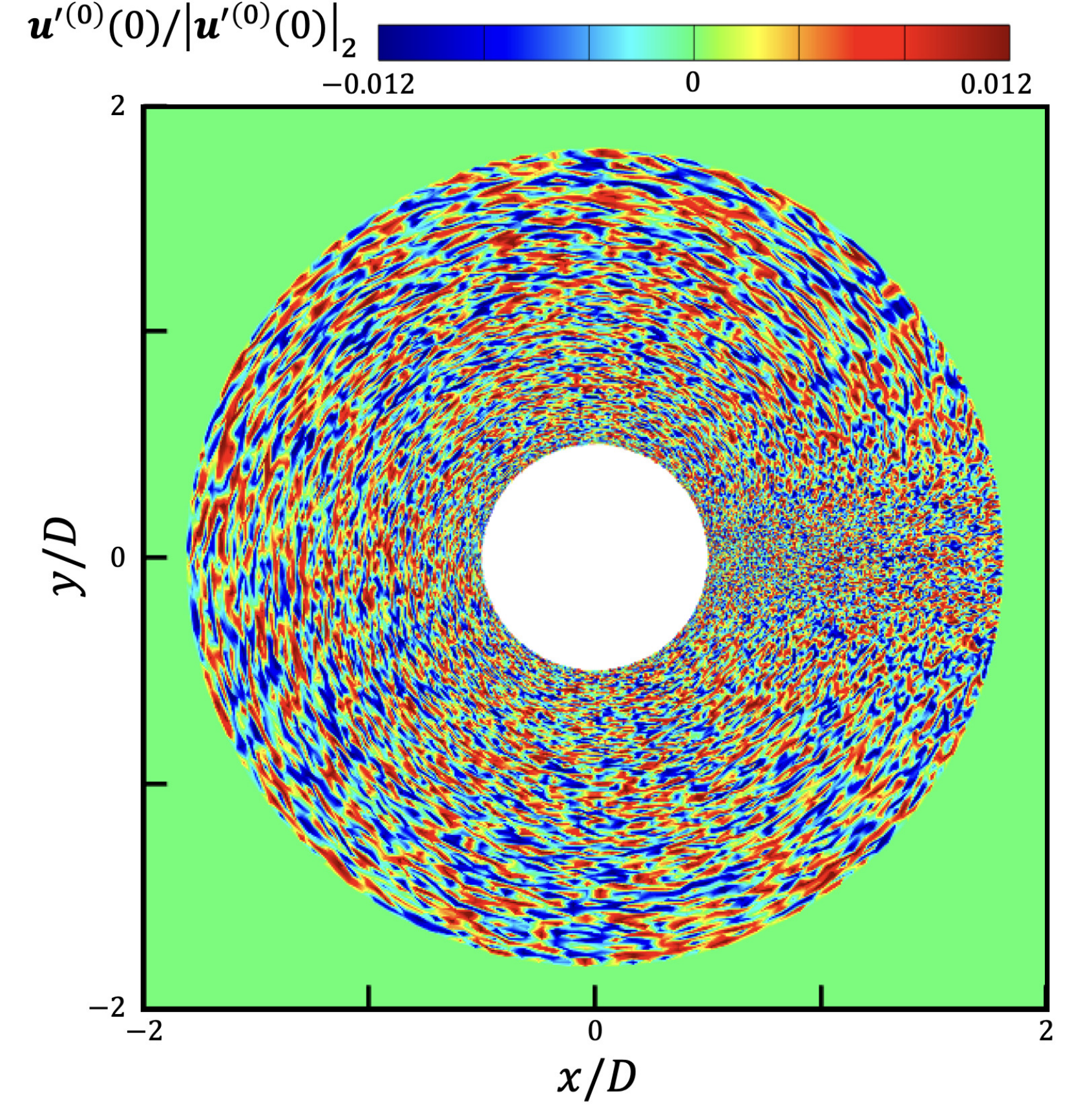}
  \captionsetup{justification=raggedright,singlelinecheck=false}
  \caption{Initial disturbance for streamwise-velocity component at time-stepping LSA.}
 \label{figb1}
\end{figure}

\begin{figure}
	\centering\includegraphics[width=12cm,keepaspectratio]{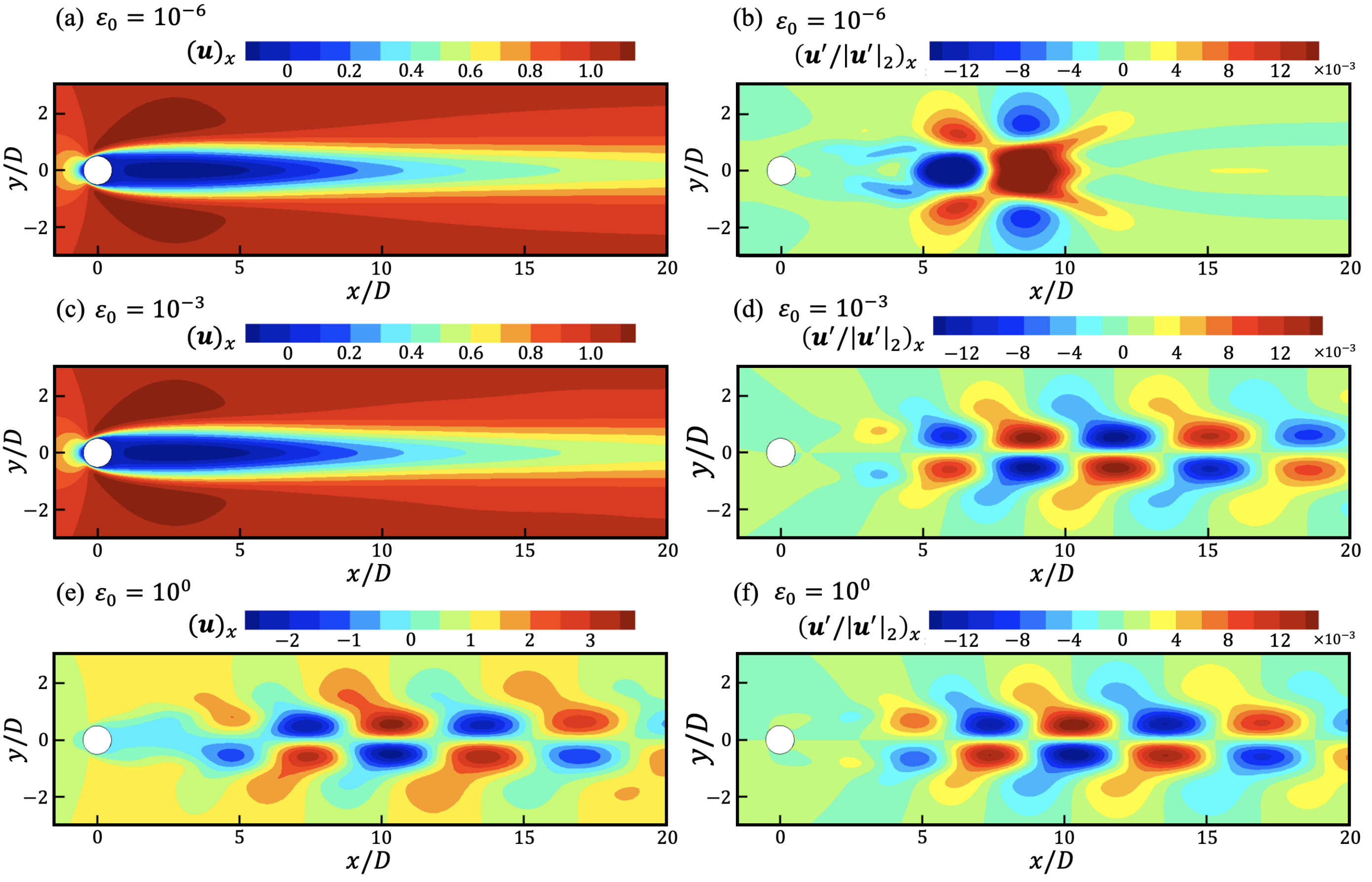}
  \captionsetup{justification=raggedright,singlelinecheck=false}
  \caption{Instantenious streamwise component of velocity and perturbation at $Re=100$ and $t=50$ in the time-stepping approach: (a) velocity with ($\epsilon_0=10^{-6}$), (b) perturbation with ($\epsilon_0=10^{-6}$), (c) velocity with ($\epsilon_0=10^{-3}$), (d) perturbation with ($\epsilon_0=10^{-3}$), (e) velocity with ($\epsilon_0=10^{0}$), and (f) perturbation with ($\epsilon_0=10^{0}$).}
 \label{fig3}
\end{figure}

For investigate the effect of $\epsilon_0$, time-stepping LSA was performed on $\epsilon_0=10^{-6}, 10^{-3},$ and $10^{0}$ at $\Rey=100$.
Figure \ref{figb1} shows the initial disturbance at $\epsilon_0=10^{-3}$. Following \citet{Ranjan_2020}, the initial disturbance was generated only near the cylinder. Figure \ref{fig3} shows instantaneous streamwise components of velocity and perturbation. For $\epsilon_0=10^{-6}$ and $10^{-3}$, the distribution of the streamwise component is an almost steady field. This is because the amount of perturbation, whose absolute value is determined by $\epsilon_0$, is smaller than that of the steady field. In contrast, when $\epsilon_0=10^{0}$, the magnitude of the perturbation is larger than the magnitude of the steady field, and the flow distribution is unphysical. In addition, for small values of $\epsilon_0=10^{-6}$, the perturbation distribution is symmetric to $y=0$, and no asymmetric Karman vortex is formed. This is due to the suppression of fluctuations by a small $\epsilon_0$ during the development process of the disturbance, and the disturbance does not grow to the formation of the Karman vortex. In the time-stepping approach, the flow field is numerically integrated with perturbation and the base flow. As a result, slight variations in the base flow are inherently allowed. When $\epsilon_0$ is small, these base-flow variations can become dominant over the imposed perturbation.

\begin{figure}
	\centering\includegraphics[width=13cm,keepaspectratio]{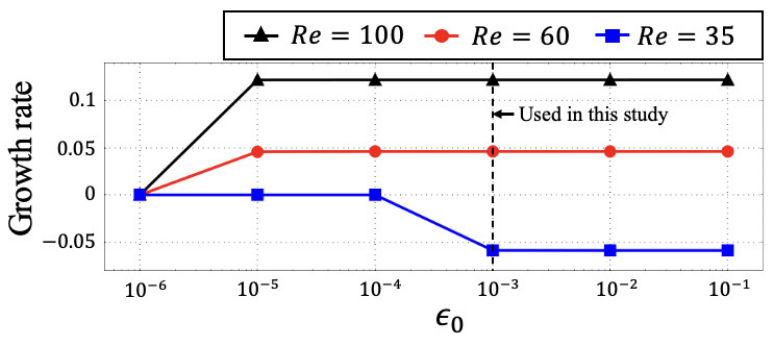}
  \captionsetup{justification=raggedright,singlelinecheck=false}
  \caption{Variation of growth rate for various $\epsilon_0$ with $\Rey=35, 60, 100$.}
 \label{figb3}
\end{figure}

To quantitatively evaluate the effect of $\epsilon_0$, growth rates were computed for several $\epsilon_0$ and $\Rey$ cases. Figure \ref{figb3} shows the variation of growth rate with respect to  $\epsilon_0$ at $\Rey=35, 60,$ and $100$. The growth rate is constant for all $\Rey$ cases when $\epsilon_0$ is sufficiently large. However, in general, time-stepping LSA must be performed within a range where $\epsilon_0$ is small; otherwise, the flow fields grow beyond the linear growth region. Therefore, this study uses  $\epsilon_0=10^{-3}$ in all cases.

\subsection{\label{apena_2}Validation for Reynolds number dependance}
Since the comparison with previous studies on the Reynolds number dependence is mentioned in figure \ref{fig_growth_LSA}, here we compare our results with those of ordinary CFD results. However, due to the difficulty of comparing the results of CFD and LSA under conditions of rapidly developing conditions at a relatively high $\Rey$, the results are compared near the bifurcation point. The growth rates and frequencies near the bifurcation point are shown in table \ref{table_bifurcation}. Between $Re=46.8$ and $46.9$, the growth rate changes from negative to positive. Since the absolute value of the growth rate is minimum at $Re = 46.8$, the bifurcation point is roughly $46.8$.

\begin{table}
 \centering
  \begin{tabular}{ccc}
   $Re$ & Growth rate & Frequency\\[3pt]
   $45.0$ & $-7.79034 \times 10^{-3}$ & $0.115833$ \\
   $46.5$ & $-1.37338 \times 10^{-3}$ & $0.116221$ \\
   $46.6$ & $-9.54785 \times 10^{-4}$ & $0.116246$ \\
   $46.7$ & $-5.37557 \times 10^{-4}$ & $0.116269$ \\
   $46.8$ & $-1.21642 \times 10^{-4}$ & $0.116294$ \\
   $46.9$ & $\,2.93599 \times 10^{-4}$ & $0.116319$ \\
   $47.0$ & $\,7.07521 \times 10^{-4}$ & $0.116343$ \\
   $50.0$ & $\,1.26154 \times 10^{-2}$ & $0.116973$ \\
  \end{tabular}
    \captionsetup{justification=raggedright,singlelinecheck=false}
   \caption{Growth rate and frequency near the bifurcation point.}
 \label{table_bifurcation}
\end{table}

\begin{figure}
	\centering\includegraphics[width=14cm,keepaspectratio]{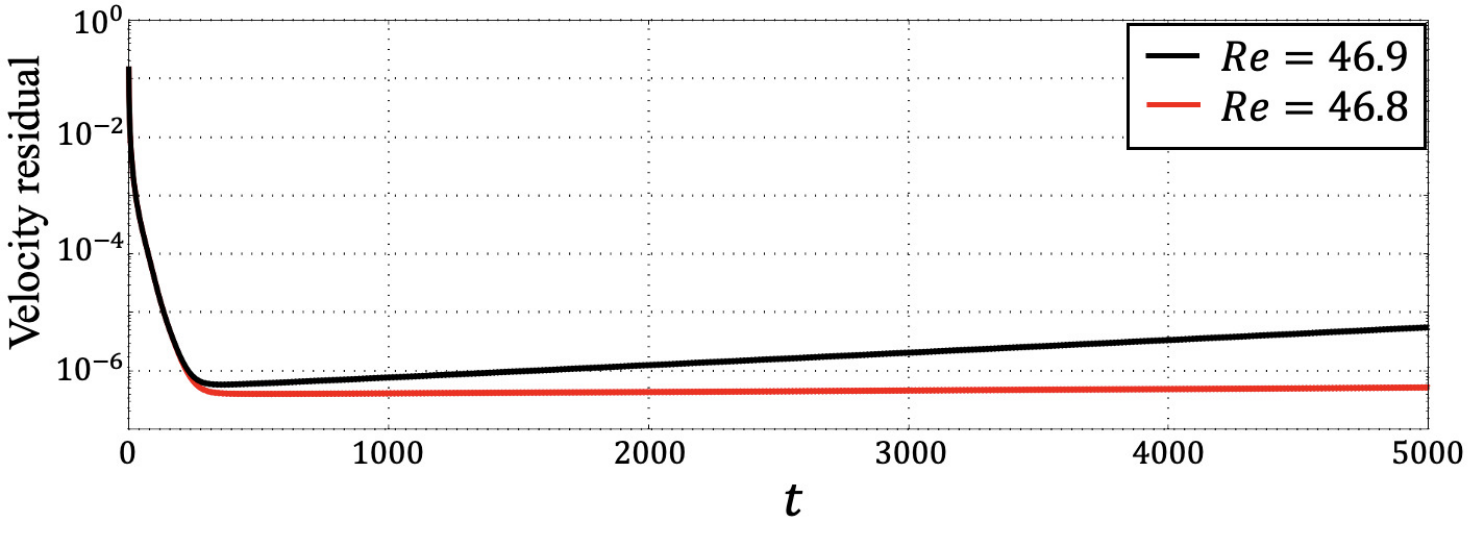}
  \captionsetup{justification=raggedright,singlelinecheck=false}
  \caption{Velocity residuals for the numerical simulation at $\Rey=46.8$ and $46.9$. In the present simulation, critical $\Rey$ is determined to be $46.8$, as the velocity residual remains at a constant order, which is consistent with the LSA result.}
 \label{figa4}
\end{figure}

The development of flow fields with $\Rey = 46.8$ and $46.9$ are computed by CFD without explicitly including disturbances. Figure \ref{figa4} shows the residuals of the velocity vector in the flow field obtained by CFD. Here, the velocity residual is computed by
\begin{equation}
(\text{Velocity residual})= \frac{1}{\Delta T}|\boldsymbol{u}(\boldsymbol{x},t+\Delta T)-\boldsymbol{u}(\boldsymbol{x},t)|_2,
   \label{residual}
\end{equation}
where $\boldsymbol{x}_i$ represents $i$th grid points. No increase in residuals is confirmed for $Re=46.8$, but for $Re=46.9$, the residuals gradually increase. An increase in residuals implies a sign of non-stationarity. Hence, the bifurcation point in the CFD obtained by the LSA is in close agreement. To compare the frequency of growing modes from CFD, DMD is applied to $t = 1000$--$2000$ at $\Rey = 46.9$. The obtained frequecy is $0.116305$, and growth rate is $4.87011 \times 10^{-4}$. Therefore, the time-stepping LSA presents the linear growing eigenmodes in the CFD.

\section{Derivation of time-derivative term with time-varying eigenmode}\label{dynamicalyorthognal}
We demonstrate that the temporal evolution of eigenmodes computed numerically via the tDMDpc algorithm agrees with equation (\ref{LSAeq_timederive_new}). First, the time-derivative term can be written as
\begin{eqnarray}
\frac{d}{dt}\left(a_{f_l} e^{i\alpha^{f_l}} \boldsymbol{\varphi}_{f_l}\right)&=& e^{i\alpha^{f_l}}\left(\frac{da_{f_l}}{dt}\boldsymbol{\varphi}_{f_l}+a_{f_l}\frac{d\boldsymbol{\varphi}_{f_l}}{dt}\right).
\end{eqnarray}
Take the inner product with $a_{f_k}e^{i\alpha^{f_k}}\boldsymbol{\varphi}_{f_k}$, and phase-averaging, it becomes
\begin{eqnarray}
&&\sum_{l=-\infty}^{\infty}\overline{\left\langle
e^{i\alpha^{f_l}}\!\left(\frac{da_{f_l}}{dt}\boldsymbol{\varphi}_{f_l}
     + a_{f_l}\frac{d\boldsymbol{\varphi}_{f_l}}{dt}\right),\,
a_{f_k}e^{i\alpha^{f_k}}\boldsymbol{\varphi}_{f_k} \right\rangle}^{\alpha} \nonumber \\
&&\,\,\,\,\,\,\,\,\,= \sum_{l=-\infty}^{\infty}\overline{e^{i(\alpha^{f_l}-\alpha^{f_k})}}^{\alpha}
    \frac{da_{f_l}}{dt}\,a^*_{f_k}
    \left\langle \boldsymbol{\varphi}_{f_l}, \boldsymbol{\varphi}_{f_k} \right\rangle + \sum_{l=-\infty}^{\infty}\overline{e^{i(\alpha^{f_l}-\alpha^{f_k})}}^{\alpha}
    a_{f_l}a^*_{f_k}
    \left\langle \frac{d\boldsymbol{\varphi}_{f_l}}{dt}, \boldsymbol{\varphi}_{f_k} \right\rangle
 \nonumber \\
&&\,\,\,\,\,\,\,\,\,
= \frac{da_{f_k}}{dt}\,a^*_{f_k}
    \left\langle \boldsymbol{\varphi}_{f_k}, \boldsymbol{\varphi}_{f_k} \right\rangle
 \,+\, a_{f_k}a^*_{f_k}
    \left\langle \frac{d\boldsymbol{\varphi}_{f_k}}{dt}, \boldsymbol{\varphi}_{f_k} \right\rangle\nonumber \\
&&\,\,\,\,\,\,\,\,\,
 = \frac{da_{f_k}}{dt}a^*_{f_k}
+\, |a_{f_k}|^2
    \left\langle \frac{d\boldsymbol{\varphi}_{f_k}}{dt}, 
        \boldsymbol{\varphi}_{f_k} \right\rangle.
\end{eqnarray}

For the amplitude coefficient in tDMDpc, the first term takes the form
\begin{eqnarray}
\frac{da_{f_k}}{dt}a^*_{f_k}
   &=& \frac{d|a_{f_k}|e^{\arg{(a_{f_k})}i}}{dt}|a_{f_k}|e^{-\arg{(a_{f_k})}i} \nonumber \\
   &=& |a_{f_k}|\frac{d|a_{f_k}|}{dt}+i|a_{f_k}|^2\frac{d\arg{(a_{f_k})}}{dt} \nonumber \\
   &=& (\sigma_{f_k} + 2\pi f_ki) |a_{f_k}|^2.
\end{eqnarray}
For the real part of the second term, the following holds:
\begin{eqnarray}
\text{Real}\left(|a_{f_k}|^2
    \left\langle \frac{d\boldsymbol{\varphi}_{f_k}}{dt}, 
        \boldsymbol{\varphi}_{f_k} \right\rangle\right) 
   &=& \frac{|a_{f_k}|^2}{2}\left(
    \left\langle \frac{d\boldsymbol{\varphi}_{f_k}}{dt}, 
        \boldsymbol{\varphi}_{f_k} \right\rangle
    + \left\langle \frac{d\boldsymbol{\varphi}_{f_k}}{dt}, 
        \boldsymbol{\varphi}_{f_k} \right\rangle^* \right) \nonumber \\
      &=& \frac{|a_{f_k}|^2}{2}\left(
    \left\langle \frac{d\boldsymbol{\varphi}_{f_k}}{dt}, 
        \boldsymbol{\varphi}_{f_k} \right\rangle
    + \left\langle \boldsymbol{\varphi}_{f_k}, 
    	\frac{d\boldsymbol{\varphi}_{f_k}}{dt} \right\rangle \right) \nonumber \\
     &=& \frac{|a_{f_k}|^2}{2}\frac{d}{dt}\left(
    \left\langle \boldsymbol{\varphi}_{f_k}, 
    	\boldsymbol{\varphi}_{f_k} \right\rangle \right) = 0.
\end{eqnarray}
The last line holds because, in tDMDpc, the eigenmodes are normalized, i.e., $\left\langle \boldsymbol{\varphi}_{f_k},\boldsymbol{\varphi}_{f_k} \right\rangle=1$ at arbitrary time.
For the imaginary part, it becomes
\begin{eqnarray}
\text{Imag}&&\left(|a_{f_k}|^2
    \left\langle \frac{d\boldsymbol{\varphi}_{f_k}}{dt}, 
        \boldsymbol{\varphi}_{f_k} \right\rangle\right) \nonumber \\
&=& 
|a_{f_k}|^2\text{Imag}\left(
    \left\langle \frac{\boldsymbol{\varphi}_{f_k}(\boldsymbol{x},t+\Delta T) - \boldsymbol{\varphi}_{f_k}(\boldsymbol{x},t)}{\Delta T}, 
        \boldsymbol{\varphi}_{f_k}(\boldsymbol{x},t) \right\rangle\right) \nonumber \\
&=& \frac{|a_{f_k}|^2}{\Delta T}\text{Imag}
\left(
    \left\langle \boldsymbol{\varphi}_{f_k}(\boldsymbol{x},t+\Delta T), \boldsymbol{\varphi}_{f_k}(\boldsymbol{x},t) \right\rangle
   - \left\langle \boldsymbol{\varphi}_{f_k}(\boldsymbol{x},t), 
        \boldsymbol{\varphi}_{f_k}(\boldsymbol{x},t) \right\rangle\right) \nonumber \\
&=& \frac{|a_{f_k}|^2}{\Delta T}\text{Imag}
\left(
    \left\langle \boldsymbol{\varphi}_{f_k}(\boldsymbol{x},t+\Delta T),\boldsymbol{\varphi}_{f_k}(\boldsymbol{x},t) \right\rangle
   - 1\right).
   \label{imagorhonomal}
\end{eqnarray}
Here, since the tDMDpc modes are transformed by the solution of equation~(\ref{Aargmin_tDMDpc}), the following holds:
\begin{eqnarray}
&&\left\langle \boldsymbol{\varphi}_{f_k}(\boldsymbol{x},t+\Delta T),\boldsymbol{\varphi}_{f_k}(\boldsymbol{x},t) \right\rangle \nonumber \\
     &=& 
        \left\langle e^{z_n} \boldsymbol{\check{\varphi}}_{f_k}(\boldsymbol{x},t+\Delta T),e^{z_{n-1}}\boldsymbol{\check{\varphi}}_{f_k}(\boldsymbol{x},t)  \right\rangle \nonumber \\
         &=&  e^{z_{n-1}}  \left\langle \boldsymbol{\check{\varphi}}_{f_k}(\boldsymbol{x},t), \boldsymbol{\check{\varphi}}_{f_k}(\boldsymbol{x},t+\Delta T)   \right\rangle
    e^{-z_{n-1}}   \left\langle \boldsymbol{\check{\varphi}}_{f_k}(\boldsymbol{x},t+\Delta T),\boldsymbol{\check{\varphi}}_{f_k}(\boldsymbol{x},t)   \right\rangle \nonumber \\
    &=&\left( \left\langle \boldsymbol{\check{\varphi}}_{f_k}(\boldsymbol{x},t+\Delta T),\boldsymbol{\check{\varphi}}_{f_k}(\boldsymbol{x},t)   \right\rangle \right)^*
\left\langle \boldsymbol{\check{\varphi}}_{f_k}(\boldsymbol{x},t+\Delta T),\boldsymbol{\check{\varphi}}_{f_k}(\boldsymbol{x},t) \right\rangle \nonumber \\
&=&\left| \left\langle \boldsymbol{\check{\varphi}}_{f_k}(\boldsymbol{x},t+\Delta T),\boldsymbol{\check{\varphi}}_{f_k}(\boldsymbol{x},t)   \right\rangle \right|^2, 
\end{eqnarray}
where $\boldsymbol{\check{\varphi}}_{f_k}$ represents arbitary eigenmodes not necessary to satisfy equation~(\ref{Aargmin_tDMDpc}). Therefore, $\left\langle \boldsymbol{\varphi}_{f_k}(\boldsymbol{x},t+\Delta T),\boldsymbol{\varphi}_{f_k}(\boldsymbol{x},t) \right\rangle$ is real. Consequently, equation (\ref{imagorhonomal}) vanishes (i.e., equals zero).

Therefore, by imposing constraints on the tDMDpc modes, we obtain
\begin{eqnarray}
\left\langle \frac{d{\boldsymbol{\varphi}_{f_l}}}{d{t}}, \boldsymbol{\varphi}_{f_k} \right\rangle = 0,
\end{eqnarray}
and equation (\ref{LSAeq_timederive_new}) is satisfied in tDMDpc modes. This condition coincides with the dynamically orthogonal condition introduced by \citet{Sapsis_2009} for stochastic, time-dependent partial differential equations.

\section{Orthotgonality of phase domain for phase-averaged energy}\label{energyorthognal}
We show that the amplitude of the DMD mode is directly linked to the phase-averaged kinetic energy. This relationship holds only when variations in amplitude, growth rate, frequency, and eigenmodes along the phase direction are absent, or when sufficient averaging renders their influence negligible. We begin by defining the kinetic energy as the squared velocity norm, $\langle \boldsymbol{u}-\boldsymbol{u}_t, \boldsymbol{u}-\boldsymbol{u}_t \rangle$. With the DMD mode representation, the energy is
\begin{eqnarray}
\langle \boldsymbol{u}-\boldsymbol{u}_t, \boldsymbol{u}-\boldsymbol{u}_t \rangle = \left\langle \sum_{l=-\infty}^{\infty}a_{f_l} e^{\alpha^{f_l}i}\boldsymbol{\varphi}_{f_l}, \sum_{n=-\infty}^{\infty}a_{f_n} e^{\alpha^{f_n}i}\boldsymbol{\varphi}_{f_n} \right\rangle \nonumber \\
= \sum_{l=-\infty}^{\infty}\sum_{n=-\infty}^{\infty} a_{f_l}a^*_{f_n}  e^{\alpha^{f_l}i} e^{- \alpha^{f_n}i}\langle\boldsymbol{\varphi}_{f_l}, \boldsymbol{\varphi}_{f_n} \rangle.
 \label{LSAeq_average_energy1_new}
\end{eqnarray}
In general, $\langle\boldsymbol{\varphi}_{f_l}, \boldsymbol{\varphi}_{f_n} \rangle \neq \delta_{jl}$. Hence, the sum of squared modal amplitudes does not equal $\langle \boldsymbol{u}-\boldsymbol{u}_t, \boldsymbol{u}-\boldsymbol{u}_t \rangle$, because cross terms $a_{f_l}a^*_{f_n}$ ($l \neq j$) also contribute to the energy.
To resolve this, we consider the phase-averaged energy $\overline{\langle \boldsymbol{u}-\boldsymbol{u}_t, \boldsymbol{u}-\boldsymbol{u}_t \rangle}^{\alpha}$, which can be expressed as
 \begin{eqnarray}
\overline{\langle \boldsymbol{u}-\boldsymbol{u}_t, \boldsymbol{u}-\boldsymbol{u}_t \rangle}^{\alpha} &=& \sum_{l=-\infty}^{\infty}\sum_{n=-\infty}^{\infty} \overline{a_{f_l}a^*_{f_n}  e^{\alpha^{f_l}i} e^{- \alpha^{f_n}i}\langle\boldsymbol{\varphi}_{f_l}, \boldsymbol{\varphi}_{f_n} \rangle}^{\alpha}\nonumber \\
&=& \sum_{l=-\infty}^{\infty}\sum_{n=-\infty}^{\infty} a_{f_l}a^*_{f_n} \overline{e^{(\alpha^{f_l} - \alpha^{f_n})i}}^{\alpha}\langle\boldsymbol{\varphi}_{f_l}, \boldsymbol{\varphi}_{f_n} \rangle\nonumber \\
&=& \sum_{l=-\infty}^{\infty}\sum_{n=-\infty}^{\infty} a_{f_l}a^*_{f_n}\delta_{ln}\langle\boldsymbol{\varphi}_{f_l}, \boldsymbol{\varphi}_{f_n} \rangle\nonumber \\
&=&\sum_{l=-\infty}^{\infty}|a_{f_l}|^2\langle\boldsymbol{\varphi}_{f_l}, \boldsymbol{\varphi}_{f_l} \rangle.
 \label{LSAeq_average_energy2_new}
\end{eqnarray}
By choosing DMD modes normalized such that $\langle\boldsymbol{\varphi}_{f_l}, \boldsymbol{\varphi}_{f_l} \rangle = 1$, the sum of modal amplitudes becomes equal to the phase-averaged kinetic energy.

\section{Recipient-donor framework for transfer term}\label{recipient-donor}
In this Appendix, we outline the recipient-donor framework for energy transfer associated with the nonlinear term. Specifically, we consider a general nonlinear term
\begin{equation}
\left\langle (\boldsymbol{\hat{u}}_{f_{k-l}} \cdot \nabla)\boldsymbol{\hat{u}}_{f_{l}},\, \boldsymbol{\hat{u}}_{f_{k}} \right\rangle = \int_{\Omega} \boldsymbol{\hat{u}}^H_{f_{k}}\left(\boldsymbol{\hat{u}}_{f_{k-l}} \cdot \nabla\right)\boldsymbol{\hat{u}}_{f_{l}}\, d\boldsymbol{x},
\end{equation}
where $\boldsymbol{\hat{u}}_{f_l}$, $\boldsymbol{\hat{u}}_{f_{l-k}}$, and $\boldsymbol{\hat{u}}_{f_{k}}$ are frequency components with relation to triadic interaction.

For arbitrary vector fields $\boldsymbol{a}$, $\boldsymbol{b}$, $\boldsymbol{c}$ the following pointwise identity holds:
\begin{equation}
\boldsymbol{b}^H(\boldsymbol{c}\cdot\nabla)\boldsymbol{a}
= \nabla\cdot\!\big( (\boldsymbol{b}^H \boldsymbol{a})\,\boldsymbol{c} \big)
 - (\nabla\cdot\boldsymbol{c})\,(\boldsymbol{b}^H\boldsymbol{a})
 - \boldsymbol{a}^H(\boldsymbol{c}^*\cdot\nabla)\boldsymbol{b}.
\label{eq:key_identity_pointwise}
\end{equation}
This follows from the product rule applied to $\nabla\cdot\!\big( (\boldsymbol{a}\cdot\boldsymbol{b})\,\boldsymbol{c} \big)$.
Integrating \eqref{eq:key_identity_pointwise} over $\Omega$ and using the divergence theorem gives
\begin{align}
\int_{\Omega} \boldsymbol{b}^H(\boldsymbol{c}\cdot\nabla)\boldsymbol{a}\,d\boldsymbol{x}
&= \int_{\partial\Omega} (\boldsymbol{c}\cdot\boldsymbol{n})(\boldsymbol{b}^H\boldsymbol{a})\,d\boldsymbol{s}
   - \int_{\Omega} (\nabla\cdot\boldsymbol{c})\,(\boldsymbol{b}^H\boldsymbol{a})\,d\boldsymbol{x}
   - \int_{\Omega} \boldsymbol{a}^H(\boldsymbol{c}^*\cdot\nabla)\boldsymbol{b}\,d\boldsymbol{x},
\label{eq:key_identity_integrated}
\end{align}
where $\partial\Omega$ denotes boundary of $\Omega$, $\boldsymbol{n}$ is normal vector of boundary, and $d\boldsymbol{s}$ denotes the line element.
Setting $\boldsymbol{a}=\boldsymbol{\hat{u}}_{f_l}$, $\boldsymbol{b}=\boldsymbol{\hat{u}}_{f_k}$ $\boldsymbol{c}=\boldsymbol{\hat{u}}_{f_{k-l}}$ in \eqref{eq:key_identity_integrated} yields
\begin{align}
\int_{\Omega} \boldsymbol{\hat{u}}_{f_k}^H(\boldsymbol{\hat{u}}_{f_{k-l}}\cdot\nabla)\boldsymbol{\hat{u}}_{f_l}\,d\boldsymbol{x}
&= \int_{\partial\Omega} (\boldsymbol{\hat{u}}_{f_{k-l}}\cdot\boldsymbol{n})(\boldsymbol{\hat{u}}_{f_l}^H\boldsymbol{\hat{u}}_{f_k})\,d\boldsymbol{s}\nonumber \\
   - \int_{\Omega} (\nabla\cdot\boldsymbol{\hat{u}}_{f_{k-l}})\,&(\boldsymbol{\hat{u}}_{f_l}\cdot\boldsymbol{\hat{u}}_{f_k})\,d\boldsymbol{x}
   - \int_{\Omega} \boldsymbol{\hat{u}}_{f_l}^H(\boldsymbol{\hat{u}}_{f_{l-k}}\cdot\nabla)\boldsymbol{\hat{u}}_{f_k}\,d\boldsymbol{x}.
\label{eq:key_identity_integrated_uhat}
\end{align}

If the  is divergence free in $\Omega$, that is $\nabla\cdot\boldsymbol{\hat{u}}_{f_{k-l}}=0$, then second term of right-hand side drops out, and   \eqref{eq:key_identity_integrated_uhat} simplifies to
\begin{align}
\left\langle (\boldsymbol{\hat{u}}_{f_{k-l}} \cdot \nabla)\boldsymbol{\hat{u}}_{f_{l}},\, \boldsymbol{\hat{u}}_{f_{k}} \right\rangle
= F^b(\boldsymbol{\hat{u}}_{f_{l}}, \boldsymbol{\hat{u}}_{f_{l-k}}, \boldsymbol{\hat{u}}_{f_{k}}) 
   - \left\langle (\boldsymbol{\hat{u}}_{f_{l-k}} \cdot \nabla)\boldsymbol{\hat{u}}_{f_{k}},\, \boldsymbol{\hat{u}}_{f_{l}} \right\rangle,
\label{eq:advec_pairing_incomp}
\end{align}
where first term of right hand side is boundary flux term as follows
\begin{align}
F^b(\boldsymbol{\hat{u}}_{f_l}, \boldsymbol{\hat{u}}_{f_{k-l}}, \boldsymbol{\hat{u}}_{f_k}) = \int_{\partial\Omega} (\boldsymbol{\hat{u}}_{f_{k-l}}\cdot\boldsymbol{n})(\boldsymbol{\hat{u}}_{f_l}^H\boldsymbol{\hat{u}}_{f_k})\,d\boldsymbol{s}.
\label{eq:advec_pairing_incomp}
\end{align}

When the boundary flux term vanishes, the quantities $\left\langle (\boldsymbol{\hat{u}}_{f_{k-l}} \cdot \nabla)\boldsymbol{\hat{u}}_{f_{l}},\, \boldsymbol{\hat{u}}_{f_{k}} \right\rangle$ and $\left\langle (\boldsymbol{\hat{u}}_{f_{l-k}} \cdot \nabla)\boldsymbol{\hat{u}}_{f_{k}},\, \boldsymbol{\hat{u}}_{f_{l}} \right\rangle$are equal in magnitude but opposite in sign. This relationship between the nonlinear terms reflects the exchange of energy between $\boldsymbol{\hat{u}}_{f_{l}}$ and $\boldsymbol{\hat{u}}_{f_{k}}$, with $\boldsymbol{\hat{u}}_{f_{k-l}}$ acting as a catalyst in the process. The boundary flux contribution becomes negligibly small when the domain boundaries are taken sufficiently far, provided that $l \neq k$.

For the case $l = k$, we denote the boundary flux term by 
\begin{align}
F^b(\boldsymbol{\varphi}_{f_l}, \boldsymbol{u}_{b}, \boldsymbol{\varphi}_{f_l}) =  \int_{\partial\Omega} (\boldsymbol{u}_{b}\cdot\boldsymbol{n})(\boldsymbol{\varphi}_{f_l}^H\boldsymbol{\varphi}_{f_l})\,d\boldsymbol{s}.
\end{align}
 This term vanishes when the normal component of $\boldsymbol{\hat{u}}_{f_{k-l}}$ at the boundary is zero. Consequently, along the cylinder wall, $\boldsymbol{\hat{u}}_{f_{k-l}}\cdot\boldsymbol{n}$ vanishes for all frequencies $\boldsymbol{\hat{u}}_{f_{k-l}}$. At inflow and outflow boundaries, $\boldsymbol{\hat{u}}_{f_{k-l}}\cdot\boldsymbol{n}$ does not vanish when $\boldsymbol{\hat{u}}{f_{k-l}}$ corresponds to base flow, including the main flow, the term acquires a finite value. Thus, 
 \begin{align}
\left\langle (\boldsymbol{u}_{b} \cdot \nabla)\boldsymbol{\varphi}_{f_l},\, \boldsymbol{\varphi}_{f_l} \right\rangle
= \frac{1}{2} F^b(\boldsymbol{\varphi}_{f_l}, \boldsymbol{u}_{b},\boldsymbol{\varphi}_{f_l}).
\label{eq:advec_pairing_incomp}
\end{align}
This expression indicates that the base flow mediates the flux of $\boldsymbol{\varphi}_{f_l}^H \boldsymbol{\varphi}_{f_l}/2$ across the boundary, representing the inflow and outflow of energy associated with the $f_l$-frequency eigenmode.

\section{\label{apenc}Convergence study of tDMDpc}
For tDMDpc, $j_\text{max}$ is the number of snapshots in the eigenmode extraction using DMD. The validity of eigenmode extraction for time-dependent linear operators can be evaluated by the fact that the eigenmodes remain constant when the number of snapshots is sufficiently large. First, we check that the average field over $\alpha$ does not change as increasing  $j_\text{max}$. 

\begin{figure}
	\centering\includegraphics[width=12cm,keepaspectratio]{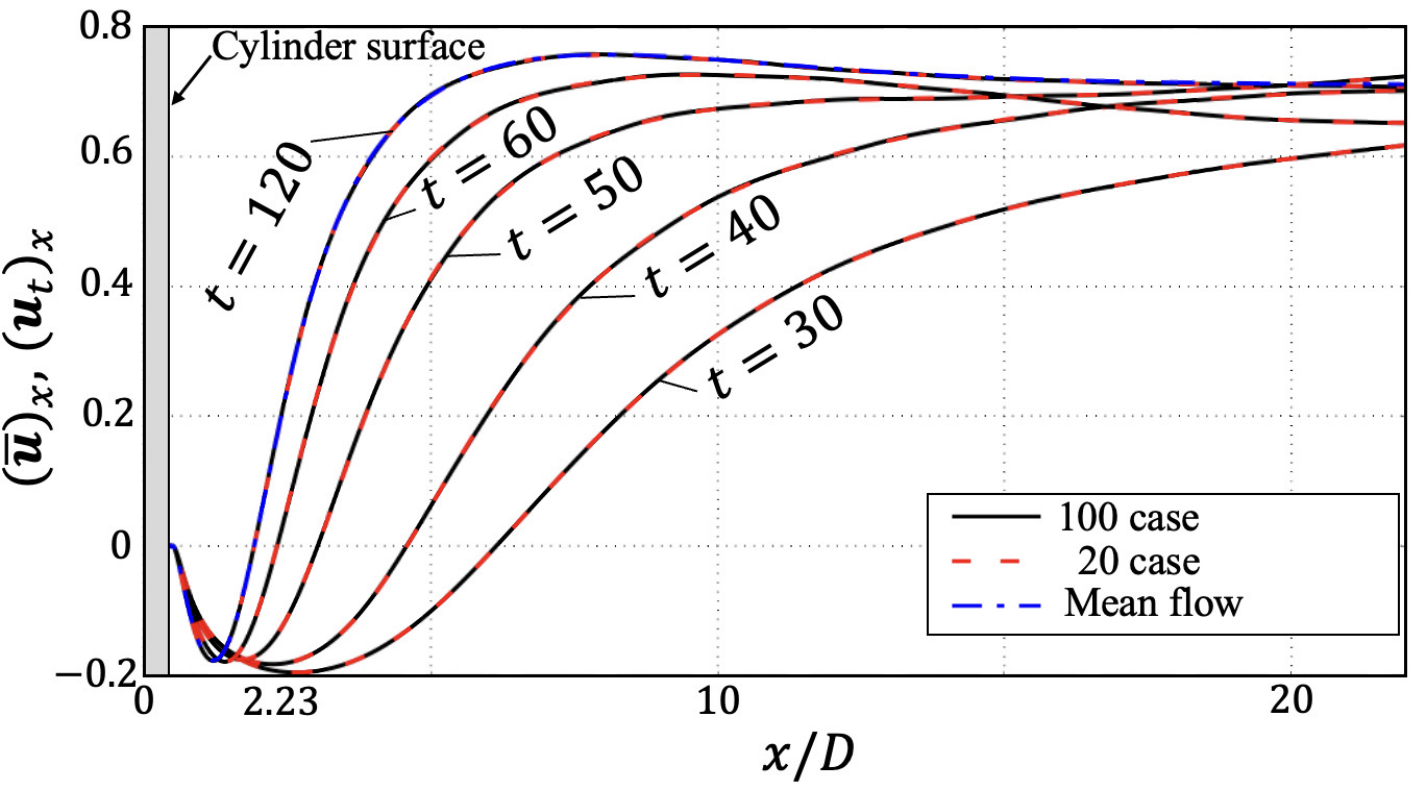}
  \captionsetup{justification=raggedright,singlelinecheck=false}
  \caption{Average-fields of phase-controlled transient flow at $\Rey=100$. Flow fields are averaged over $20$ and $100$ cases at $t=30$, $40$, $50$, $60$, $80$, and $120$. The blue dash-dot line shows the mean fields of the post-transient flow field, which is obtained from time-averaging of fully developed periodic flow fields.}
 \label{fig_tDMDpc_average}
\end{figure}

Figure \ref{fig_tDMDpc_average} shows the time variation of the average field over $\alpha$ at $y/D=0$ for $j_\text{max}=20$ and $100$. The average fields during the development process are completely consistent between $20$ and $100$ cases, and the average field for $j_\text{max}=20$ cases is well converged. To quantitatively evaluate whether the flow field reaches a periodic state after the transient development, the mean field of the fully developed periodic flow is shown by the blue dotted line. Note that the fully developed periodic flow refers to the flow fields at $t \geq 2000$, averaged over $t=2000$--$2300$, and not averaged over $\alpha$.  Moreover, after a sufficient time evolution, the average field over $\alpha$ exhibits a reasonable distribution, as it coincides with the mean-field represented by the blue dotted line.

We check the convergence of the eigenmodes of the non-$0$ frequencies of the time-dependent linear operator. Turning to the mode extraction process in the DMD, the eigenmodes are computed by projecting the eigenmodes of the low-dimensional linear operator $\tilde{A}$ by $U^T_r$. Since $U_r$ is determined by the SVD of the dataset, the convergence of the matrix projected by $U_r$ is confirmed by the singular values. Then, in a form independent of the number of snapshots, we define 
\begin{eqnarray}
\gamma_k  \stackrel{\mathrm{def}}{=}  \frac{\sigma_k^2}{(M-1)},
\label{gammak}
\end{eqnarray}
where $\sigma_k$ is the singular value of matrix $X$, thus, $k$th diagonal elements of $S_r$ in equation (\ref{SVD}), and $M$ is the number of snapshots in the dataset. Here, $\gamma_k$ equals the eigenvalue of the variance-covariance matrix $XX^T/\sqrt{M-1}$ and POD eigenvalue. 
The convergence of $\gamma_k$ with respect to snapshot number means the convergence of the major features of the snapshots that make up the matrix $X$ and ensures a sufficient number of snapshots. This also suggests the DMD modes computed using the SVD results for $X$ converge.

\begin{figure}
	\centering\includegraphics[width=12cm,keepaspectratio]{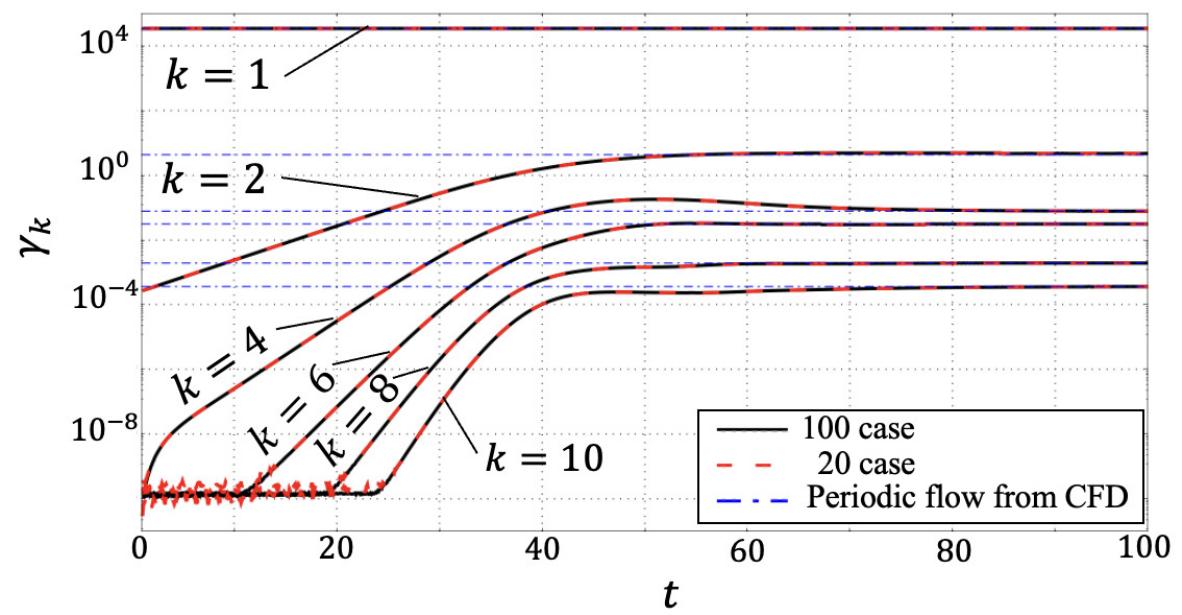}
  \captionsetup{justification=raggedright,singlelinecheck=false}
  \caption{Square quantities of the singular value obtained from the SVD for matrix $X$ in the DMD algorithm. All singular values were normalized by snapshot number, indicated in (\ref{gammak}).  The blue dash-dot line is the $\gamma_k$ computed from fully developed periodic flow fields.}
 \label{fig_singular}
\end{figure}

Figure \ref{fig_singular} shows the time variation of $\gamma_k$ for $j_\text{max}=20$ and $100$ cases. The $\gamma_k$ of DMD applied to the fully developed periodic flow is shown by the blue dotted line. Since singular values appear in pairs--$k = 2$ and $3$, $k=4$ and $5$, $\cdots$--the one with the larger is plotted in the figure. $k=1$ corresponds to a mean field with no pair $\gamma_k$ and, as shown in figure \ref{fig_tDMDpc_average}, has the same value at $j_\text{max}=20$ and $100$. The time variation of the value of $\gamma_1$ appears to be almost negligible because the energy of the main flow is very large with respect to the transient variation for the cylinder backward. $k=2$ corresponds to the most dominant eigenmode and is well converged at all times. For $k\geq4$, in the early stage of development, $\gamma_k$ value is very small and is considered to be affected by the errors in the numerical calculation of SVD. In the early stage, the spatial structure corresponding to $k\geq4$ is too small or does not exist because the mode corresponding to $k\geq4$ is not added to the disturbance at $t=0$. However, after sufficient development, $j_\text{max}=20$ and $100$ are close agreement. Therefore, $j_\text{max}=20$ is large enough for capturing the eigenmode of the time-dependent linear operator.

\backsection[Acknowledgments]{We thank Prof. Oliver T. Schmidt for pointing out the convective instability.
We gratefully acknowledge Dr. Yasuhito Okano for providing valuable knowledge about LSA. We thank Yuta Iwatani for the valuable discussions on nonlinear interactions and resolvent analysis and for pointing out the bi-orthogonality of the DMD mode. }

\backsection[Funding]{The numerical simulations were performed on the supercomputer systems “AFI-NITY” and “AFI-NITY II” at the Advanced Fluid Information Research Center, Institute of Fluid Science, Tohoku University, and JAXA Supercomputer System Generation 3 (JSS3). 
This study was partially supported by a Sasakawa Scientific Research Grant from the Japan Science Society. This study was partially supported by JST SPRING, Grant Number JPMJSP2114, Japan. }

\backsection[Declaration of interests]{The authors report no conflict of interest.}


\backsection[Author ORCIDs]{\\Y. Nakamura, https://orcid.org/0009-0008-4118-1078;\\Y. Kuroda,  https://orcid.org/0009-0002-4578-5887;\\ S. Sato, https://orcid.org/0000-0002-9979-0051;\\ N. Ohnishi, https://orcid.org/0000-0001-5895-0381
}


%

\bibliographystyle{jfm}

\end{document}